\documentclass[showpacs,showkeys,amsmath,amssymb,aps,twocolumn,prx]{revtex4-1}
\usepackage{graphicx}
\usepackage{dcolumn}
\usepackage{bm}
\begin{document}

\preprint{}

\title{Modeling magnetosensitive ion channels in 
viscoelastic environment of living cells}

\author{Igor Goychuk}
 \email{igoychuk@uni-potsdam.de}
 
\affiliation{Institute for Physics and Astronomy, University of Potsdam, 
Karl-Liebknecht-Str. 24/25, 14476 Potsdam-Golm, Germany}

\date{\today}

\begin{abstract}

We propose and study a model of hypothetical magnetosensitive ionic channels 
which are long thought to be a possible candidate to explain the influence 
of weak magnetic fields on living organisms ranging from magnetotactic bacteria
to fishes, birds, rats, bats and other mammals including humans.
The core of the model is provided by a
short chain of magnetosomes serving as a sensor which is coupled
by elastic linkers to the gating elements of ion channels forming a small cluster
in the cell membrane. The magnetic sensor is fixed by one end on 
cytoskeleton elements attached to the membrane and is 
exposed to viscoelastic cytosol. Its free end can reorient stochastically 
and subdiffusively  in viscoelastic cytosol responding to external 
magnetic field changes and open the gates of coupled 
ion channels. The sensor
dynamics is generally bistable due to bistability of the gates which can be
in two states with probabilities which depend on the sensor orientation.
For realistic parameters, it is shown that this model channel can operate in 
the magnetic
field of Earth for a small number (5 to 7) 
of single-domain magnetosomes 
constituting the sensor rod  each of which has
a typical size  found in magnetotactic bacteria and other
organisms, or even just one sufficiently large nanoparticle of a characteristic
size  also found in nature.  
It is shown that due to viscoelasticity of medium
the bistable gating dynamics generally exhibits power law and
stretched exponential distributions of the residence times
of the channels in their open and closed states. This provides a generic
physical mechanism for explanation of the origin of such anomalous
kinetics for other ionic channels whose sensors move in viscoelastic environment
provided by either cytosol or biological membrane, in a quite general
context, beyond the fascinating hypothesis of magnetosensitive ionic channels
we explore.    

\end{abstract}
\maketitle

\section{Introduction}

Evidences of the influence of weak electromagnetic fields including 
magnetic fields on many living
organisms are abound \cite{Webb,Devyatkov74,Grunder,
Grissom,BinhiBook,BinhiSavin03,Devyatkov,Barnes,Pikov}. Most spectacular and doubtless
of such manifestations are related to navigation of honeybees, fishes,
birds, rats, bats, and other animals \cite{Kirschvink85,Kirschvink81,Kirschvink92,Wilchko05,Grissom} 
in the magnetic field of Earth of about
$B_e=50\;\mu T$  in strength. Various
hypotheses related to profoundly different physical mechanisms
have been suggested to explain this influence. They range from non-thermal
quantum mechanisms related to spin-dependent electron transfer 
\cite{Schulten,Grissom},
which  circumvent the so-called "kT" problem \cite{Barnes,BinhiBook,Adair}, to 
a variety of classical mechanisms based on a widespread occurrence of
biomagnetite nanoparticles in tissues of many living organisms starting from 
magnetotactic bacteria \cite{Blackmore,Faivre,Devouard,Abracado} and ending by 
the human brain \cite{KirschvinkPNAS}. 
Magnetite ($\rm Fe_3O_4$) has a saturation magnetization of 
$M_s=4.8\cdot 10^5$ A/m, and
elongated magnetite nanoparticles  are in a single 
domain ferrimagnetic state when
their sizes range from 20 nm to 200 nm depending on the short-to-long axis aspect ratio
(shape factor) \cite{Kirschvink81}.
Hence, magnetic energy of a spherical magnetosome 
(magnetic nanoparticle dressed in a lipid-protein
membrane shell \cite{Faivre}) of the radius 100 nm with magnetic
dipole moment $\mu\approx 2{\rm \;fAm^2}$ is as large as  
$E_B=\mu B_e\sim 24.5\;k_BT$, when
its magnetic moment is aligned with the Earth magnetic field. 
The whole cells with magnetic moments in the range $\mu\sim 4-100{\rm \;fA m^2}$
were identified recently as candidate magnetoreceptor cells
in trout olfactory epithelium \cite{Eder}. Another recent study \cite{Edelman}
confirms the existence of cells possessing such large magnetic moments both
in trouts and pigeons. However, it expresses doubts that this magnetic
moment is caused by biomagnetite,
and not by magnetic nanoparticles polluting environment, which are
absorbed to the cell membranes. In this respect, the presence
of iron-rich organelles filled by ferrihydrate in the hair cells of pigeons \cite{Edelman}
can spell out in support of biological origin of magnetite therein, rather than on contrary.
Even though the biochemical pathway of biomagnetite
synthesis remains still controversial \cite{Faivre}, the presence of ferrihydrate indirectly
supports one of biochemical schemes earlier suggested \cite{Kirschvink85}.
Furthermore, 
about 10\% of biomagnetite
particles found in human brain \cite{KirschvinkPNAS,Kirschvink92} 
(about $10^6$ or even $10^8$ per one gram of tissue 
in grey matter and in meninges, correspondingly, and about 50 ng/g in 
hippocampus on average \cite{Dobson})
are about or larger than 100 nm in size. In bacteria,
the length of elongated magnetoparticles can reach $110$ nm with
shape factor $0.8-0.9$ \cite{Devouard} and even larger, up to 200 nm \cite{Abracado}.
Important is that genes encoding magnetosome specific proteins in some
bacteria were identified \cite{Faivre}. This provides one of the strongest argumentation
in the favor of biological origin of magnetite in biological cells.  

Importantly, the magnitude
of own magnetic field produced by a spherical magnetosome can reach (at peak of a
 highly anisotropic distribution)
$B_{\rm mag}\sim 402\;{\rm mT}$ near to its surface (see in Appendix \ref{append1}),
independently of its radius. This is about $8000\times$ larger than $B_e$.
Thus, quantum magnetic effects can also be mediated by the magnetic near-field
of a magnetosome reorienting in external magnetic field and positioned
nearby an electron-transferring magneto-sensitive molecular complex, 
rather than directly by an external
field itself \cite{Binhi05}.
The particles of intermediate size $55\times 44\times 44$ nm, typical for magnetosomes
met in magnetotactic bacteria \cite{Vander} have $E_B\sim 0.623\;k_BT$, and they easily
make chains joined by magnetic cohesion force, being rigid enough \cite{Shcherbakov} 
in not too strong magnetic fields, smaller than $30\;{\rm mT}$ \cite{Koernig}.
Hence, a rod consisting of 5 such nanoparticles with magnetic moments 
aligned has $E_B\sim 3.12\;k_BT$, and it can easily serve as a quite classical, compass-like
sensor element for magneto-sensitive ion channels, as it will be shown in this paper. 
When external field becomes  compatible with $B_{\rm mag}$ it can disrupt
the magnetosome chain, or make it unstable \cite{Koernig}. This is why such a 
sensor will not work in too strong fields. It will be literally broken
in pieces, presenting a disordered cluster of nanoparticles. For another typical size 
$103\times 85\times 85$ nm  also commonly met in bacteria
 \cite{Devouard}, $E_B\sim 4.23\;k_BT$ for a single magnetosome.
Hence, sensor can consist also of a single nanoparticle. 

The idea that a magnetic nanorod can serve as a sensor and transducer of magnetic 
field torque has been first suggested by Yorke \cite{Yorke}.
Kirschvink \cite{Kirschvink81,Kirschvink92} proposed that it can be a magnetosensitive
ion channel involved with a spherical magnetosome attached to a cytoskeleton element 
nearby an ion channel in biological membrane and coupled by an elastic linker
to the ion channel gating machinery. External magnetic field creates a torque
on the magnetosome, which rotates and opens the ion channel. The sensor is considered to 
be essentially monostable in the Kirschvink model. This model has been refined recently
for the chain of magnetosomes serving as sensor \cite{Winklhofer}, but remained 
monostable as in the original proposal. 
Binhi and Chernavsky proposed a different model 
\cite{Binhi05}, based on bistability of magnetosome rotations induced 
by magnetic field for a spherical  magnetosome elastically coupled to cytoskeleton.
It is not related to gating of ionic channels. Rather a change 
in distribution of the magnetic field induced by magnetosome is of interest,
in the context of a related quantum mechanism \cite{Binhi05,Yan}. Furthermore, 
it has been  shown recently
that streptavidin-linked magnetite
nanoparticles of $50$ nm size (in radius on average) can 
induce ion channel like activity
being absorbed on phospholipid bilayer \cite{Mohanta}. The corresponding
ion current recordings remind somewhat the ion channel activity induced in electric fields 
by alamethicin peptides inserted into the membrane \cite{Bezrukov}.

In this paper a further generalization of the model by Kirschvink \textit{et al.} 
is suggested and studied. The generalization consists in several profound 
aspects. First,
stochastic motion of  sensor is considered to be  bistable because of bistability
of ion channel gate to which the sensor is coupled. Such a bistability is a common
point in describing stochastic dynamics of ionic channels \cite{JacksonBio}. To include it in the 
sensor dynamics we adopt a gating spring model \cite{Hudspeth} 
assuming that
the gate can take on just two conformations, open and closed. 
Similar model has originally been suggested in relation
to the hair cell dynamics. 
Second, we consider
a possibility that magnetic sensor can be coupled to gates of several ionic
channels making a cluster, i.e. that a compact cluster of ionic channels and 
a magnetic nanoparticle serving as sensor make a magneto-sensitive complex
in biological membrane. This can explain why such magnetosensitive ion channels,
as separate units,
were not found thus far. 
Third, and most important in a more general context of gating stochastic dynamics
of ionic channels: We consider the influence of viscoelasticity of the
environment in which sensor is moving on the sensor dynamics.
We show that viscoelasticity alone can result in profoundly non-exponential residence
time distributions of the channels in their open and closed states
such as stretched exponential distribution \cite{Liebovich,Croxton} and power law distributions
\cite{Liebovich,Lauger,Millhauser,Sansom,Gorczynska,Qu}. This explanation of unusual gating kinetics 
is different from other
physical theories suggested thus far which are based, in particular, on a complex
free energy landscape for sensor or conformational 
dynamics of the whole ion channel 
with huge many multiply degenerated minima and maxima 
(glassy like dynamics) \cite{Austin,Frauenfelder,Rappaport}. 
The latter one can be modeled in a simplest possible fashion as a continuous
normal diffusion in a potential
box, which already allows to explain \cite{GoychukPNAS} the origin of $-3/2$ power law
in the distribution of closed times \cite{Sansom,Millhauser} in conjunction with the origin of
Hodgkin-Huxley voltage dependence \cite{GoychukPNAS,JacksonBio}. 
Such normal diffusion can become also anomalously slow
(fractional diffusion) -- a modeling pathway explored in \cite{GoychukPRE04}. 

Recent work \cite{Goychuk09,Goychuk12}
suggests, however, that the discussed  anomalous kinetics can also result from 
standard bistable
dynamics of sensor, commonly assumed in biophysics textbooks \cite{JacksonBio},
as a result of memory effects caused by viscoelasticity
of the environment. This is a very appealing and simple physical
explanation indeed. In this respect, both cytosol and plasma membrane
are viscoelastic \cite{Goychuk12,Jeon12}. Hence viscoelasticity is considered as a major
cause of experimentally observed anomalous diffusion in crowded colloidal and polymer 
solutions \cite{Mason,Amblard,Caspi,Mizuno,Waigh,Szymanski,Holek}, and living cells
\cite{Weiss,Weber,Wilhelm,Bruno}, as well as in single
protein molecules \cite{Kou}. 
To study such effects, the approach of 
generalized Langevin equation (GLE) characterized by
 a power law scaling memory kernel and power law correlated thermal noise 
 of environment, which are 
 related by fluctuation-dissipation theorem (FDT), provides a major well-established
 theoretical framework in the case of linear dynamics \cite{Mason,Waigh}. 
It is not easy to generalize this framework 
 towards nonlinear dynamics in bi- and multistable potentials.
 For example, a corresponding \textit{exact} Fokker-Planck description 
 which would mirror and
 complement the GLE
 approach, like in the case of memoryless dynamics,
  is simply not developed thus far for potentials other than 
 linear and parabolic \cite{Adelman}. In other words, any nonlinearity creates a problem for Fokker-Planck
 description of such a dynamics with memory \cite{GH07,Goychuk09,Goychuk12}. 
 Recently, we bypassed such
 difficulties within the GLE approach using the road of multi-dimensional Markovian
 embedding of GLE dynamics within a generalized Maxwell-Langevin model of 
 viscoelasticity \cite{Goychuk09,Goychuk12}. It has also been generalized 
 to include negative correlations of stochastic force and corresponding memory effects
 leading to superdiffusion and supertransport \cite{Siegle}.
 The utility of this approach
 has been demonstrated on various basic models of nonlinear stochastic
 dynamics such as bistable dynamics \cite{Goychuk09}, washboard dynamics \cite{Goychuk09,GH11,Goychuk12},
 anomalous rocking ratchets \cite{Goychuk10,Goychuk12,GKh12,GKh13,KhG13}, anomalous
 flashing  ratchets \cite{KhG12}, and also  in applications to molecular motor
 dynamics in viscoelastic cytosol \cite{GKhMet14a,GKhMet14b,Goychuk15}. 
 
 In the context of magnetosomes dynamics in viscoelastic cytosol Kirschvink 
 \textit{et al.} have repeatedly taken the cytosol influence into account
 by enhancement of the coefficient of  \textit{normal} viscous friction
 experienced by magnetosome by a factor 
 of about $100$ \cite{Kirschvink92,Eder}. 
 This was a standard way to think about influence of viscoelasticity and 
 crowding in cytosol in biological applications until recently \cite{Luby}. 
 However, recent results on anomalous diffusion of nano- and 
 submicron particles in living cells suggest that this enhancement can be much larger, of the
 order of $1000$, and even larger \cite{Weiss,GoychukPRE12,GKhMet14a}, depending, in particular,
 on the size of particle \cite{Odijk}. Cytosol seem to normally operate
  at the edge of a phase transition from liquid-like state to a solid-like 
  state with broken ergodicity \cite{Parry}.  
 Moreover, even for the enhancement factor
 $100$ the bistable orientational dynamics of magnetic nanoparticles 
 would be so slow, as we show in this paper, that it would
 be completely out of interest within a biological context. However,
 a major effect, which introduces viscoelasticity, is emergence 
 of transient subdiffusion  which is much faster than 
 the asymptotic limit of normal diffusion.
 In fact, this asymptotic regime can become completely
 irrelevant for mesoscopic dynamics. This paradoxical fact also follows from 
 the non-Markovian rate theory developed beyond the standard memoryless Kramers theory 
 \cite{Kramers,Grote,HanggiMojtabai,Nitzan,Pollak}, 
 see \cite{HTB90} for a review, and below.
 Namely this ``much faster, not slower''
 \cite{GoychukFNL,GoychukPRE12,GKh14PRL}, paradoxically due to subdiffusion,
  makes operating of such bistable magnetic sensors possible in viscoelastic
  cytosol.  This provides one of most important results of this work, which lends
 a further support to the idea of magnetic field sensing with classical 
  dynamics of sufficiently large biomagnetite nanoparticles.

\section{Model and Theory}
 
\begin{figure}
\vspace{1cm}
\resizebox{0.95\columnwidth}{!}{\includegraphics{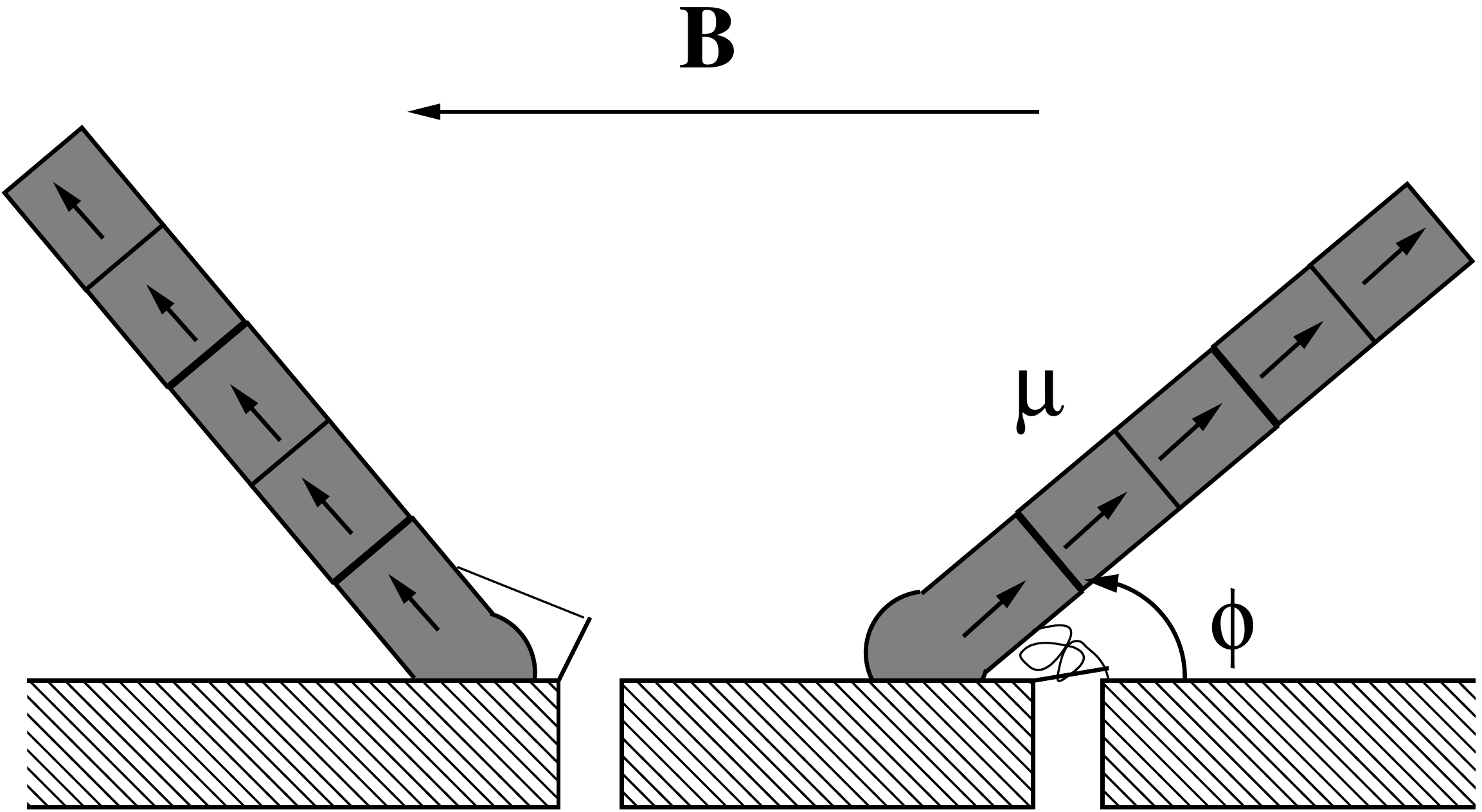}}
\caption{Cartoon of the considered model of magnetosensitive ion channel
expressing the idea. Proportions are not held. Rotation of nanorod is coupled
via flexible linkers to gates of ion channels, which can form a cluster. One
channel is shown for simplicity. }
\label{Fig1}       
\end{figure}

\begin{figure}
\vspace{1cm}
\resizebox{0.95\columnwidth}{!}{\includegraphics{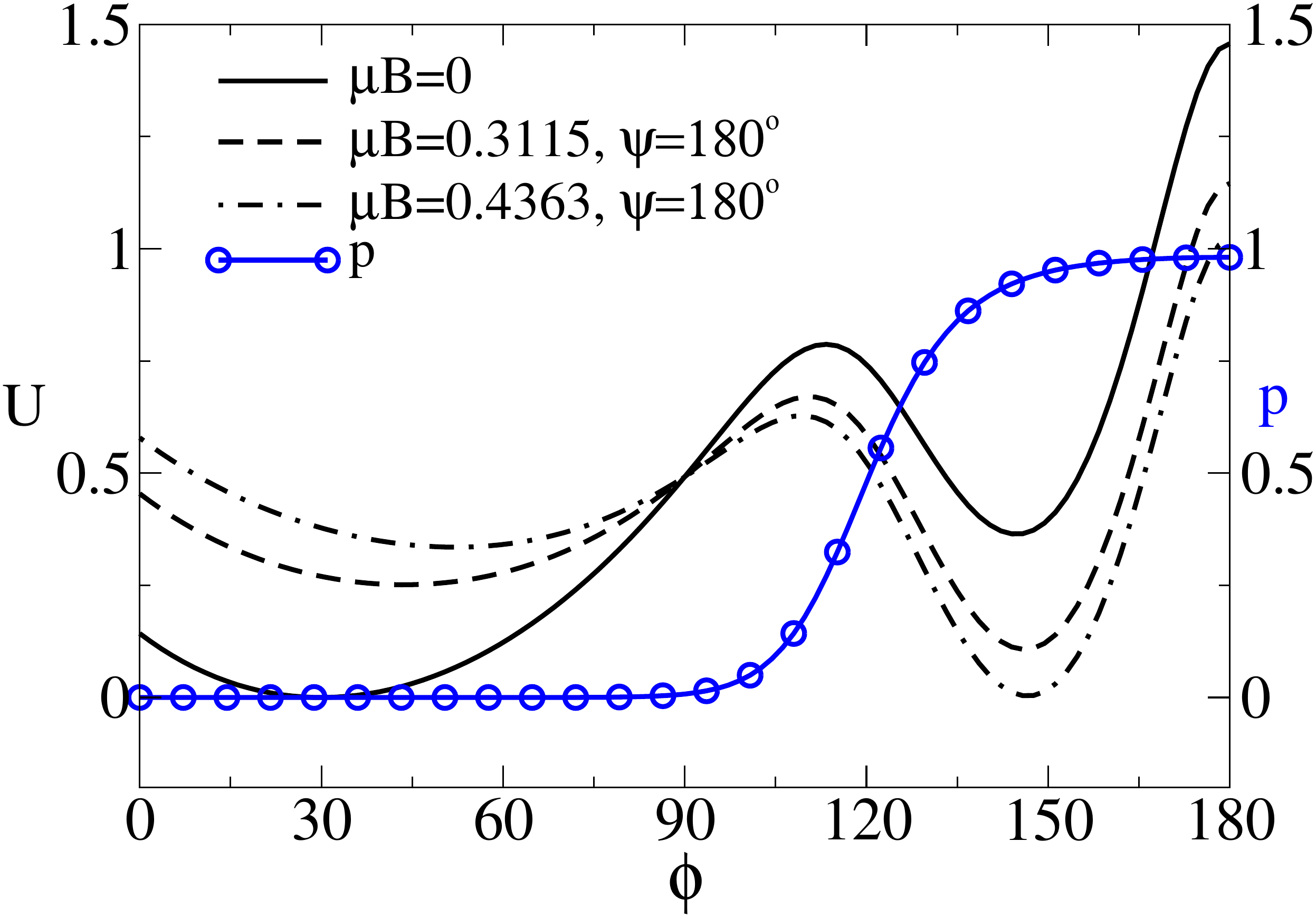}}
\caption{(color online) Orientational potential $U(\phi)$ and opening probability 
$p(\phi)$ (blue line with circles) of ion channel versus angle $\phi$ in degrees. Potential box
walls are imposed for $\phi<0$ and $\phi>\pi$ restricting motion to one
side of membrane. Magnetic field is applied
at the angle $\psi=180^{\rm o}$, and $\phi_0=30^{\rm o}$. The model parameters: $T=0.1$, 
$l_{\rm max}=1.5$, $f_0=1.5$, $l_0=1.22$, and $m=7$ channels in the sensor cluster.
The gating energy difference $\epsilon_2-\epsilon_1=f_0 l_0=1.83$, or $18.3$ in units 
of $k_BT_r$.}
\label{Fig2}       
\end{figure}

We consider the following model. Biomagnetite rod made of a chain of magnetosomes
of total length $L$, 
or a single elongated magnetic nanoparticle can rotate  with one end fixed e.g. on
on a cytoskeleton meshwork attached to the cell membrane inside the cell (see Fig. \ref{Fig1} for an idea).
It is also elastically  attached to the gates of ionic channels (one is shown) by flexible linkers, 
which are modeled here within a finite extensible nonlinear elastic (FENE) model \cite{FENE}. 
The channel gate or rather a molecular latch, which fixes the gate
in its either open or closed state, can be in two states.  The closed state is 
characterized by the 
energy $\epsilon_1$, and the open one has the energy $\epsilon_2-f_0 x$, 
which depends
on the linker elongation $x$, where $f_0$ is a force constant characterizing
the strength of coupling (force exerted by the linker on the gate).
Elastic energy is $U_{\rm FENE}(x)=
-\frac{1}{2}k l_{\rm max}^2\ln(1-x^2/l_{\rm max}^2)$ within FENE model,
where $k$ is elastic spring constant, and $l_{\rm max}$ is the maximal extension
length of the linker, when it is fully stretched.
Statistical mean force exerted by the channel gate  on the linker can
be found as $f(x)=-dG(x)/dx$ from the potential
of mean force  $G(x)=-k_BT\ln Z(x)$,  where $Z(x)=\exp[-\beta\epsilon_1]+
\exp[-\beta(\epsilon_2-f_0x)]$ is the statistical sum of gate, and $\beta=1/(k_BT)$
is inverse temperature. Here, one implicitly assumes that the own gate/latch dynamics 
(transitions between
two states) is very fast, and, hence the actual gating dynamics is enslaved by the sensor dynamics
and reflects the latter one.
Mean force is 
$f(x)=f_0p(x)$, where
\begin{eqnarray}\label{prob}
p(x)=\frac{1}{1+\exp[-f_0(x-l_0)/(k_BT)]},
\end{eqnarray}
 is probability of the gate to be open and $l_0=(\epsilon_2-\epsilon_1)/f_0$. In order to define 
some $x_0$ as equilibrium point, we
following  \cite{Hudspeth} redefine mean force by a shift  as 
$f(x)=f_0[p(x)-p(x_0)]$. The motion of rod is assumed to be restricted to the plane orthogonal
to the membrane, and
characterized by the angle $\phi$, $0\leq \phi\leq \pi$ counted from the 
membrane plane in 
the counter-clockwise direction. The linker elongation is approximated as
$x(\phi)=2l[\sin(\phi/2)-\sin(\phi_0/2)]$, where $l$ is the rotation arm,
and $\phi_0$ is an equilibrium angle. The external magnetic field $B$ is directed
at the angle $\psi$ within the plane of motion. The potential of mean force
 (torque) acting on the rod
in our model is 
\begin{eqnarray}\label{potential}
U(\phi)&=&
-\frac{1}{2}k l_{\rm max}^2\ln \left\{ 1-\left[ x(\phi)/l_{\rm max}\right ]^2\right\}
\nonumber \\
&- &k_BT m \ln \left \{ 1+  \exp[f_0( x(\phi)-l_0)/(k_BT) ]\right\} \nonumber \\
&+& mf_0p(\phi_0) x(\phi)-\mu B\cos(\psi-\phi),
\end{eqnarray}
where $p(\phi_0)=p(x=2l\sin(\phi_0/2))$, $\mu$ is the magnetic moment of the rod,
and $m$ is the number of ion channels
coupled to sensor and treated in a mean-field fashion (all gates move
synchronously enslaved by the same sensor).
We shall scale the energy in units of $U_0=kl^2$, temperature in the units of 
$U_0/k_B$, distances in units of $l$, and forces in units of $f_u=U_0/l$. $U_0$ will
be fixed to $U_0=10\;k_BT_r \approx{\rm 41 pN\cdot nm}\approx 0.25\;\rm eV$,
where $T_r$ is a typical room temperature, $T_r\approx 297$ K. For a typical
$k=0.3\; \rm pN/nm$ \cite{Kojima}, this corresponds to $l\approx 11.69$ nm and force units 
$f_u\approx 3.51$ pN. For the purpose of illustration,
magnetic nanorod is assumed to be made of magnetosomes of size 
$55\times 44\times 44\;\rm nm^3$.  For the magnetosome core made of magnetite with saturating
magnetization of $M_s=480$ kA/m its elementary magnetic moment is 
$\mu_1\approx 0.0511\;{\rm fAm^2}$ and its energy
in the magnetic field of Earth taken to be $B=B_e=50\;\mu\rm T$ is $\mu_1 B\approx\; 
0.623\; k_BT_r$,
when it is aligned with the field direction. The sensor can be operable already for
$n=5$ nanoparticles in the rod with $\mu_5 B=5\mu_1 B\approx \;3.115\; k_BT_r$, 
and a reliable
operation can be achieved for $n=7$ with $\mu_7 B=7\mu_1 B\approx\; 4.363\; k_BT_r$. For these
two values we shall do illustrative calculations below, noting, however, once again that the
sensor can consist also of just one sufficiently large ferro- or ferrimagnetic nanoparticle. For example,
a particle with
the size $103\times 85\times 85\;\rm nm^3$ will have about the same $\mu B\approx 
4.36\; k_BT_r$ as our rod with $n=7$, and such nanoparticles are also customly
found in living species. 
Some examples of $U(\phi)$ for various magnetic energies and 
the corresponding $p(\phi)$ are plotted in Fig. \ref{Fig2}.

Consider first the $B=0$ 
case (full line), and notice, 
that when $U(\phi)$ arrives at its maximum at $\phi_{\rm max}\approx 103.3^{\rm o}$,
$p(\phi_{\rm max})\approx 0.269$. For $\phi\geq \phi_{\rm max}$, a further increase
of $\phi$ introduces a negative stiffness instability and the rod rotates to a new 
metastable minimum at $\phi_{\rm min,2}\approx 144.81^{\rm o}$, where the channel opening
probability becomes $p(\phi_{\rm min,2})\approx 0.926$. At the first metastable
minimum,  $\phi_{\rm min,1}=\phi_0=30^{\rm o}$, 
$p(\phi_{\rm min,1})\approx 10^{-8}$. The corresponding energy differences 
between two metastable minima is $\Delta U=U(\phi_{\rm min,2})-U(\phi_{\rm min,1})
\approx 0.3639=3.639\;k_BT_r$, and the energy barriers are 
$\Delta U_1=U(\phi_{\rm max})-U(\phi_{\rm min,1})
\approx 0.787=7.87\;k_BT_r$, and $\Delta U_2=U(\phi_{\rm max})-U(\phi_{\rm min,2})
\approx 0.4232=4.232\;k_BT_r$. Being attached to a common sensor the channels will fluctuate
stochastically but synchronously between their closed and open states following the
sensor motion, with the averaged opening probability which can be roughly estimated as
 $\langle p\rangle \sim 1/[1+\exp(\Delta U/(k_BT_r))]\approx 0.026$ (at room temperatures),
i.e. the channels are closed most of time. Actually, the opening probability
will be lower that this rough estimate because the first minimum is shallower that
the second one, i.e. it is also entropically preferred.
Judging from the value of $\Delta U$, one can 
expect that for a sufficiently strong magnetic field at a proper angle $\psi$ such that $\mu B\sim \Delta U$,
the second metastable minimum can be made lower relative the first one and the channel
will become open on average. Indeed, this is the case
already for $\mu B=0.3115 U_0=3.115\; k_BT_r$ at $\psi=\pi$, see Fig. \ref{Fig2}.
The entropic effects may, however, compensate somewhat for the field-induced negative $\Delta U$, and the
channels are indeed about half-open on average, see below. For a larger 
$\mu B=0.4363 U_0=4.363\; k_BT_r$ in Fig. \ref{Fig2}, the channels will predominantly be open.

\subsection{Averaged open probability and ionic current as function of field direction}

\begin{figure}
\vspace{1cm}
\resizebox{0.95\columnwidth}{!}{\includegraphics{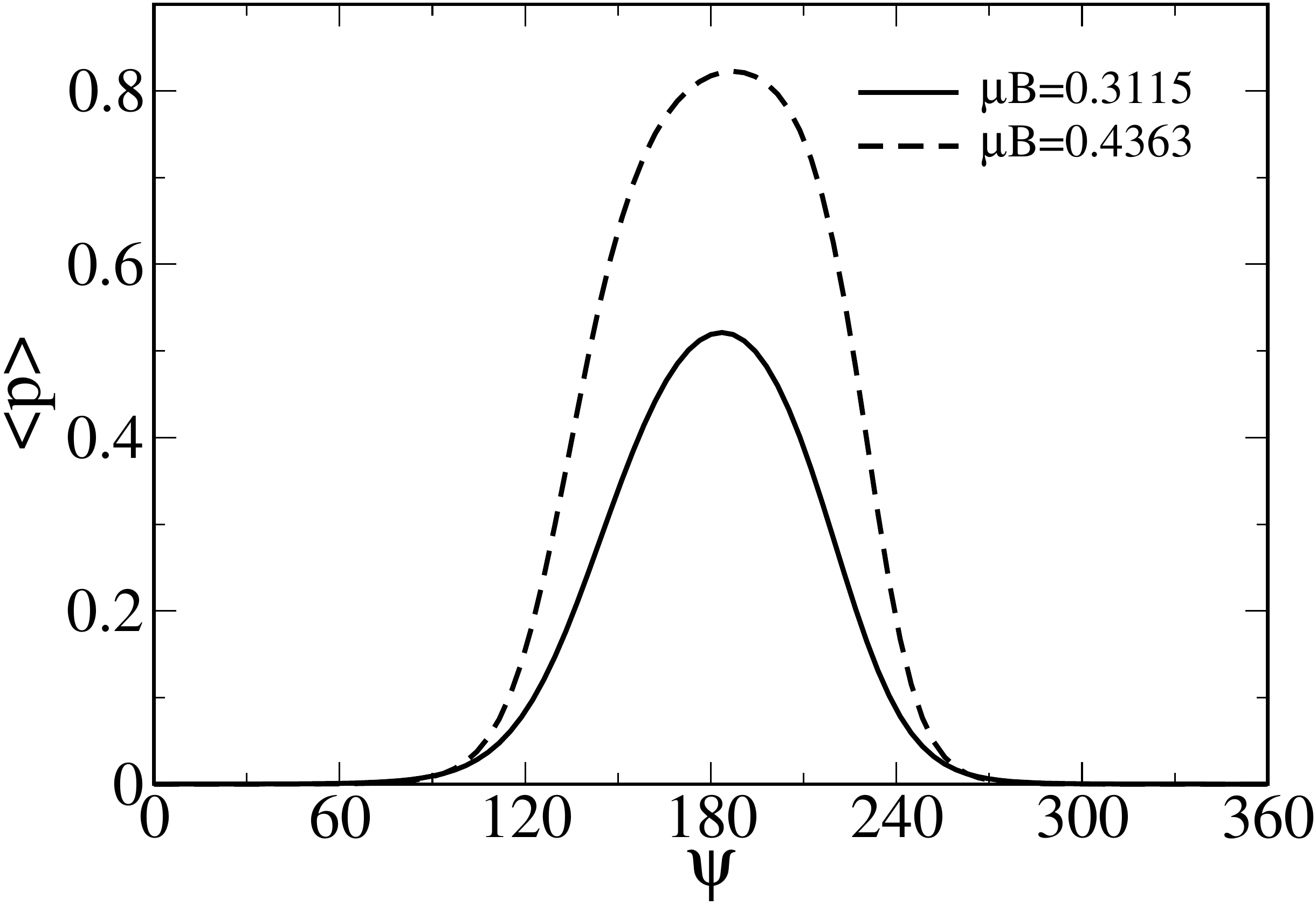}}
\caption{Averaged probability of the channels to be open as function
of magnetic field orientation for two values of $\mu B$ shown in the plot.
  The model parameters: $\phi_0=30^{\rm o}$, $T=0.1$, 
$l_{\rm max}=1.5$, $f_0=1.5$, $l_0=1.22$, and $m=7$ channels in the sensor cluster.}
\label{Fig3}       
\end{figure}

The averaged probability of the channel to be open can be found as
\begin{eqnarray}\label{pens}
\langle p(B,\psi)\rangle=\int_0^\pi p(\phi)e^{-U(\phi)/(k_BT)}d\phi/Z,
\end{eqnarray}
where $Z=\int_0^\pi e^{-U(\phi)/(k_BT)}d\phi$ is the corresponding statistical
sum (integral). It accounts also for entropic effects.
Unfortunately, this expression cannot be found in a closed compact analytic form
for the model considered. However, its numerical evaluation can be easily done. 
The corresponding results are shown in Fig. \ref{Fig3} and reveal that the direction
of magnetic field can be detected by a bell-shaped increase of the opening probability 
within the angle $\Delta \psi \sim 180^{\rm o}\pm  60^{\rm o}$. Further sharpening of the detection
of the field direction can be achieved via an adjusted threshold of excitation in the sensory
cell.

The averaged ionic current conducted by this sensory complex is thus 
$\langle I \rangle=m i_0\langle p(B,\psi)\rangle$, where $i_0$ is unitary
current through one open channel. Let us estimate it as $i_0=50\;\rm pA$ being a
typical value for large conductance cation channels \cite{JacksonBio}. 
These should be either sodium, or calcium channels
in order to cause depolarization of the cell membrane by their opening, 
given typically small out-of-equilibrium
innercell concentrations of sodium and calcium ions with respect to the cell exterior.  
Then, for $m=7$ and $\langle p\rangle =0.5$, $\langle I \rangle=175\;\rm pA$. 
Estimating the whole cell membrane resistance at rest
as $R=100\;\rm M\Omega$ \cite{Koch}, the transmembrane potential change is estimated as 
$\Delta V=R\langle I \rangle\sim 17.5$ mV. This can already be sufficient to
depolarize an excitable cell
membrane and to trigger a spiking activity mediating further information
about external magnetic field changes. However, a detailed elaboration of related excitable
cell model in the spirit of the Hodgkin-Huxley type conductances based approach 
is beyond the scope of the present work and is left for a future.

\subsection{Stochastic dynamics}

In order to operate as sensor, the opening-closing stochastic 
dynamics of the considered
magnetic sensor should also be sufficiently fast. Clearly, if it would take 
minutes on average to accomplish transitions to the open state, such a sensor would
simply be too slow to be of any relevance in biology. The motion of sensor 
occurs in viscoelastic cytosol. It is considered to be overdamped with
the inertial effects neglected.
Apart from the mean torque $f(\phi)=- d U(\phi)/d \phi$,
it is subjected to viscoelastic 
memory-friction torque  
$-\int_0^{t}\eta_{\rm mem} (t-t')\dot \phi(t') dt'$, which acts in addition 
to the viscous Stokes friction torque, $-\eta_0\dot \phi$, 
caused by the primary water component of cytosol. The dissipative forces are
complemented by the corresponding zero-mean Gaussian thermal random noises of the environment
at temperature $T$, 
$\xi_{\rm mem}(t)$ and $\xi_0(t)$, correspondingly. 
The friction and thermal
noise are related by the (second) fluctuation-dissipation theorem (FDT) by Kubo 
\cite{Kubo,Ford,Zwanzig}
\begin{eqnarray}\label{2FDT}
\langle \xi_{\rm mem}(t)\xi_{\rm mem}(t') \rangle &= & k_B T \eta_{\rm mem}(|t-t'|),\\
\langle \xi_0(t)\xi_0(t') \rangle &= & 2 k_B T \eta_0 \delta(t-t')\;.
\end{eqnarray}
This ensures thermal detailed balance in the absence of external driving. Stochastic dynamics
is described by the Generalized Langevin Equation (GLE) \cite{Kubo,Ford,Zwanzig,CoffeyBook}
\begin{eqnarray}\label{GLE}
\eta_0\dot \phi=f(\phi)-\int_0^{t}\eta_{\rm mem} (t-t')\dot \phi(t') dt'\\
+\xi_{\rm mem}(t)+\xi_0(t)\;. \nonumber
\end{eqnarray}
The memory kernel reflecting viscoelasticity of complex polymeric fluids such as
cytosol has often intermediate power law
scaling $\eta_{\rm mem}(t)=\eta_\alpha t^{-\alpha}/\Gamma(1-\alpha)$,  between two
memory cutoffs, $\tau_{l}$ and $\tau_{h}$, with $0<\alpha<1$. It corresponds to a complex
shear modulus $G^*(\omega) \propto \omega^\alpha$ \cite{Mason,Waigh},  at intermediate
frequencies in accordance with a huge body of rheology \cite{Gemant,Russel} 
and microrheology research \cite{Amblard,Caspi,Weiss,Mizuno,Waigh,Holek}. 
$\eta_{\alpha}$ is a fractional friction coefficient \cite{Metzler,Goychuk12}
corresponding to a fractional viscosity coefficient $\zeta_{\alpha}$ \cite{Gemant}, 
$\eta_\alpha\propto \zeta_\alpha$.
A strict power law is clearly an idealization and cutoffs must be present on physical
grounds.
 The short time cutoff $\tau_{l}$ reflects molecular size effects 
or highest vibrational modes present in the environment (which cannot be captured
by any continuous medium type approximation). 
The long time cutoff $\tau_{h}$   must also be present in any fluid-like environment making the 
overall integral $\eta_{\rm eff}=\int_0^{\infty}\eta_{\rm mem}(t)dt$ finite. This
reflects a finite macroscopic viscosity $\zeta_{\rm eff}\sim \eta_{\rm eff}$ 
of such complex fluids
on a large time scale $t\gg \tau_{h}$. The model with $\tau_l\to 0$, 
$\tau_h\to \infty$ (strict power law scaling) corresponds to the so-called fractional
Langevin equation (FLE) upon  using
the formalism of fractional time derivatives \cite{Mainardi,Lutz,GH07,KhG13,CoffeyBook},
or a strict sub-Ohmic memory friction within dynamical approach to generalized 
Brownian motion \cite{Ford,WeissBook}. Then, the solution of a potential-free FLE
(\ref{GLE}), $f(\phi)=0$, with the Stokes friction neglected, $\eta_0\to 0$, and 
for $\phi$  regarded as a linear, rather than cyclic variable,  is
fractional Brownian motion (fBm) \cite{GH07,Goychuk12}.  It presents a Gaussian process with stationary increments 
and a  long-range memory, which is completely characterized by its variance, 
$\langle \delta \phi^2 \rangle =2D_{\alpha}t^\alpha/\Gamma(1+\alpha)$, growing
sublinearly. The fractional (orientational) diffusion coefficient 
$D_{\alpha}$ is related to fractional friction coefficient $\eta_{\alpha}$ by
the generalized Einstein relation, $D_{\alpha}=k_B T/\eta_{\alpha}$.
Upon taking the  Stokes friction into account it becomes \cite{KhG13},
\begin{eqnarray}\label{exact}
\langle \delta \phi^2(t)\rangle =2D_{0} t E_{1-\alpha,2}[-(t/\tau_{in})^{1-\alpha}],
\end{eqnarray}
where $E_{a,b}(z):=\sum_0^{\infty}z^n/\Gamma(an+b)$ is generalized Mittag-Leffler function,
and  $D_0=k_BT/\eta_{0}$ is a normal diffusion coefficient. 
Furthermore, $\tau_{in}=(\eta_0/\eta_{\alpha})^{1/(1-\alpha)}$
is a transient time constant. For $t\ll \tau_{in}$, diffusion is initially normal,
$\langle \delta \phi^2(t)\rangle \approx 2D_{0} t$. It becomes anomalously
slow, $\langle (\delta \phi)^2 \rangle =2D_{\alpha}t^\alpha/\Gamma(1+\alpha)$,  
for $t\gg \tau_{in}$. 
We take further the advantage of approximation of the power-law scaling memory kernel by 
a sum of exponentials \cite{Goychuk09,Goychuk12}
\begin{eqnarray}\label{kernel2}
\eta_{\rm mem}(t)=\sum_{i=1}^N k_i \exp(-\nu_i t),
\end{eqnarray}
with a fractal scaling of relaxation rates 
$\nu_i=\nu_0/b^{i-1}$ and weights $k_i\propto \nu_i^\alpha$ (having physical dimension of energy
in the present case) 
to embed non-Markovian dynamics of $\phi(t)$ as a component or 
projection of $N+1$-dimensional Markovian dynamics
\begin{eqnarray}
\label{embedding}
\eta_0\dot{\phi}&=& f(\phi)-\sum_{i=1}^{N}k_i(\phi-y_i)+\xi_0(t),
\nonumber\\
\eta_i\dot{y_i}&=&k_i(\phi-y_i)+\xi_i(t),
\end{eqnarray}
where $y_i$ are nondimensional linear auxiliary variables, $\eta_i=k_i/\nu_i$, and 
$\xi_i(t)$ are independent auxiliary white Gaussian noises
obeying
\begin{eqnarray}
\langle \xi_i(t)\xi_j(t') \rangle &= & 2 \delta_{ij}k_B T \eta_j \delta(t-t'),
\end{eqnarray} 
and also independent of $\xi_0(t)$. 
The initial positions $y_i(0)$ are sampled from a Gaussian
distribution centered around $\phi(0)$, $\langle y_i(0)\rangle=\phi(0)$ with variances
$\langle[y_i(0)-\phi(0)]^2\rangle=k_BT/k_i$, in order to have complete equivalence
with the corresponding GLE in Eqs. (\ref{2FDT})-(\ref{GLE}), (\ref{kernel2}) 
in the ensemble sense \cite{Goychuk12}. It is convenient to choose
\begin{eqnarray}
k_i=\nu_0\eta_{\rm eff}\frac{b^{1-\alpha}-1}{b^{(i-1)\alpha}[b^{N(1-\alpha)}-1]},
\end{eqnarray}
where $\nu_0=1/\tau_l$ is the largest relaxation rate of environment equal to the inverse
small time cutoff, $\tau_l\ll \tau_{in}$. Diffusion becomes again normal on the
time scale $t\gg \tau_h=\tau_l b^{N-1}$, and 
$\eta_\alpha=\eta_{\rm eff}\tau_h^{\alpha-1}/g_\alpha$, with a proportionality
coefficient $g_{\alpha}$ about unity, $g_{\alpha}\sim 1$ \cite{GKhMet14a}. 
For example, $g_{\alpha}\approx 0.93$, for $\alpha= 0.4$ and $N \geq 5$ \cite{GKhMet14b} . 
The scaling coefficient $b$
controls the quality of approximation of the power law dependence between two time cutoffs.
Relative error is about 4\% only 
already for a crude decade scaling with $b=10$, which suffices in most studies,
 and improves further to 0.01\% for
$b=2$ \cite{GKh13}. Interestingly, 
$\tau_h/\tau_{in}=(\eta_{\rm eff}/\eta_0)^{1/(1-\alpha)}$ independently
of $b$, which allows to estimate the time duration of intermediate subdiffusion
in units of $\tau_{in}$ from merely the knowledge of $\alpha$ and an effective enhancement of
friction in cytosol relative one in water in the long-time normal diffusion limit. 
For example, if $\tau_{in}\sim 1$ msec and
$\tilde \eta_{\rm eff}=\eta_{\rm eff}/\eta_0=10^3$ for $\alpha=0.5$, intermediate subdiffusion will last until 
$\tau_h\sim 10^3$ sec, i.e. it extends over six time decades in units 
of $\tau_{in}$. Such a consideration can be very useful to estimate $\tilde \eta_{\rm eff}$  from experimental
data. We scale time in the units of
$\tau_{sc}=\eta_0/U_0$. For $U_0=10\;k_BT_r$ and for the rod of length
$L=275$ nm consisting
of $5$ magnetosomes it is estimated as $\tau_{sc}\approx 0.404$ msec, and
for the rod of length $385$ nm consisting of $7$ magnetosomes 
$\tau_{sc}\approx 0.905$ msec (see Appendix \ref{append}).
This corresponds to the rotational diffusion
coefficients $D_0=0.248 \;{\rm rad^2/msec }$, and 
$D_0=0.110 \;{\rm rad^2/msec}$, respectively.

\subsection{Relaxation within a potential well}

\begin{figure}
\vspace{1cm}
\resizebox{0.95\columnwidth}{!}{\includegraphics{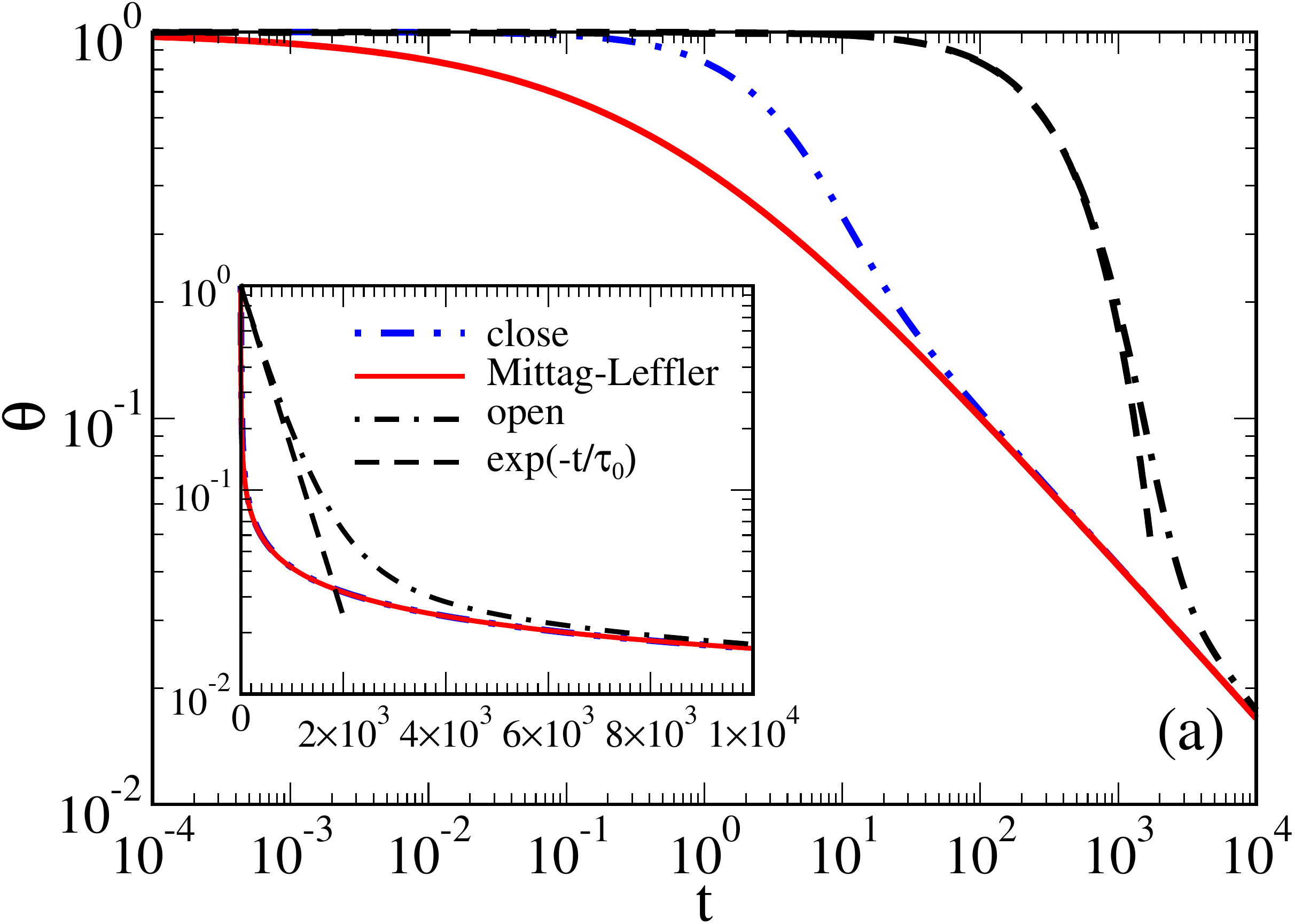}}\vfill \vspace{0.5cm}
\resizebox{0.95\columnwidth}{!}{\includegraphics{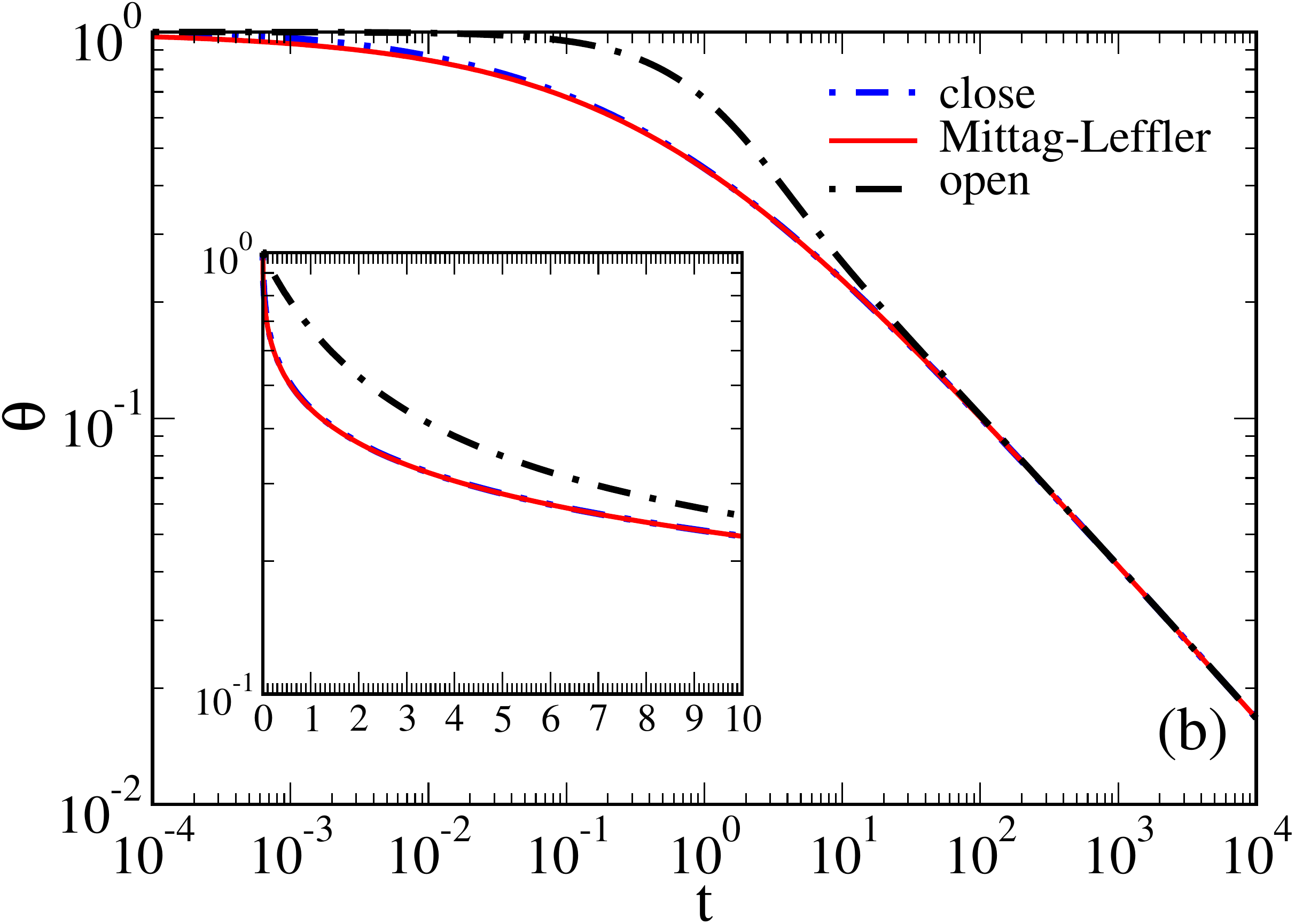}}
\caption{(color online) Non-Markovian relaxation function against time in units of $\tau_r$ 
for two different potential wells corresponding to
a potential $U(\phi)$ realization in Fig. \ref{Fig2} at $\mu B=0.3115$ for $\alpha=0.4$ 
in parabolic approximation and in neglect of the memory cutoff effects.
Time constant $\tau_r$  has different absolute values (in units of $\tau_{sc}=\eta_0/U_0$) 
for different potential wells.
Parts (a) and (b) correspond to  two different values of fractional
friction coefficient $\eta_{\alpha}\sim\eta_{\rm eff}\tau_h^{\alpha-1}$.
In (a), $\tau_h=10^4$ and $\eta_{\rm eff}=100\eta_0$, 
 $z_1=\tau_0/\tau_r=4.838$ (for the relaxation in the first potential well
which corresponds to closed times, see dash-double-dotted blue curve), 
and $z_2=\tau_0/\tau_r=536.83$ 
(for the relaxation in the second potential well
which corresponds to open times, dash-dotted black curve). Dashed black line
depicts single-exponential approximation.  Full red curve is 
the Mittag-Leffler relaxation
function $E_\alpha[-(t/\tau_r)^\alpha]$, which corresponds to the Cole-Cole dielectric
response and $\beta$-relaxation in glass-like materials. The tail of relaxation
is universally a power law, $\theta(t)\sim t^{-\alpha}$. Inset in (a) shows the same plot
on semi-logarithmic scale. It reveals that nearly
70\% of initial  relaxation in the second well occurs in the exponential regime. 
In (b), $\eta_{\rm eff}$ is increased to 
$\eta_{\rm eff}=1000$, with other parameters kept the same, which corresponds
to a tenfold larger $\eta_{\alpha}$, as compare with the part (a).
Here, $z_1=0.01593$ (dash-double-dotted blue curve, which is difficult to 
see because it almost coincides with the full red line 
corresponding to the Mittag-Leffler relaxation), and $z_2=1.6976$ (dash-dotted black curve).
The relaxation in the first potential well is excellently described by the Mittag-Leffler relaxation
function, and the relaxation in the second potential well is also clearly non-exponential
all the time, as inset shows. 
Numerical results are obtained by numerical inversion of the Laplace-transform in Eq. (\ref{rfun})
with Stehfest-Gaver method as described in Refs. \cite{Stehfest,GH06}.
}
\label{Fig4}       
\end{figure}

Nonlinear viscoelastic dynamics in a bistable potential is rather intricate 
\cite{Goychuk09}. To understand its main features, it is important to realize first 
the character
of relaxation process in one potential well. This can be done
within a parabolic well approximation, $U(\phi)\approx \kappa_{1,2}(\phi-\phi_{\rm min,1,2})^2/2$,
where $\kappa_{1,2}=d^2 U(\phi)/d\phi^2|_{\phi=\phi_{\rm min,1,2}}$. The relaxation 
of an initial fluctuation, $\delta \phi(0)=\phi(0)-\phi_{\rm min,1,2}$, within
a parabolic potential follows the relaxation law 
$\langle \delta \phi(t)\rangle =\delta \phi(0)\theta (t)$, with relaxation
function $\theta(t)$ whose Laplace-transform reads \cite{Goychuk07Rapid,Goychuk12}
\begin{eqnarray}
\tilde \theta(s)=\frac{\tilde \eta(s)}{\kappa +s\tilde \eta(s)}
\end{eqnarray}
for arbitrary memory kernel $\eta(t)$. Here we omitted subindexis $1,2$
at $\kappa$ for simplicity. The relaxation function  coincides with the normalized 
\textit{stationary} autocorrelation function of fluctuations or ACF, 
$\langle \delta \phi (t_0)\delta \phi (t_0+t)\rangle_{eq}/
\langle \delta \phi^2\rangle_{eq}$, with $\langle \delta \phi^2\rangle_{eq}=k_BT/\kappa$,
 which does not depend
on time shift $t_0$. In other words,
the Onsager regression property holds generically within this model, as shown in \cite{Goychuk07Rapid}. 
However,  ACF depends generally
on both time arguments and displays aging phenomenon \cite{Goychuk07Rapid,Goychuk12,Schulz14}. 
In the present
case, $\tilde \eta(s)=\eta_0+\eta_\alpha s^{\alpha-1}$, and
\begin{eqnarray}\label{rfun}
\tilde \theta(s)&=&\frac{\tau_0+\tau_r (s\tau_r)^{\alpha-1}}{s\tau_0+1+(s\tau_r)^\alpha}
=\frac{r+r_{1-\alpha}s^{\alpha-1}}{s+r+r_{1-\alpha}s^\alpha},
\end{eqnarray}
where we denote $\tau_0=\eta_0/\kappa$, $\tau_r=(\eta_\alpha/\kappa)^{1/\alpha}$,
and $r=1/\tau_0$, $r_{1-\alpha}=\eta_\alpha/\eta_0=\tau_{in}^{\alpha-1}$. Upon change $\alpha\to 1-\alpha$,
i.e. identifying our present $\alpha$ with $1-\alpha$ in \cite{Goychuk14CTP}
and identifying our present $\tau_0$ with $\langle \tau\rangle$ therein
one can see that this result coincides (up to a normalization factor) with one
obtained in Ref. \cite{Goychuk14CTP} for the stationary ACF of fluctuations
in a very different model based on CTRW approach to relaxation phenomena. This another
approach  is featured
by two parallel relaxation channels characterized by normal rate $r$, and
fractional rate $r_{1-\alpha}$, correspondingly, and by
a \textit{finite} mean residence time $\langle \tau\rangle=1/r$, see in 
\cite{GoychukPRE12,Goychuk14CTP} for basic formulations and details. 
In \cite{Goychuk14CTP}, the corresponding spectral power of fluctuations, $S(\omega)$,  
and the response function $\chi(\omega)$ are also presented and discussed.
The relaxation behavior of $\theta(t)$ depends very strongly on the relationship
between $\tau_0$ and $\tau_r$. The time constant $\tau_r$ can be expressed through
$\tau_0$ and the above $\tau_{in}$, as 
\begin{eqnarray}
\tau_r=\tau_0\left (\frac{\tau_0}{\tau_{in}} \right )^{1/\alpha}\;.
\end{eqnarray}
A salient feature is that $\tau_r$ depends
on the potential curvature $\kappa$ only through $\tau_0$, and $\tau_{in}$
that characterizes free diffusion. The Laplace-transform of relaxation
function can be inverted exactly to the time domain for a special case $\alpha=0.5$.
Then, it reads
\begin{eqnarray}
\theta(t)=\frac{1}{2}\left ( 1+\frac{1}{\sqrt{1-4z}}\right )
e^{(1-\sqrt{1-4z})^2t/(4z^2\tau_r)} \nonumber \\
\times {\rm erfc}\left [ (1-\sqrt{1-4z})\sqrt{t/(4z^2\tau_r)} \right ] \nonumber \\
+\frac{1}{2}\left ( 1-\frac{1}{\sqrt{1-4z}}\right )e^{(1+\sqrt{1-4z})^2t/(4z^2\tau_r)} \nonumber \\
\times {\rm erfc}\left [ (1+\sqrt{1-4z})\sqrt{t/(4z^2\tau_r)} \right ] \; ,
\end{eqnarray} 
where $z=\tau_0/\tau_r$, and $\rm erfc$ is complementary error function.
Furthermore, for any $0<\alpha<1$,
if $\tau_0\ll \tau_r$, then 
relaxation follows approximately  
$\theta(t)\approx E_{\alpha}[-(t/\tau_r)^\alpha]$, see numerical results in Fig. \ref{Fig4} (b),
for $\alpha=0.4$ and the relaxation in the first potential well (closed times), within the
parabolic approximation. 
 It is initially stretched
exponential for $t\ll \tau_r$, and then a power law, $\theta(t)\propto t^{-\alpha}$.
The spectral power of intrawell fluctuations is 
$S(\omega)\propto 1/\omega^{1-\alpha}$, at low frequencies, and such an intrawell
motion is featured by the Cole-Cole response to external fields \cite{Goychuk07Rapid}.
It is measured in many biological tissues \cite{Barnes}, and corresponds
to the so-called $\beta-$ relaxation in glassy systems.
Clearly, for a sufficient small $\kappa$ and $\tau_{in}$ such a regime can always be realized.
Another regime with $\tau_0\gg \tau_r$
can also be very important.
 For a given $\tau_{in}$, in can  always be realized for a sufficiently large 
 $\kappa$, i.e. sufficiently stiff trapping potential.
Then, the
relaxation of $\theta(t)$ starts universally from an 
exponential regime, $\theta(t)=\exp(-t/\tau_0)$, which changes 
into a power law tail $\theta(t)\propto 1/t^\alpha$, 
see in Fig. \ref{Fig4} (a), for $\alpha=0.4$ (open times), and also 
in Figs. 2 and 3 in \cite{Goychuk14CTP} for $\alpha=0.5$. The larger the ratio 
$\tau_0/\tau_r$, the smaller is the weight of a heavy tail. In other words,
the major part of $\theta(t)$ relaxation occurs exponentially for $z\gg 1$. 
The power law tail starts then from some $\theta_c\ll 1$. 
For example, in Fig. \ref{Fig4} (a), about 70\%  of the whole relaxation 
in the second potential well
(open times) occurs in the exponential regime.
The power law tail of open times has just a few percent weight therein,
as the main double-logarithmic plot in Fig. \ref{Fig4} (a) reveals.
And nevertheless it can be very important yielding
 to the $S(\omega)\propto 1/\omega^{1+\alpha}$ feature of the fluctuations power
spectrum  for $\tau_0^{-1}\ll \omega\ll \tau_r^{-1}$. It has been detected e.g. 
experimentally in Ref. 
\cite{Bruno} for transversal fluctuations of  cargo moving along a microtubule. 
Moreover, slow residual relaxation in the potential wells leads generally to a
breakdown of the rate theory and validates the physical picture of slowly (on a characteristic
time scale of transitions) fluctuating barriers and fluctuating rates.

\subsection{Thermally activated transitions between metastable states of sensor}

Despite transitions between the potentials wells cannot be described as a rate
process for the biophysically most important case of intermediate barriers (less than about 
$10\;k_BT$ depending on $\alpha$), in the case of power law memory kernels, 
the rate theory has been shown to be able
to predict, at least in some cases, the most probable logarithm of residence times in the potential 
wells \cite{Goychuk09}. Moreover, the rate description is restored in the
limit of very high barriers. Then, the transition rates $R_{1,2}$ follow as 
\begin{eqnarray} \label{rate}
R_{1,2}(\mu)=\frac{\omega_{1,2}}{2\pi}\Xi(\mu)\exp(-\beta \Delta U_{1,2}),
\end{eqnarray}
where $\omega_{1,2}=\sqrt{\kappa_{1,2}/J}$ are circular attempt frequencies, 
and $0\le \Xi(\mu)\le 1$
is the so-called transmission coefficient, which for the considered 
intermediate-to-strong friction limit is
found \cite{Grote,HTB90} as $\Xi(\mu)=\mu/\omega_b$. Here, $\omega_b=\sqrt{\kappa_b/J}$ is
the imaginary barrier frequency, $\kappa_{b}=-d^2 U(\phi)/d\phi^2|_{\phi=\phi_{\rm max}}$, 
and $\mu$ is the renormalized barrier frequency. The latter one is
found at the largest positive root of a transcendental equation taking (in the considered overdamped
limit, $J\to 0$) the general form $\mu \tilde \eta(\mu)=\kappa_b$, and  for the memory
kernel considered,  $\mu \eta_0+
\eta_{\alpha}\mu^{\alpha}=\kappa_b$. By introducing $\tau_0^{(b)}=\eta_0/\kappa_b$,
and $\tau_r^{(b)}=(\eta_\alpha/\kappa_b)^{1/\alpha}$, we can write it as
\begin{eqnarray}\label{disp1}
\tau_0^{(b)}\mu+(\tau_r^{(b)}\mu)^\alpha=1. 
\end{eqnarray}
For the memory kernel expanded into a sum of exponentials it takes on the form
\begin{eqnarray}\label{disp2}
\mu\left (\eta_0+\sum_{i=1}^N \frac{k_i}{\mu+\nu_i}\right )=\kappa_b, 
\end{eqnarray}
which can be rewritten as an algebraic equation of $N+1$ degree for the unknown $\mu$. 
For a special case $\alpha=0.5$, Eq. (\ref{disp1}) can
be solved exactly to yield $\mu=4[\tau_r^{(b)}]^{-1}/\left [1+\sqrt{1+4\tau_0^{(b)}/\tau_r^{(b)}}
\right ]^2$,
and
\begin{eqnarray} 
R_{1,2}=\frac{1}{2\pi}\sqrt{\frac{\kappa_{1,2}}{\kappa_b}}
\frac{1}{\tau_r^{(b)}}\frac{4\exp(-\beta \Delta U_{1,2})}{\left [1+\sqrt{1+4\tau_0^{(b)}/\tau_r^{(b)}}
\right ]^2}\;.
\end{eqnarray}
This insightful result can be expressed in terms of the normal diffusion ($\eta_\alpha\to 0$) 
overdamped Kramers rate
\begin{eqnarray}\label{rate_normal} 
R_{1,2}^{(0)}& = &\frac{1}{2\pi}\sqrt{\frac{\kappa_{1,2}}{\kappa_b}}
\frac{1}{\tau_0^{(b)}}\exp(-\beta \Delta U_{1,2})\;\\
& = &\frac{1}{2\pi}
\frac{1}{\tau_0^{(b)}}\exp(-\beta \Delta G_{1,2})\nonumber
\end{eqnarray}
as
\begin{eqnarray}\label{rate_memory} 
R_{1,2}=R_{1,2}^{(0)}F(z_b=\tau_0^{(b)}/\tau_r^{(b)}),
\end{eqnarray}
where 
\begin{eqnarray}\label{F} 
F(z)=\frac{4z}{[1+\sqrt{1+4z}]^2}\;.
\end{eqnarray}
Notice that in the second line of Eq. (\ref{rate_normal}), we incorporated 
the difference of the curvatures $\kappa_1$ and $\kappa_2$ as \textit{additional} 
entropic contributions
to the free energy barriers, $\Delta G_{1,2}=\Delta U_{1,2}-T\Delta S_{1,2}^{(ad)}$
with entropy differences $\Delta S_{1,2}^{(ad)}=k_B\ln (\kappa_{1,2}/\kappa_b)/2$.
Generally, reduction of a multidimensional dynamics to a two-state dynamics contains
such important additional entropic contributions. 
It should be mentioned, however,  that in our model 
$U(\phi)$ in Eq. (\ref{potential}) is also temperature-dependent, i.e. is in fact
(Gibbs) free energy profile. 
We consider, however, a fixed value of temperature throughout the paper.
The correct separation of $\Delta G_{1,2}$ into the internal energy 
(or rather enthalpic) part $\Delta H_{1,2}$, and entropic part $-T\Delta S_{1,2}$
must always be done using fundamental thermodynamic relation
$\Delta H=\Delta G+T\Delta S=\Delta G-T(\partial \Delta G/\partial T)_P$ \cite{Israelachvili}.

Furthermore, for $z\ll 1$ in (\ref{F}), $F(z)\approx z$, and for $z\gg 1$, $F(z)$ approaches unity,
$F(z)\to 1$. 
Hence, in the parameter regime $\tau_0^{(b)}\gg \tau_r^{(b)}$, which correlates with 
$\tau_0\gg \tau_r$, see above, $R_{1,2}\approx R_{1,2}^{(0)}$,
i.e. the rate is practically not affected by subdiffusion.
This is a very important result, which is valid also for other values of $\alpha$.
It is very different from the suppression by the factor $\eta_{\rm eff}/\eta_0$,
which some unjustified intuition might suggest. 
This result shows that subdiffusion does not necessary suppress the activation
rates, in accordance with our earlier results in \cite{Goychuk09} obtained
in the presence of inertial effects. Here, the inertial effects were, however,
entirely neglected. Moreover, it will be shown elsewhere that in the presence
of inertial effects the transmission factor for subdiffusive dynamics 
can arrive at its maximal value of unity, in the case of cusp-like potential barriers
$\kappa_b\gg \kappa_{1,2}$. The combination of nonlinearity and viscoelastic subdiffusion
is really counterintuitive and paradoxical! An essential suppression of rate
occurs in the regime $\tau_0^{(b)}\ll \tau_r^{(b)}$, which correlates with 
$\tau_0\ll \tau_r$, i.e. when the Mittag-Leffler relaxation within potential
wells covers most of time.  Then, we obtain (for arbitrary $\alpha$)
\begin{eqnarray} 
R_{1,2}\approx  \frac{1}{2\pi}\sqrt{\frac{\kappa_{1,2}}{\kappa_b}}
\frac{1}{\tau_r^{(b)}}\exp(-\beta \Delta U_{1,2}).
\end{eqnarray}
Notice that it is smaller than $R_{1,2}^{(0)}$
by the factor $\tau_0^{(b)}/\tau_r^{(b)}$.

\section{Results}

\subsection{Markovian dynamics}

\begin{figure}
\vspace{1cm}
\resizebox{1\columnwidth}{!}{\includegraphics{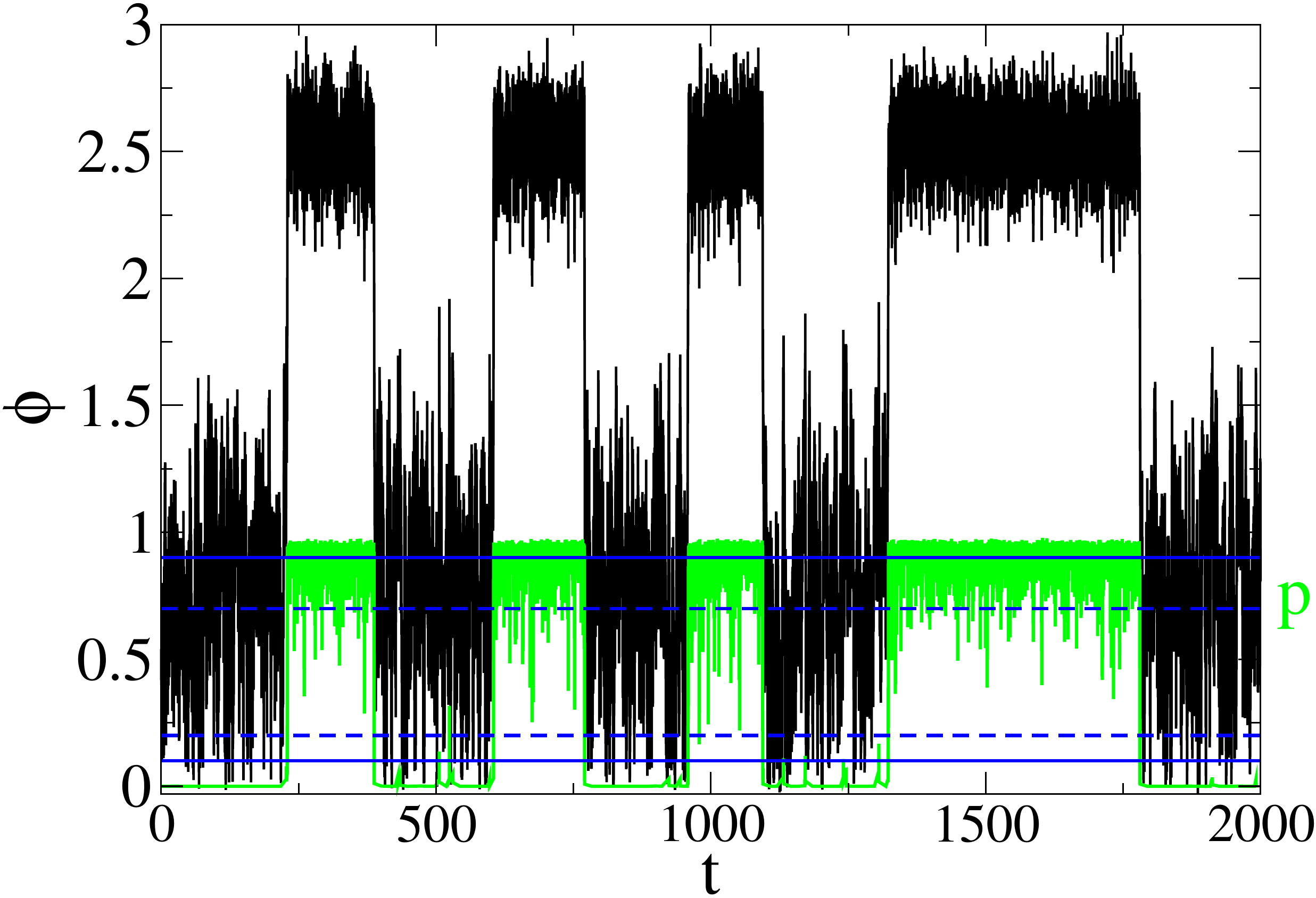}}
\caption{(color online) Trajectory realization of Markovian memoryless dynamics.
Black line describes stochastic rotational fluctuations of sensor, while the green
(light gray) line depicts 
fluctuations
of the probability of the gate to be open (and normalized 
ion current fluctuations). The blue horizontal 
lines depict some detection thresholds (two possible 
sets are shown).
The model parameters: $T=0.1$, $\mu B=0.3115$, $\phi_0=\pi/6$, $\psi=\pi$,
$l_{\rm max}=1.5$, $f_0=1.5$, $l_0=1.22$, and $m=7$ channels in the sensor cluster.}
\label{Fig5}       
\end{figure}

We performed first stochastic simulations of Markovian memoryless dynamics 
($\eta_{\alpha}\to 0$) for the same parameters
as in Figs. \ref{Fig2}, \ref{Fig3}, $\psi=\pi$, 
with time step $\delta t=2\cdot 10^{-6}$
using stochastic Heun method, in accordance with our previous studies. 
A typical trajectory of bistable fluctuations
for $\mu B\approx 3.12\;k_BT_r$ is shown in Fig. \ref{Fig5}. One can see characteristic
bistable fluctuations of sensor orientation. It flips between two metastable positions 
(interwell fluctuations) and exhibits also profound fluctuations within potential wells 
(intrawell fluctuations).
The open-shut dynamics reveals a new flickering feature, 
apart from large amplitude fluctuations reflecting bistability of sensor.
Flickering comes about from intrawell fluctuations of sensor corresponding
to its open metastable state, see below.

The statistics of transitions can be determined from one very long single
trajectory by setting different detection thresholds,
e.g. $p_1=0.2$ and $p_2=0.7$ (set 1), or $p_1=0.1$ and $p_2=0.9$ (set 2). 
One can also relate thresholds to the minima of $U(\phi)$ (set 3). For
example, for $\mu B=0.3115$ in Fig. \ref{Fig2}, 
$\phi_{\rm min,1}\approx 0.762\approx 43.69^{\rm o}$, 
$\phi_{\rm min,2}\approx 2.551\approx 146.13^{\rm o}$, and 
for $\mu B=0.4363$ in Fig. \ref{Fig2}, 
$\phi_{\rm min,1}\approx 0.910\approx 52.14^{\rm o}$, 
$\phi_{\rm min,2}\approx 2.559\approx 146.63^{\rm o}$. With this choice, 
$\phi_{\rm min,1}$ corresponds to a  very small $p_1\sim 10^{-8}-10^{-6}$,
and $\phi_{\rm min,2}$ to $p_2\sim 0.92-0.95$.
Yet, an experimentalist can prefer the choice $p_1=0.5$
and some $p_2>p_1$, e.g. $p_2=0.9$ (set 4).
It must be emphasized that the statistics of transitions can very essentially
depend on the thresholds, i.e. on how the current signal is detected.
This is because there are very fast events even within the purely Markovian
version of the considered dynamics, where very short outbursts occur from
the open state to the closed state, when the probability of the channel
to be open becomes briefly less than one-half. The lower the low detection threshold 
$0<p_1\leq 0.5$ in Fig. \ref{Fig5}, the larger is the number of missed short closure events. 
This is why the
averaged residence times in the states do increase with lowering
$p_1$, when the bursting events are increasingly disregarded. 
The explanation of a highly bursting character of dynamics even in the absence
of memory effects can easily be grasped from Fig. \ref{Fig2}. Indeed the maximum
of the potential $U(\phi)$ corresponds to the opening probability 
$p\sim 0.17-0.27$, depending on $\mu B$ value in  Fig. \ref{Fig2}. Hence, an essential
part of closure dynamics occurs when the sensor moves within the second metastable
minimum of sensor not reaching the transition state of sensor. 
This observation is also quite generally of a great importance  within the context of 
gating dynamics
of other ionic channels. It emphasizes the importance of a very complex molecular structure
of ionic channels \cite{Pollard}, where the sensory part and the gating part 
are generally not the 
same and their mechanical coupling can be very important in all fine details. 
Our treatment makes this implicit complexity quite obvious within a simplest
model setting. The opening probability
of the channel defined as a time average, 
$p=\langle \tau_o\rangle/(\langle \tau_o\rangle+\langle \tau_c\rangle)$, 
is, however, weakly sensitive to the choice of the detection thresholds,
and agrees reasonably well with the ensemble averaged result depicted
in Fig. \ref{Fig3}, where $p\approx 0.52$ at $\mu B=0.3115$
and $\psi=\pi$.

\begin{figure*}
\vspace{1cm}
\resizebox{0.95\columnwidth}{!}{\includegraphics{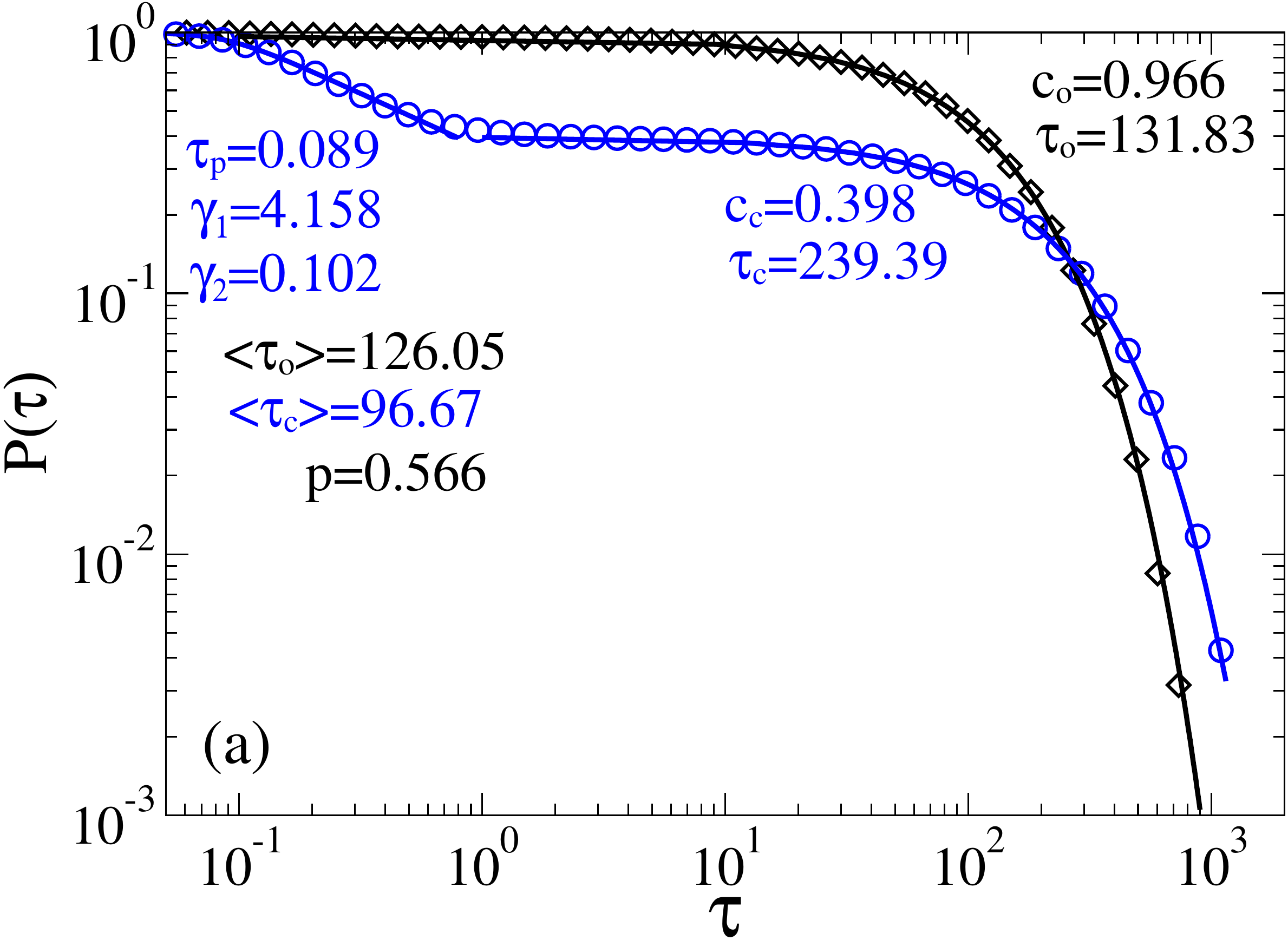}} \hspace{0.5cm}
\resizebox{0.95\columnwidth}{!}{\includegraphics{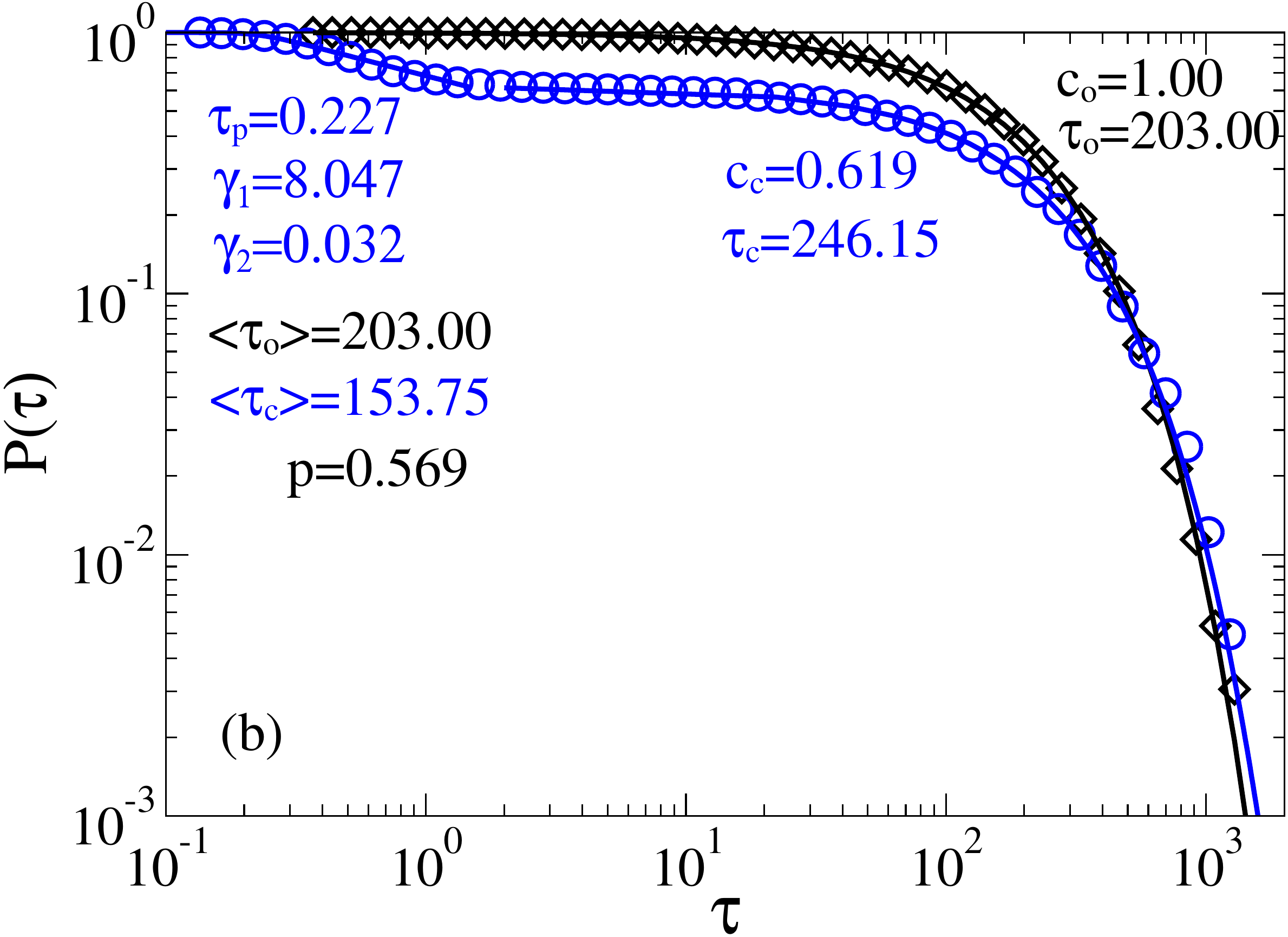}}\\ \vspace{1cm}
\resizebox{0.95\columnwidth}{!}{\includegraphics{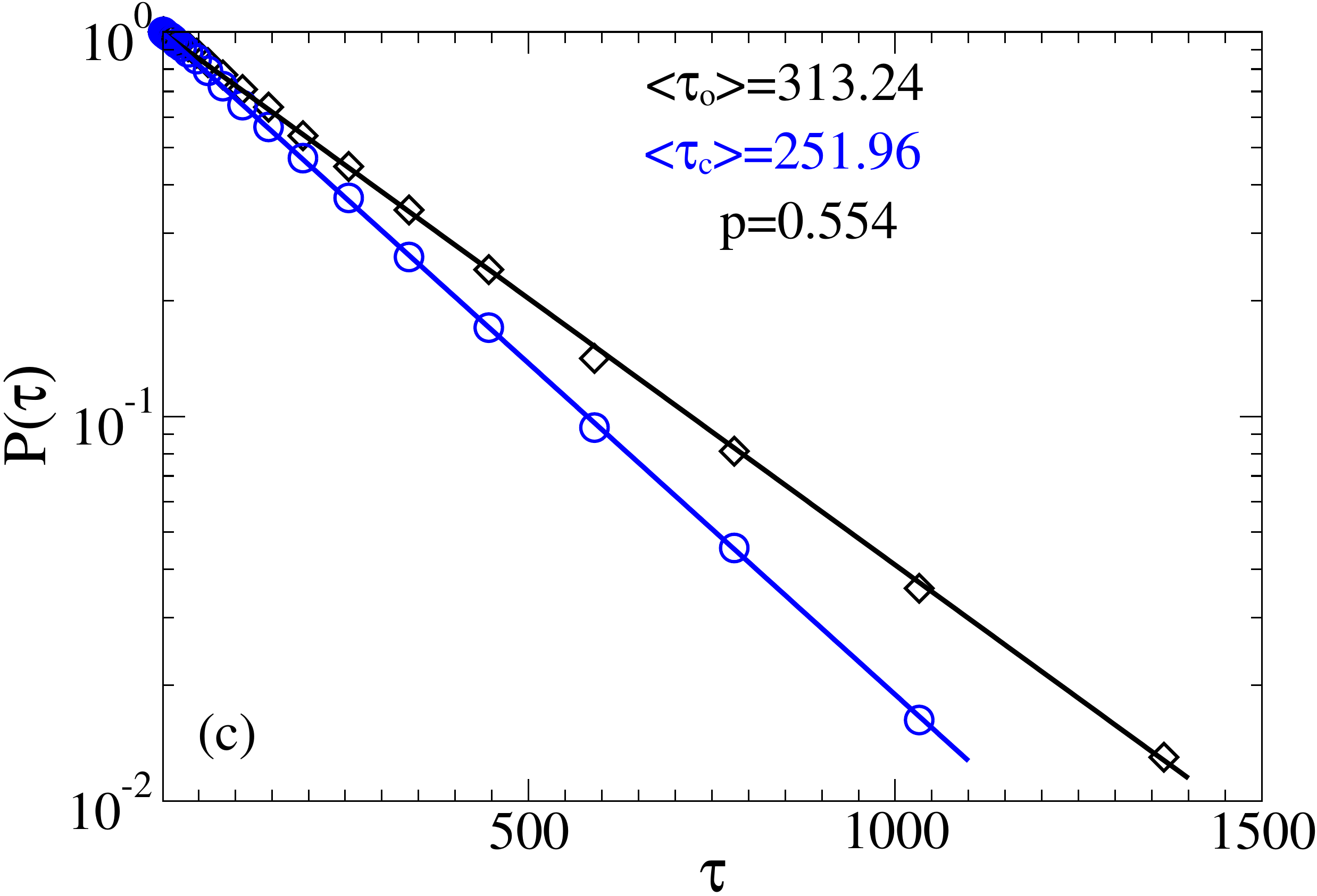}}\hspace{0.5cm}
\resizebox{0.95\columnwidth}{!}{\includegraphics{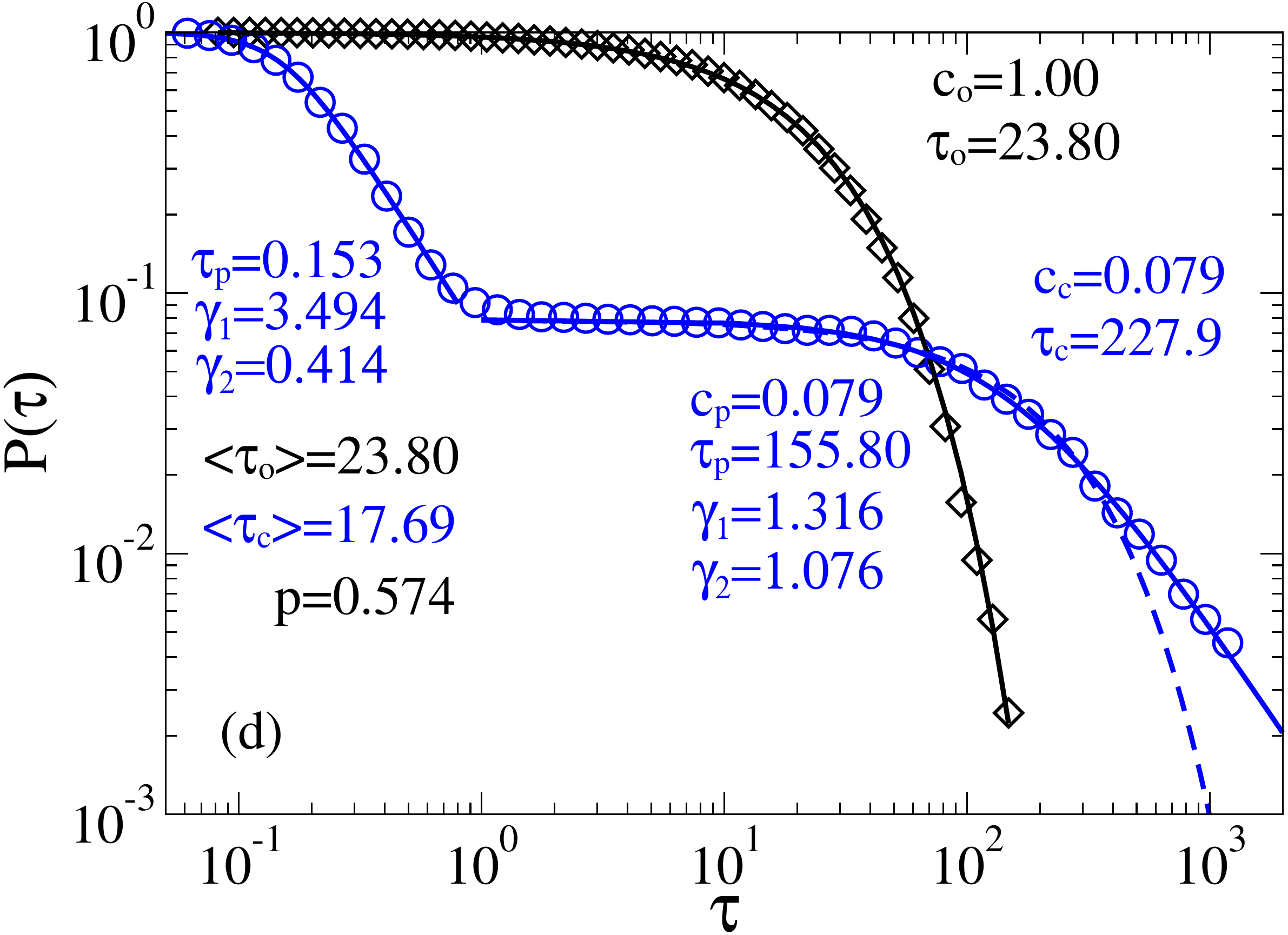}}
\caption{(color online) 
Survival probabilities of open (black diamond symbols), and
closed (blue open circles) times derived from numerical data for the case $\mu B=0.3115$, $\psi=\pi$,
$\phi_0=\pi/6$,
 using different sets 
of detection thresholds: (a) $p_1=0.2$, $p_2=0.7$; (b) $p_1=0.1$, $p_2=0.9$,
(c) for the minima of $U(\phi)$ used as detection thresholds, $p_1=p(\phi_{\rm min,1})$,
 $p_2=p(\phi_{\rm min,2})$; (d) 
$p_1=0.5$, $p_2=0.9$.
Lines present the corresponding exponential, $c_i\exp(-\tau/\tau_i)$, or
Pareto law (\ref{Pareto}) fits
with the parameters shown in the plots.
The mean residence times, as well as the opening
probability $p=\langle \tau_o\rangle/(\langle \tau_o\rangle+\langle \tau_c\rangle)$ are also shown. 
Notice a strong dependence of these fits on the detection thresholds used.
Other parameters:
 $T=0.1$,
$l_{\rm max}=1.5$, $f_0=1.5$, $r_0=1.22$, and $m=7$ channels in the sensor cluster.}
\label{Fig6}       
\end{figure*}

\begin{figure*}
\vspace{1cm}
\resizebox{0.95\columnwidth}{!}{\includegraphics{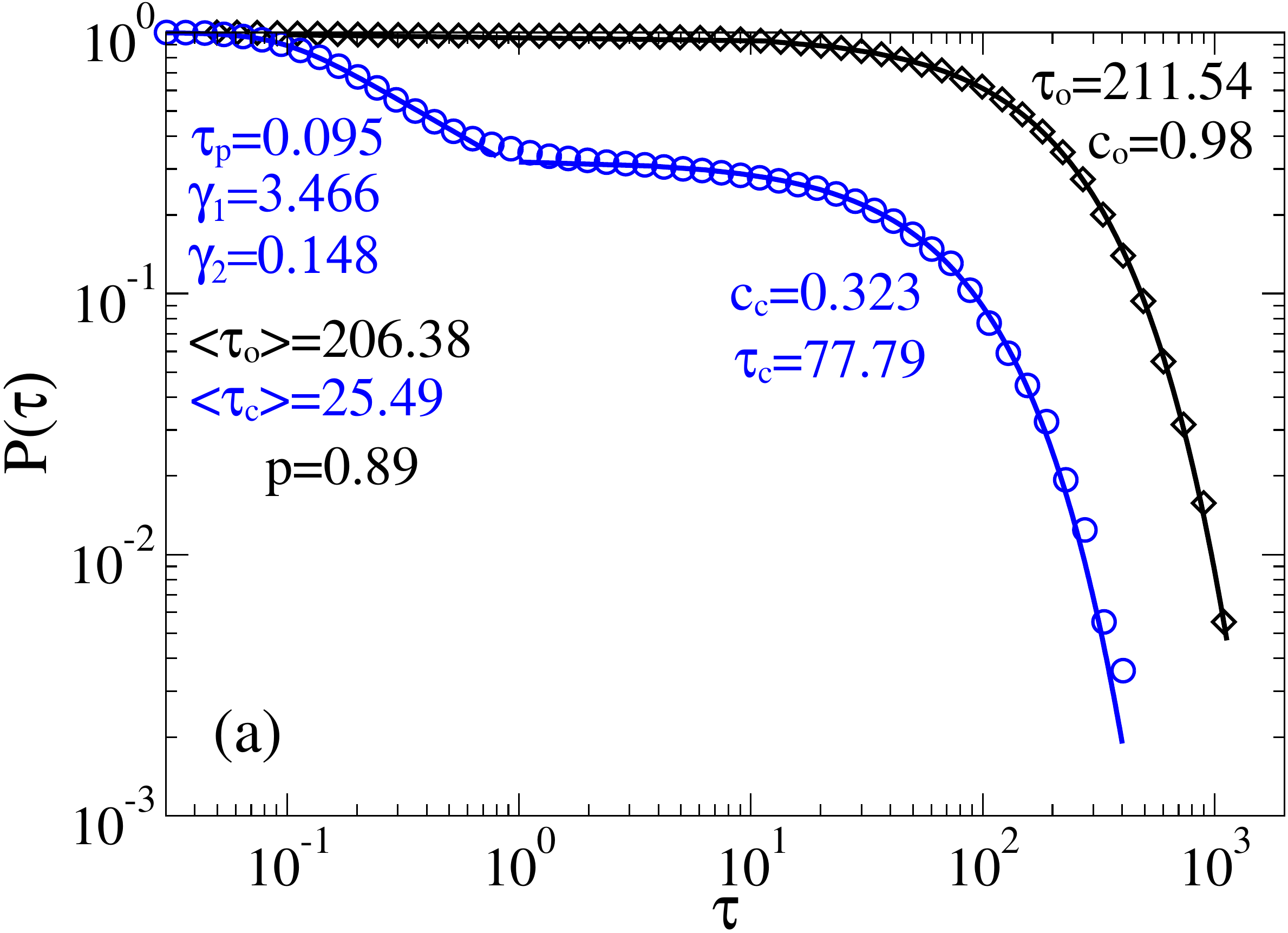}} \hspace{0.5cm}
\resizebox{0.95\columnwidth}{!}{\includegraphics{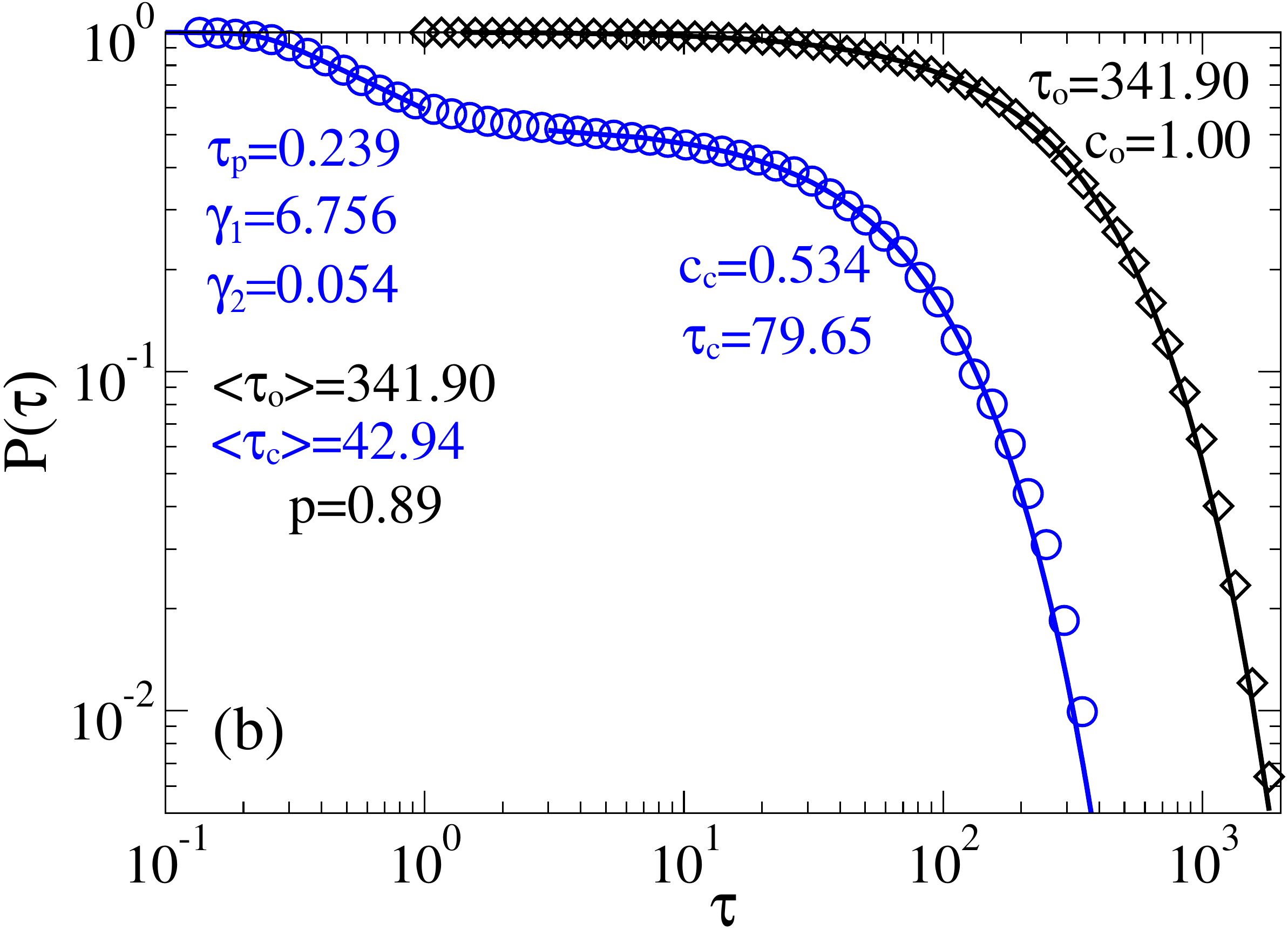}}\\ \vspace{0.9cm}
\resizebox{0.95\columnwidth}{!}{\includegraphics{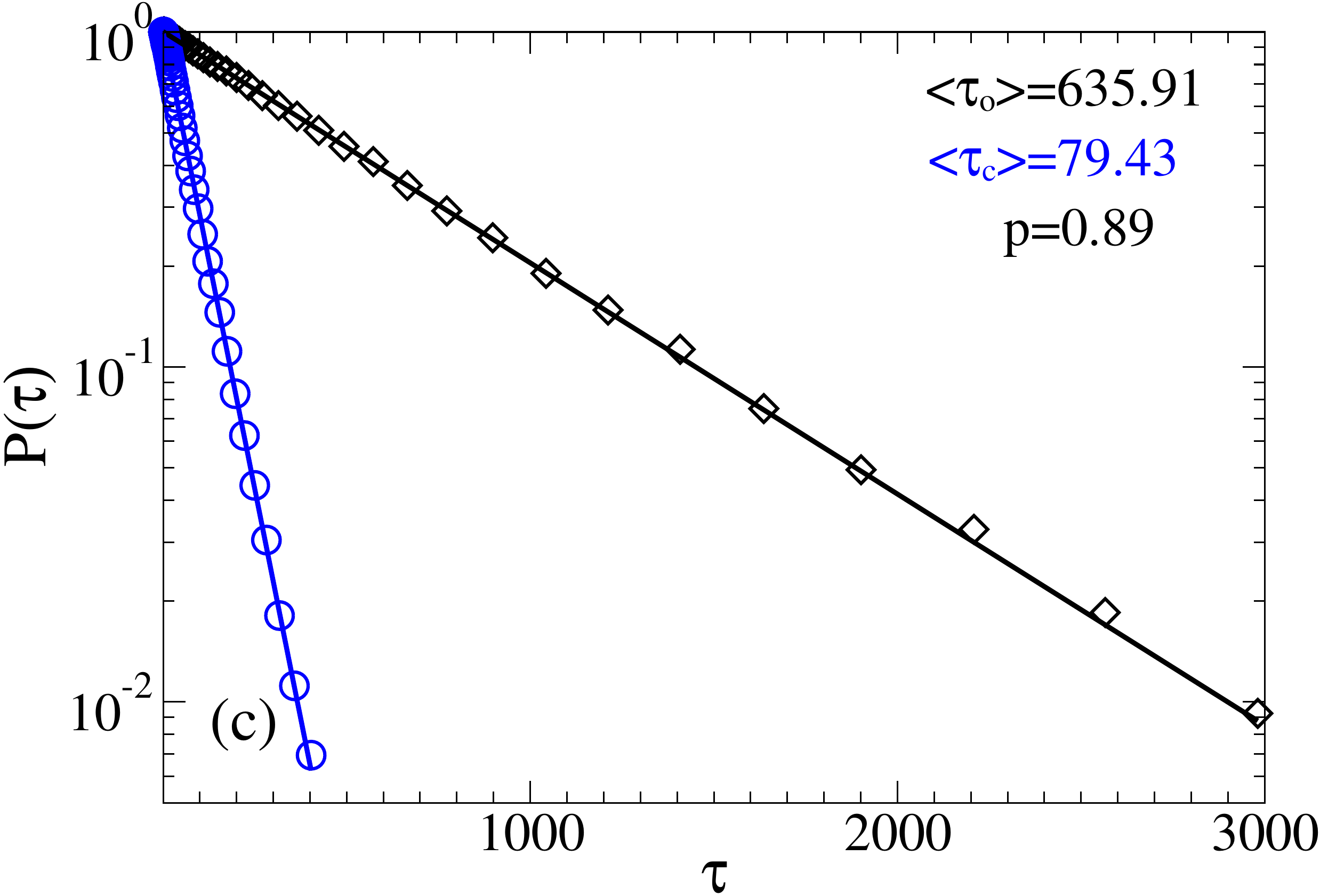}}\hspace{0.5cm}
\resizebox{0.95\columnwidth}{!}{\includegraphics{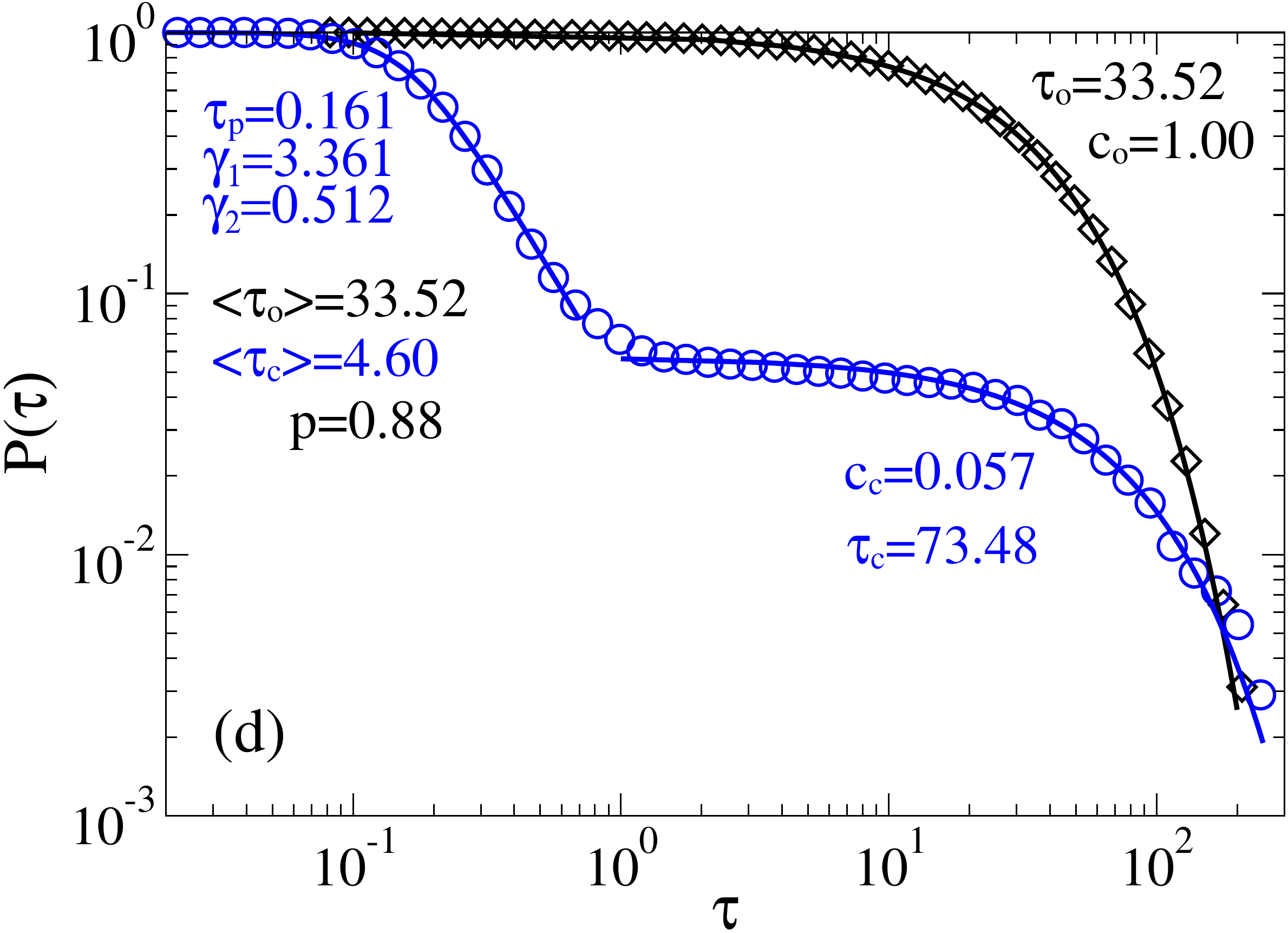}}
\caption{(color online) Survival probabilities of open (black diamond symbols), and
closed (blue open circles) times derived from numerical data 
for the case $\mu B=0.4363$, $\psi=\pi$
using different sets 
of detection thresholds: (a) $p_1=0.2$, $p_2=0.7$; (b) $p_1=0.1$, $p_2=0.9$,
(c) for the minima of $U(\phi)$ used as detection thresholds, $p_1=p(\phi_{\rm min,1})$,
 $p_2=p(\phi_{\rm min,2})$; (d) 
$p_1=0.5$, $p_2=0.9$.
Lines present the corresponding exponential, $c_i\exp(-\tau/\tau_i)$, and the 
Pareto law (\ref{Pareto}) fits  with the parameters shown in the plots.
The mean residence times, as well as the opening
probability 
$p=\langle \tau_o\rangle/(\langle \tau_o\rangle+\langle \tau_c\rangle)$ are also 
displayed. 
Notice a strong dependence of these fits on the detection thresholds used.
Other parameters:
 $T=0.1$,
$l_{\rm max}=1.5$, $f_0=1.5$, $r_0=1.22$, and $m=7$ channels in the sensor cluster.}
\label{Fig7}       
\end{figure*}

Transitions  from open to closed state are regarded as accomplished
by a downward crossing of the low threshold. 
In turn, transitions from closed to open state are accomplished
by an upward crossing of the high threshold. 
In  this way, one finds pairs of random time intervals in the closed, $\tau_c$, and open, $\tau_o$,
states whose statistics is subsequently studied. For a very long trajectory
of the kind depicted in Fig. \ref{Fig5}, the survival probabilities $P_o(\tau)$ and
$P_c(\tau)$, which
correspond to residence time distributions, $\psi_i(\tau)=-dP_i(\tau)/\tau$, $i=o,c$, 
are shown in Fig. \ref{Fig6}, (a)-(d). They are derived from the same single very long trajectory by
using different sets of thresholds. As a test of ergodicity (in distribution), 
we derived probability density $\hat P(\phi)$ of sensor orientations
from a single trajectory and compared it with the ensemble equilibrium $P_{\rm eq}(\phi)\propto \exp[-U(\phi)/(k_BT)]$.
Excellent agreement (not presented) implies ergodicity.
 In Fig. \ref{Fig6}, we provide fits of the tail of distributions
with the exponential functions $c_i\exp(-\tau/\tau_i)$,  weight $c_i$ and 
time constant $\tau_i$. Furthermore, the initial part of closed
time distribution therein is nicely fitted by the 4-th type Pareto law with survival probability \cite{Pareto}
\begin{eqnarray}\label{Pareto}
P(\tau)=\frac{1}{[1+(\tau/\tau_p)^{\gamma_1}]^{\gamma_2}}\;.
\end{eqnarray}
For $\tau\gg \tau_p$, $P(\tau)\propto\tau^{-\gamma}$ with $\gamma=\gamma_1\gamma_2$.
It reflects
the short time bursting dynamics. Notice that in Fig. \ref{Fig6} (a) the open time
distribution is almost single exponential, $c_o\approx 0.966$, while the weight
of the exponential tail of closed time distribution is about $c_c\approx 0.4$ only,
and nearly 60\% of distribution is described by the Pareto law (\ref{Pareto}).
With lowering the detection threshold $p_1$ in Fig. \ref{Fig6} (b) to $p_1=0.1$ the weight
of exponential tail increases to 
$c_c\approx 0.62$. This is because less bursting events are detected. 
The variation of the time constant $\tau_c$ describing mean residence time in the closed
state with a variation of detection thresholds n Fig. \ref{Fig6} (a)-(c)
is not statistically significant. Notice, that a two-stage relaxation of the closed times
$P(\tau)$ cannot be described by a simple sequential Markovian  scheme with just two closed substates,
$C_2\leftrightarrow C_1\to O$. This is because: (i)  the relative weight of two stages
depends very essentially on the detection threshold, (ii) initial stage is described by the Pareto
 power law (\ref{Pareto}) and not by an exponential. 
 
Furthermore, if to detect
transitions by crossing $p_{1,2}$ levels corresponding to the minima  of $U(\phi)$ (set 3)
then the survival probabilities become practically
single exponential, see in  Fig. \ref{Fig6} (c), where the mean times coincide
with the corresponding time constants. In this case, the bursts within long opening
events are completely neglected. The mean residence
times increase accordingly. The standard Kramers  rate result in (\ref{rate_normal}) gives 
$R_1^{(0)}\approx 0.0038052$ and $R_2^{(0)}\approx 0.0030899$. The corresponding
inverse values $1/R_1^{(0)}\approx 262.80$ and $1/R_2^{(0)}\approx 323.63$ agree
with $\langle \tau_c\rangle \approx 251.96$ and $\langle \tau_o\rangle \approx 313.24$, 
correspondingly, in Fig. \ref{Fig6} (c) 
within a 4\% error margin. This is a typical accuracy of our stochastic
simulations. However, if to use $p_1=0.5$ for the lower threshold,
what experimentalists can prefer, and to keep $p_2=0.9$, a more complex picture
emerges for the closed residence times, see in Fig. \ref{Fig6} (d). It reveals two
characteristic power law regimes described by Pareto laws. The initial Pareto law 
 describes almost 90\% of the probability decay with $\gamma=\gamma_1\gamma_2=1.405$.
 It corresponds to the residence time distribution $\psi(\tau)\propto 
 \tau^{-\delta}$, with $\delta=1+\gamma=2.405$. Interestingly, similar power laws
 were indeed derived from experimental recordings of large conductance 
 BK ion channels \cite{Gorczynska}, which resemble the open-shut dynamics in our Fig. \ref{Fig5}. However, 
 a very different phenomenological theory has been
 proposed earlier for such a gating dynamics \cite{GoychukPRE04}. The second
Pareto law part (\ref{Pareto}) in Fig. \ref{Fig6} (d) has 
weight $c_p\approx 0.079$ and describes thus about last 8\% of decay.
Unexpectedly, an exponential tail fit of the data for the closed times in 
Fig. \ref{Fig6} (d) is worser, although its time constant correlates
with one in Fig. \ref{Fig6} (c), as expected. 
The distribution of open times remains, however, practically
single exponential independently of thresholds with a threshold-dependent time constant. 
Our results show that 
a continuous-state Markovian bistable dynamics can explain such experimental non-Markovian 
features (within a contracted two-state non-Markovian dynamics),
as bursting and power law distributions of the residence times. They can be caused by a nontrivial interplay
of coupled sensor and gate dynamics, as well as threshold levels used for
detection. What we measure depends really on how we detect.
This is a signature of complex dynamics. 
Notice that the opening probability within the two-state interpretation of continuous
state dynamics is rather robust with respect to the choice of the detection
thresholds.  It  agrees fairly with the ensemble average in (\ref{pens}) done for
the continuous-state dynamics.

With increasing the number of magnetosomes in the sensor rod to $n=7$ (or for a correspondingly
larger single nanoparticle used as sensor), the probability
of channels to be open in the magnetic field of the Earth 
increases to over
$p=0.8$ at a proper field orientation, see in Fig. \ref{Fig3}. The  residence times statistics
displays similar features, see in Fig. \ref{Fig7} (a)-(d). Namely,
the open times are nearly exponentially distributed, whereas the distribution
of closed times depends strongly on the lower detection
threshold. The lower the threshold, the closer  is distribution to a single exponential.
The rate theory yields $R_1^{(0)}\approx 0.012306$, and the corresponding
$1/R_1^{(0)}\approx 81.26$ agrees well, within $2.25$ \% error margin, with
the numerical value of mean close time $\langle \tau_c\rangle\approx 79.43$ in Fig. \ref{Fig7} (c). 
For the corresponding open times, the rate theory yields $1/R_2^{(0)}\approx 708.11$,
which agrees in this case somewhat worser, within  $10.2\%$ error margin
with numerical $\langle \tau_o\rangle\approx 635.91$  in Fig. \ref{Fig7} (c).
However, experimentalists  can reveal power law features related to bursting, if 
they set $p_1=0.5$, or somewhat lower.
So, in Fig. \ref{Fig7} (d) Pareto  law with $\gamma=\gamma_1\gamma_2\approx 1.721$ covers about
94\% of the survival probability decay of the closed time distribution, 
which ends by an exponential tail with the weight of about 
5.7\% and decay time constant $\tau_c$ which weakly depends on the choice of threshold,
see in Fig. \ref{Fig7}, from (a) to (d). The Pareto law describes durations
of closed time intervals within a burst, whereas the time constant $\tau_c$ corresponds
approximately to the mean time interval between the bursts, when the channels are well closed.
Clearly, the time constant of open times $\tau_o$ depends very essentially on the
choice of lower detection threshold.

Our results show that if the sensor motion would occur in water
its dynamics would be reasonably fast being in hundreds of milliseconds range. 
Hence, it could serve as a detector for quasi-static or slowly changing magnetic fields.
However, in cytosol the effective friction $\eta_{\rm eff}$ is enhanced dramatically 
for the particles
of a typical size of magnetosomes. Kirschvink \textit{et al.} estimated $\tilde \eta_{\rm eff}=
\eta_{\rm eff}/\eta_0$ by a factor of $\tilde \eta_{\rm eff}=100$ \cite{Kirschvink92,Eder}. If to just renormalize
the Stokes friction by this factor, the effective time scale $\tau_{sc}$ would enlarge
accordingly, $\tau_{sc}\to ( \tilde\eta_{\rm eff}+1)\tau_{sc}$. Then, our sensor would be really too slow
to be functional. However, accounting for non-Markovian memory effects by a naive renormalization of the
friction coefficient is patently wrong within the context of thermally activated dynamics,
as it is well-recognized within non-Markovian generalizations of Kramers 
rate theory \cite{Grote,HanggiMojtabai,Nitzan,Pollak,HTB90}.
Therefore, it is compelling to clarify the role of low-dimensional 
non-Markovian memory effects.

\begin{figure}
\vspace{1cm}
\resizebox{0.95\columnwidth}{!}{\includegraphics{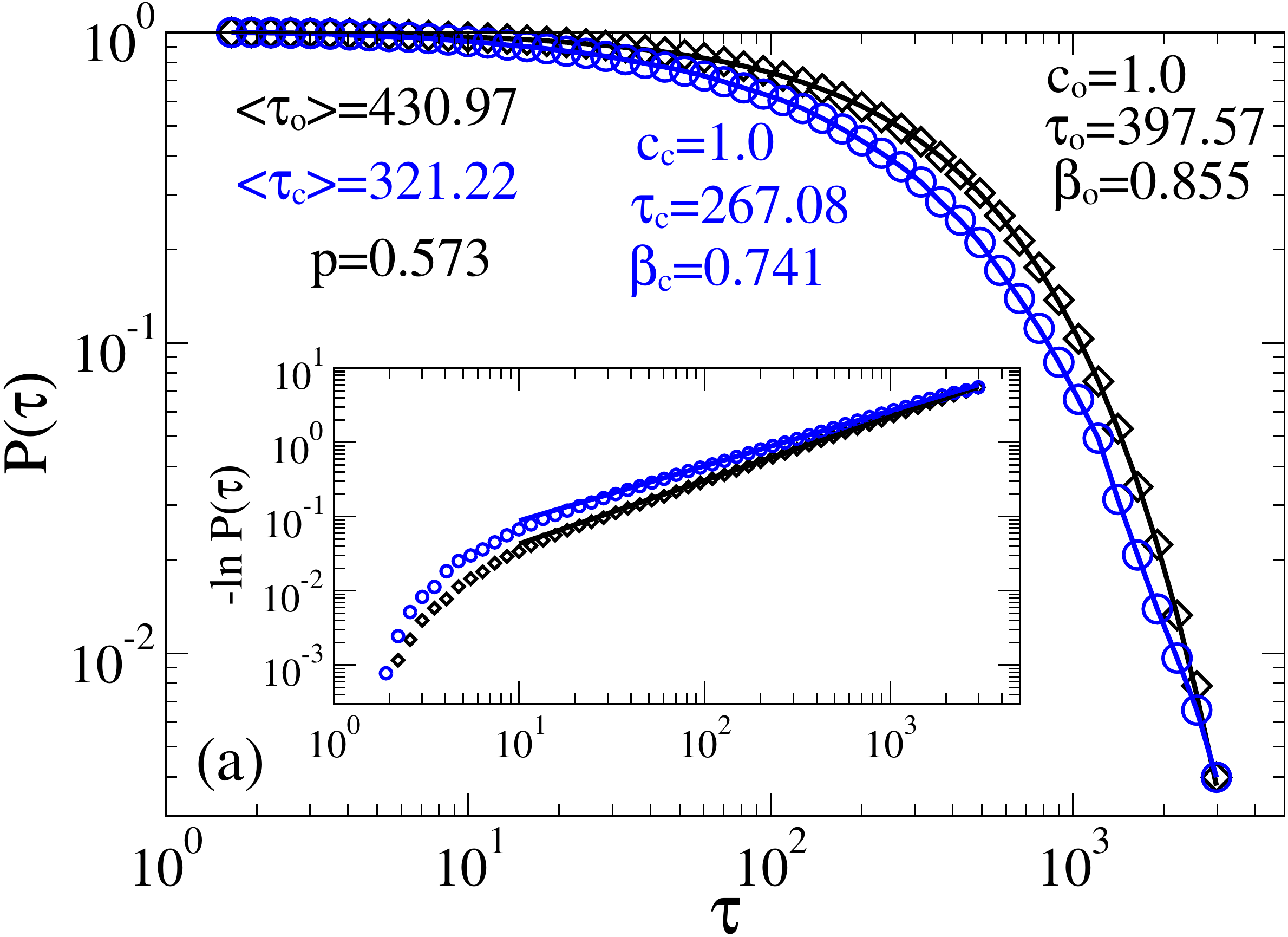}}\\ \vspace{1cm}
\resizebox{0.95\columnwidth}{!}{\includegraphics{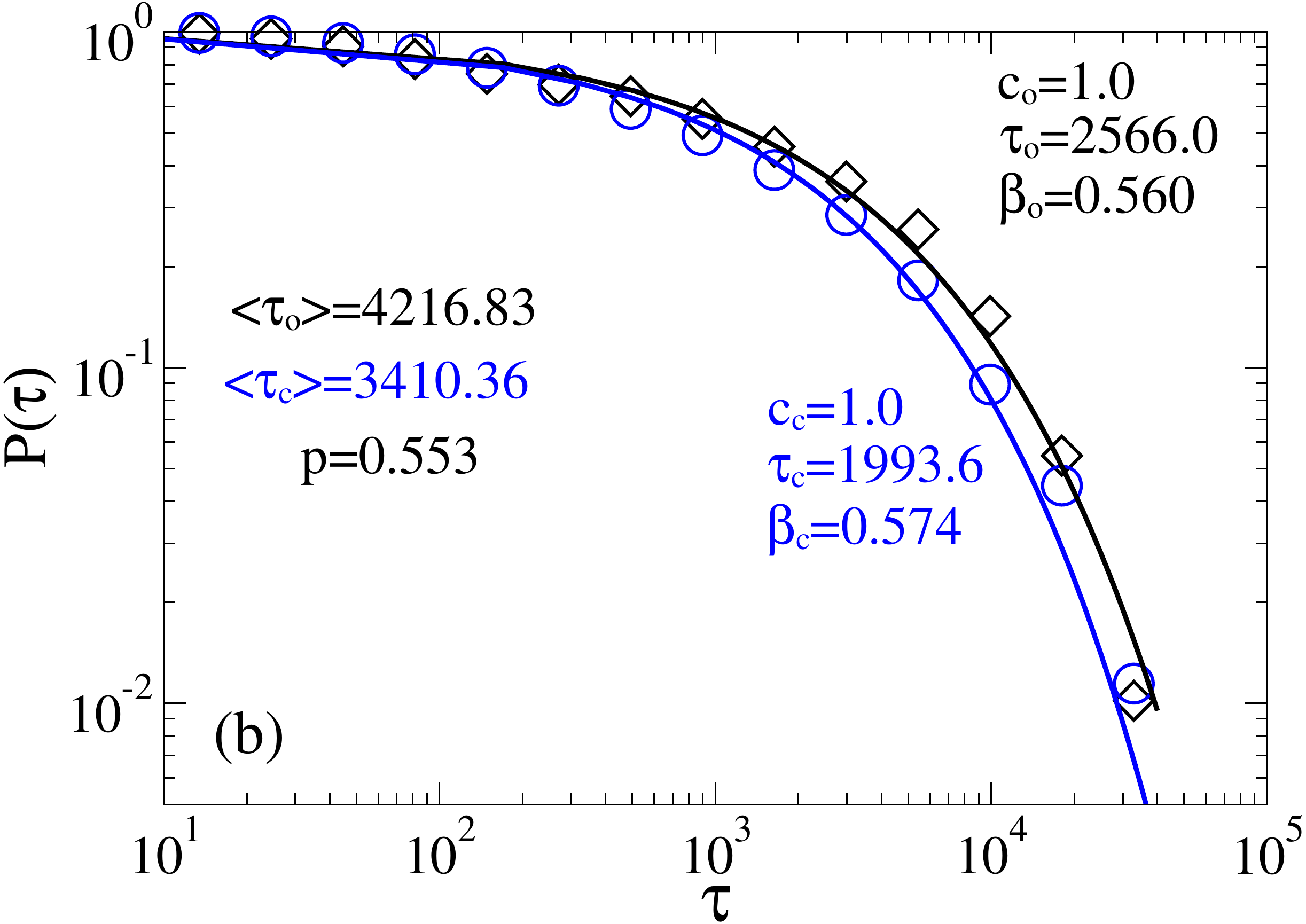}}
\caption{(color online) Survival probabilities of open (black diamond symbols), and
closed (blue open circles) times derived from numerical data 
for the case $\mu B=0.3115$, $\psi=\pi$, 
using  the detection thresholds defined by the minima of $U(\phi)$, $p_1=p(\phi_{\rm min,1})$,
 $p_2=p(\phi_{\rm min,2})$, for (a) $\eta_{\rm eff}=100$,
and (b) $\eta_{\rm eff}=1000$.
Lines present the corresponding stretched exponential, $c_i\exp[-(\tau/\tau_i)^\beta_i]$,
 fits with the parameters shown in the plots.
The mean residence times, as well as the opening
probability  
$p=\langle \tau_o\rangle/(\langle \tau_o\rangle+\langle \tau_c\rangle)$ are also displayed. 
Other parameters:
 $T=0.1$,
$l_{\rm max}=1.5$, $f_0=1.5$, $l_0=1.22$, and $m=7$ channels in the sensor cluster.
}
\label{Fig8}       
\end{figure}

\begin{figure}
\vspace{1cm}
\resizebox{0.95\columnwidth}{!}{\includegraphics{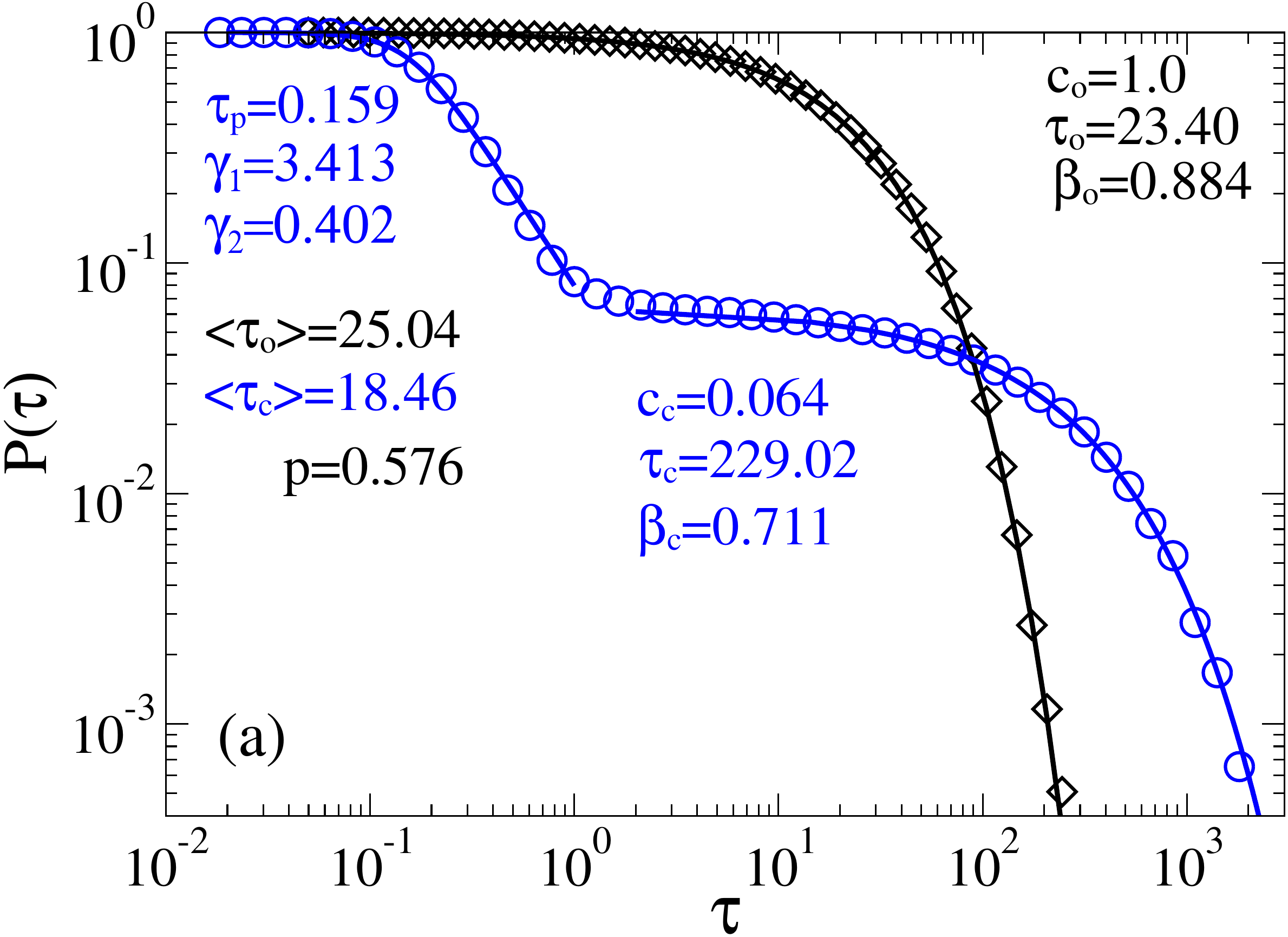}} \\ \vspace{1cm}
\resizebox{0.95\columnwidth}{!}{\includegraphics{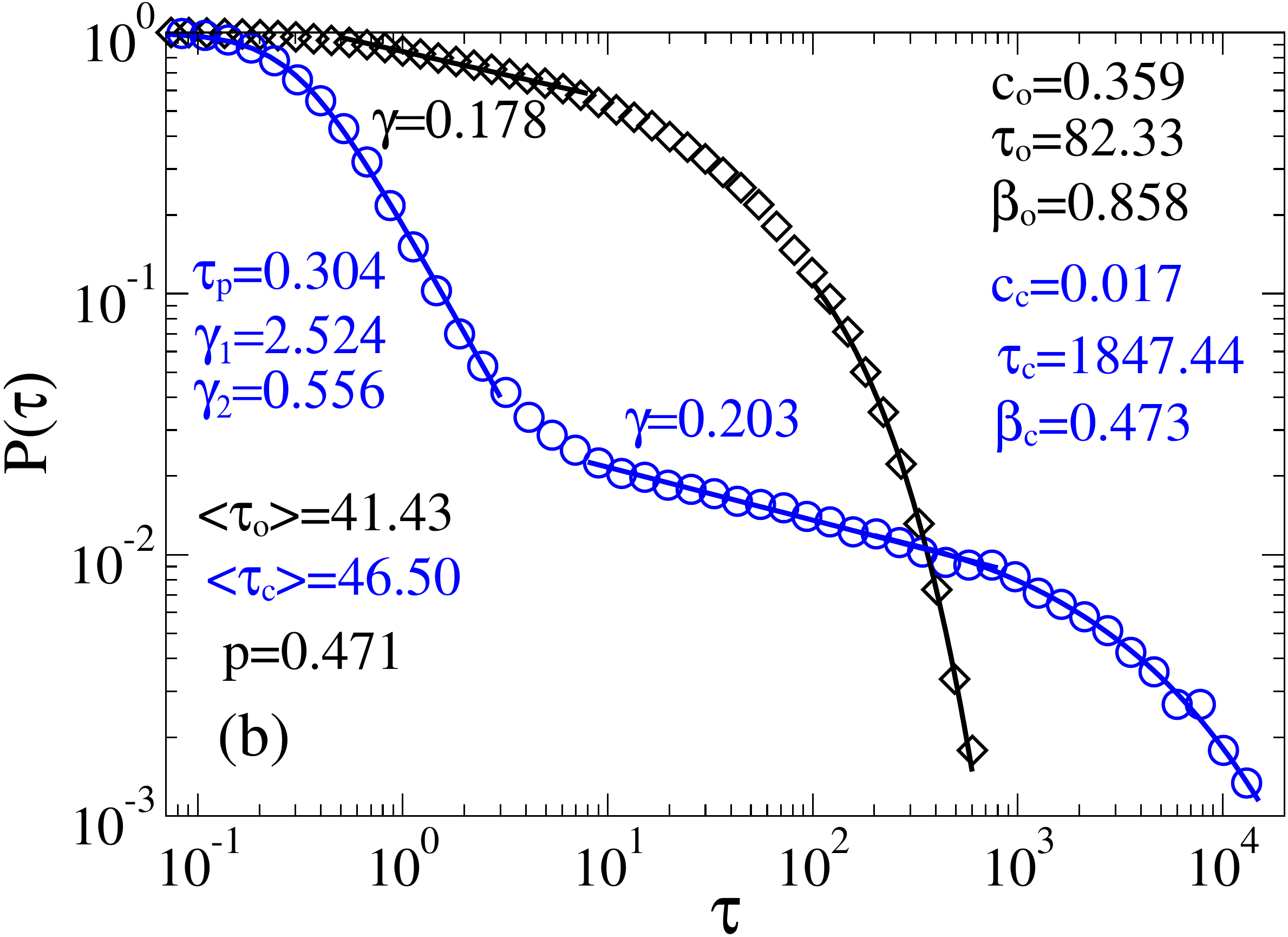}}
\caption{(color online) Survival probabilities of open (black diamond symbols), and
closed (blue open circles) times derived from numerical data 
for the case $\mu B=0.3115$, $\psi=\pi$, 
using  the detection thresholds $p_1=0.5$ and $p_2=0.9$ for (a) $\eta_{\rm eff}=100$,
and (b) $\eta_{\rm eff}=1000$.
Lines present the corresponding stretched exponential, $c_i\exp[-(\tau/\tau_i)^\beta_i]$,
 Pareto law (\ref{Pareto}),  and
power law, $\sim \tau^{-\gamma}$, fits with the parameters shown in the plots.
The mean residence times, as well as the opening
probability 
$p=\langle \tau_o\rangle/(\langle \tau_o\rangle+\langle \tau_c\rangle)$ are also displayed. 
Other parameters:
 $T=0.1$,
$l_{\rm max}=1.5$, $f_0=1.5$, $l_0=1.22$, and $m=7$ channels in the sensor cluster.
}
\label{Fig9}       
\end{figure}

\begin{figure}
\vspace{1cm}
\resizebox{0.95\columnwidth}{!}{\includegraphics{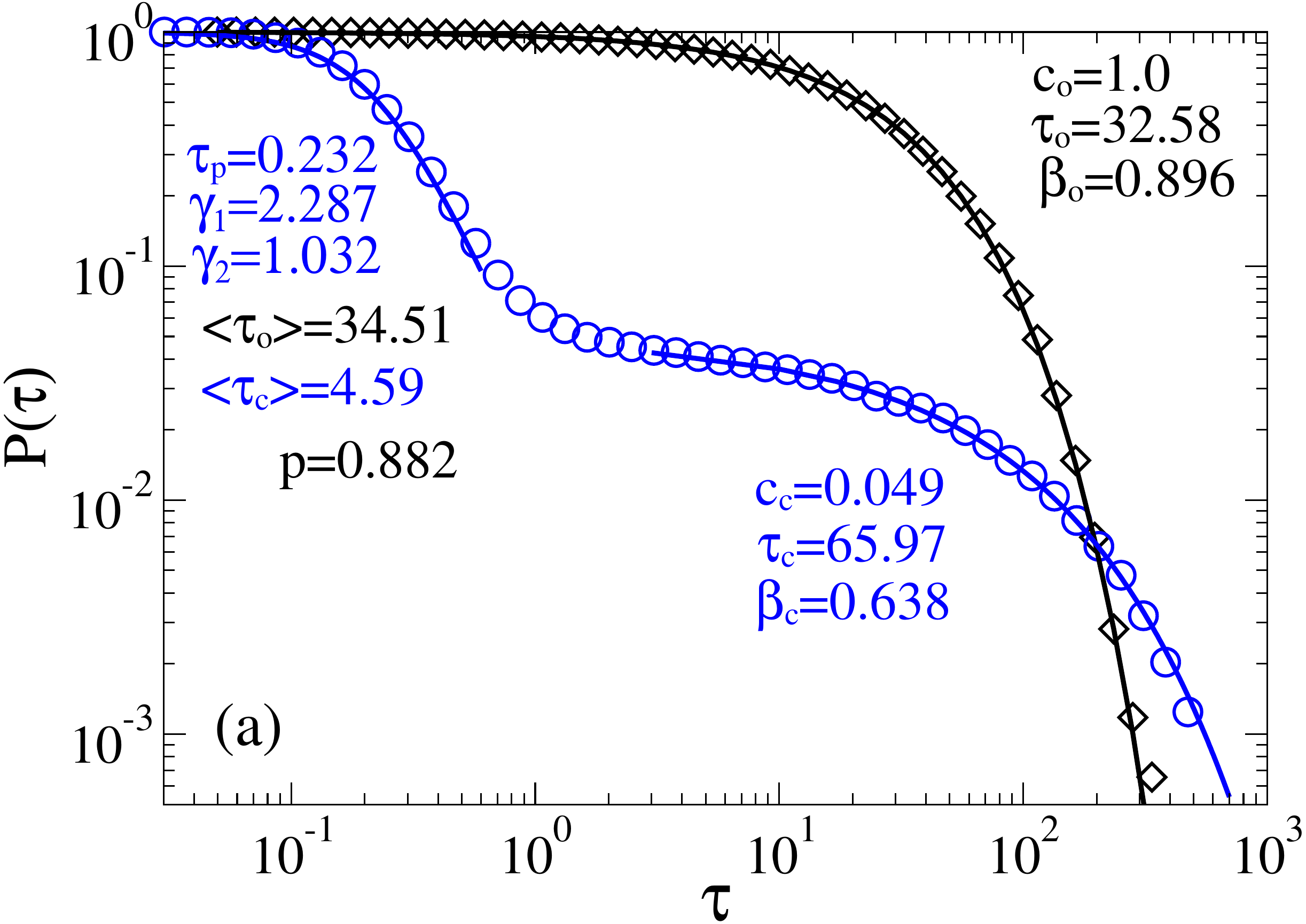}} \\ \vspace{1cm}
\resizebox{0.95\columnwidth}{!}{\includegraphics{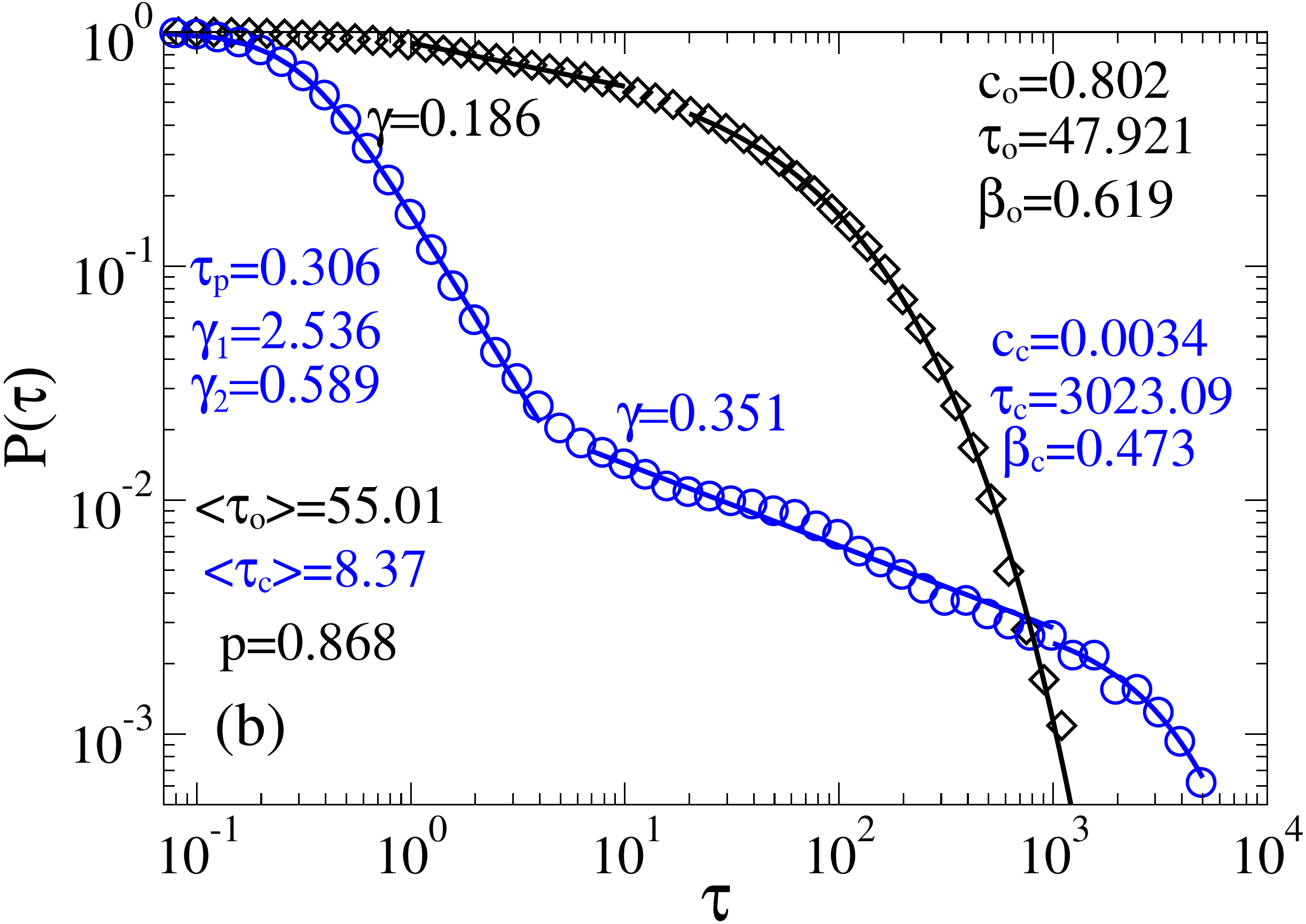}}
\caption{(color online) Survival probabilities of open (black diamond symbols), and
closed (blue open circles) times derived from numerical data 
for the case $\mu B=0.4346$, $\psi=\pi$, 
using  the detection thresholds $p_1=0.5$ and $p_2=0.9$ for (a) $\eta_{\rm eff}=100$,
and (b) $\eta_{\rm eff}=1000$.
Lines present the corresponding stretched exponential, $c_i\exp[-(\tau/\tau_i)^\beta_i]$,
 Pareto law (\ref{Pareto}),  and
power law, $\sim \tau^{-\gamma}$, fits with the parameters shown in the plots.
The mean residence times, as well as the opening
probability 
$p=\langle \tau_o\rangle/(\langle \tau_o\rangle+\langle \tau_c\rangle)$ are also displayed. 
Other parameters:
 $T=0.1$,
$l_{\rm max}=1.5$, $f_0=1.5$, $l_0=1.22$, and $m=7$ channels in the sensor cluster.
}
\label{Fig10}       
\end{figure}

\begin{figure}
\vspace{1cm}
\resizebox{0.95\columnwidth}{!}{\includegraphics{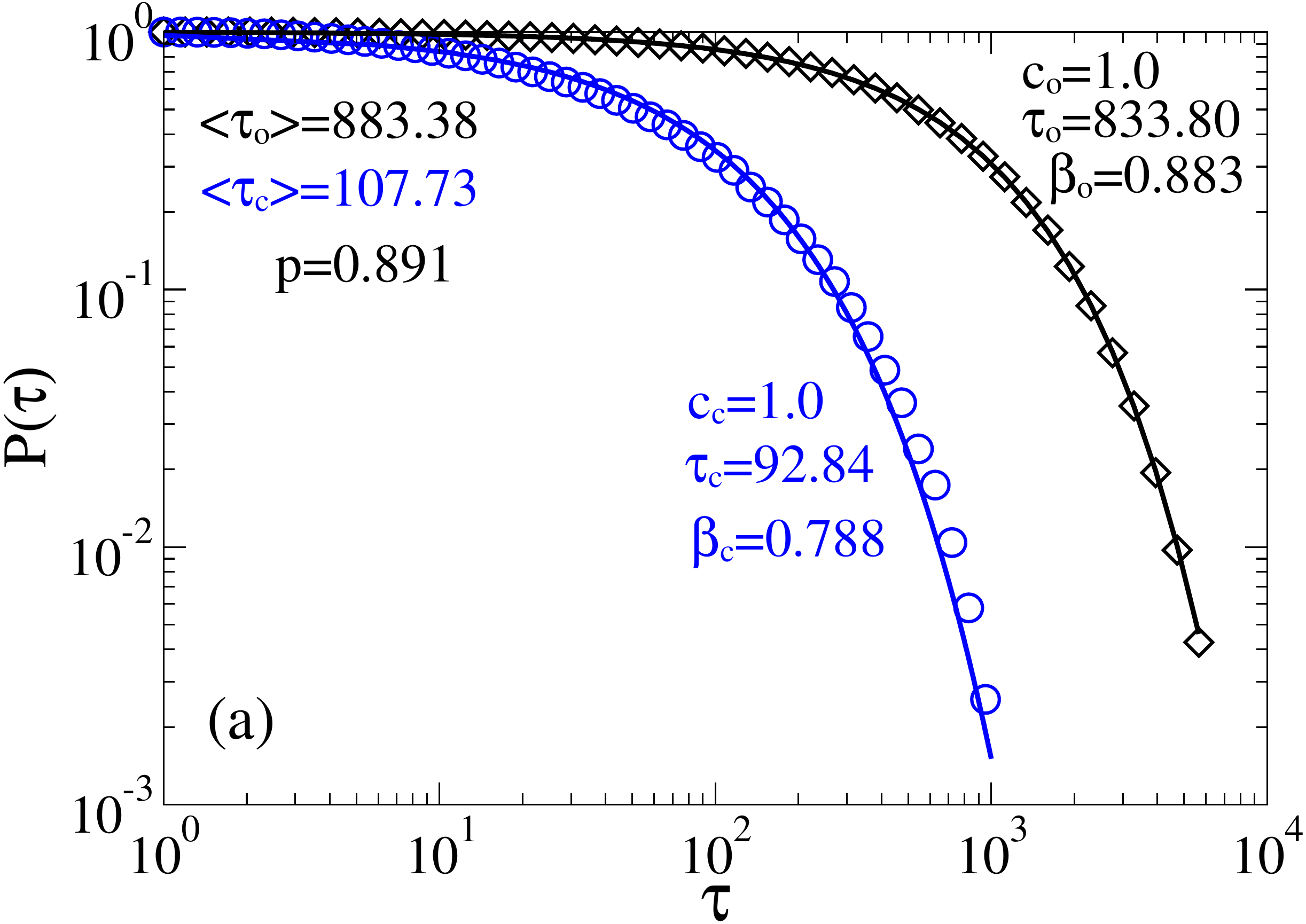}} \\ \vspace{1cm}
\resizebox{0.95\columnwidth}{!}{\includegraphics{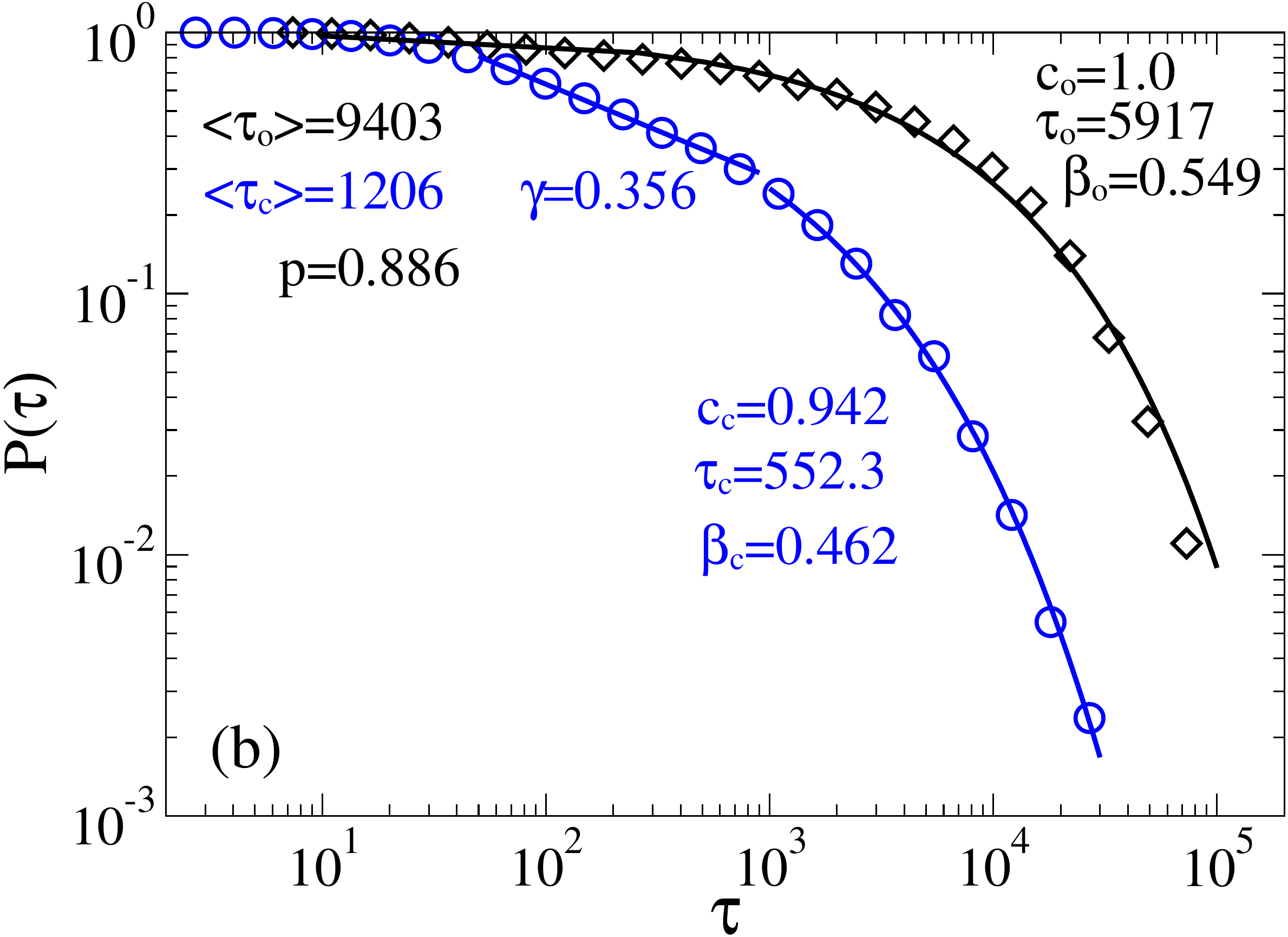}} \\
\caption{(color online) Survival probabilities of open (black diamond symbols), and
closed (blue open circles) times derived from numerical data 
for the case $\mu B=0.4363$, $\psi=\pi$, 
using  the detection thresholds defined by the minima of $U(\phi)$, $p_1=p(\phi_{\rm min,1})$,
 $p_2=p(\phi_{\rm min,2})$, for (a) 
$\eta_{\rm eff}=100$, and (b) $\eta_{\rm eff}=1000$.
Lines present the corresponding stretched exponential, $c_i\exp[-(\tau/\tau_i)^\beta_i]$,
and power law fits with the parameters shown in the plots.
The mean residence times are also shown along with the opening
probability 
$p=\langle \tau_o\rangle/(\langle \tau_o\rangle+\langle \tau_c\rangle)$. 
Other parameters:
 $T=0.1$,
$l_{\rm max}=1.5$, $f_0=1.5$, $l_0=1.22$, and $m=7$ channels in the sensor cluster.}
\label{Fig11}       
\end{figure}

\subsection{Role of memory effects}

Stochastic simulations of Eq. (\ref{embedding}) were done for the 
following additional parameters, $\alpha=0.4$ \cite{Wilhelm,Bruno,GKhMet14b,Goychuk15}, $\nu_0=10^{4}$,
$N=9$ and $b=10$. Hence, $\tau_l=10^{-4}$ and $\tau_h=10^4$. The latter time is about $4.04$ sec 
for $\tau_{sc}\approx 0.404$ msec, or about $9.05$ sec for $\tau_{sc}\approx 0.905$ msec. 
Furthermore, we used
two values of $\eta_{\rm eff}$, $\eta_{\rm eff}=100$, and $\eta_{\rm eff}=1000$, which correspond
to  two different values of 
$\eta_{\alpha}=\eta_{\rm eff}\tau_h^{\alpha-1}/g_{\alpha}$ in simulations,  and
 the rotational subdiffusion coefficients $D_\alpha\approx {\rm 0.360\; rad^2/msec^{0.4}}$
and $D_\alpha\approx {\rm 0.036\; rad^2/msec^{0.4}}$ at $\tau_{sc}\approx 0.404$ msec, 
or $D_\alpha\approx {\rm 0.260\; rad^2/msec^{0.4}}$
and $D_\alpha\approx {\rm 0.026\; rad^2/msec^{0.4}}$ at $\tau_{sc}\approx 0.905$ msec, respectively. 
In this respect, 
one can enlarge both $\eta_{\rm eff}$ and $\tau_h$ (by using a larger $N$), while keeping
$\eta_{\alpha}$ the same. For example, by taking $N=11$, and enlarging $\tau_h$ to
about $404$ sec and $905$ sec, correspondingly, we can keep $\eta_{\alpha}$ the
same by enlarging $\eta_{\rm eff}$ from $100$ to approximately $1585$ and from
$1000$ to approximately $15850$, respectively. Furthermore, for the used parameters
we have $\tau_{in}\approx 4.64$ for $\eta_{\rm eff}=100$, and  
$\tau_{in}\approx 0.01$ for $\eta_{\rm eff}=1000$. 
For $\tau_{sc}\approx 0.404$ msec, this corresponds
to $\tau_{in}\approx 1.88$ msec ($\eta_{\rm eff}=100$) or $\tau_{in}\approx 0.04$ msec 
($\eta_{\rm eff}=1000$). Furthermore, for $\tau_{sc}\approx 0.905$ msec, this corresponds
to $\tau_{in}\approx 4.20$ msec ($\eta_{\rm eff}=100$) or $\tau_{in}\approx 0.09$ msec 
($\eta_{\rm eff}=1000$).

The first striking universal feature of the influence of viscoelastic memory
effects is the emergence of a stretched exponential distribution 
$c_i\exp[-(\tau/\tau_i)^{\beta_i}], 0<\beta_i<1$ out 
of formerly exponential one, compare Fig. \ref{Fig6} (c) and Fig. \ref{Fig8} (a), where
$\eta_{\rm eff}=100$, for other parameters being the same in both figures, and detection
thresholds set at the minima of $U(\phi)$.
In Fig. \ref{Fig8} (a), the residence time distributions (RTDs) become
stretched exponential (Weibull distribution) to a good degree, $c_{o}\approx 1$, $c_{c}\approx 1$,
 except for some initial times, see in the inset, where a deviation
from linear dependence  of 
$-\ln P(\tau)\propto (\tau/\tau_i)^{\beta_i}$ plotted on the double-logarithmic scale becomes seen. 
For this choice of thresholds, we 
can compare the results of non-Markovian rate theory (NMRT), namely
the  Grote-Hynes result in 
Eqs. (\ref{rate}), (\ref{disp1}) with our
numerics for the particular memory kernel studied. As observed in Ref. \cite{Goychuk09}, the Grote-Hynes result
is capable to
describe the most probable value of the logarithmically transformed residence
times, $\ln \tau$, even beyond the strict rate regime, when RTDs become profoundly non-exponential. 
For the case of Weibull distribution characterized
by the survival probability $P_i(\tau)=\exp[-(\tau/\tau_i)^{\beta_i}]$, and
non-Markovian rate $R(\tau)=-d \ln P(\tau)/d\tau=\beta_i\tau_i^{-1}/(\tau/\tau_i)^{1-\beta_i}$,
the correspondingly  transformed distribution of $y=\ln \tau$ has its maximum at
$y_{\rm max}=\ln \tau_i$. Hence, in accordance with \cite{Goychuk09} NMRT is expected
to yield $\tau_i$ of Weibull distribution 
as $\tau_i=R_i^{-1}$. For the results depicted in Fig. \ref{Fig8} (a),  
the Grote-Hynes result yields 
$R_i=0.8169\;R_i^{(0)}$, for the exact memory kernel, 
where $R_i^{(0)}$ is the overdamped Kramers result. It has been already
discussed above in relation to our numerical results presented in Fig. \ref{Fig6} (c).
Hence, NMRT yields $\tau_{o,\rm exact}^{(NMRT)}\approx 396.17$ and 
$\tau_{c,\rm exact}^{(NMRT)}\approx 321.70$, for the exact memory kernels.
For the memory kernels approximated by a sum of exponentials,
we obtain $\tau_{o}^{(NMRT)}\approx 395.30$,
$\tau_{c}^{(NMRT)}\approx 321.00$. 
 Notice that the kernel approximation introduces an error of about 
$0.2\%$ only in theoretical NMRT results! The value $\tau_{o,\rm exact}^{(NMRT)}\approx 396.17$
 agrees very well with $\tau_o=397.57$ in Fig. \ref{Fig8} (a)
The difference is less than 0.4\%!  Furthermore,  $\tau_{c,\rm exact}^{(NMRT)}\approx 321.70$
somewhat overestimates numerical $\tau_c\approx 267.08$. However, it agrees well with
the numerical average $\langle \tau_c\rangle \approx 321.22 $.
Also for the results in Fig. \ref{Fig8} (b), where
we increase $\eta_{\rm eff}$ to $\eta_{\rm eff}=1000$, for the same other
parameters, NMRT yields $\tau_{o,\rm exact}^{(NMRT)}\approx 2531.56$ and 
$\tau_{c,\rm exact}^{(NMRT)}\approx 2055.68$, for the exact memory kernels,
and $\tau_{o}^{(NMRT)}\approx 2450.84$ and 
$\tau_{c}^{(NMRT)}\approx 1990.13$, for the approximate kernels.
The error due to the memory kernel approximation now increases to about $3.2\%$.
$\tau_{o,\rm exact}^{(NMRT)}\approx 2531.56$  agrees
with $\tau_o\approx 2566.0$ derived from the numerical data using maximum likelihood
fitting within 1.4\% error margin. Also $\tau_{c,\rm exact}^{(NMRT)}\approx 2055.68$ agrees with 
the maximum likelihood value $\tau_c\approx 1993.6$ within 3\% error margin.
Notice that deviation from the numerical mean values, $\langle \tau_o\rangle \approx 4216.83 $
and $\langle \tau_c\rangle \approx 3410.36 $ in this case is very essential. 
In this respect, NMRT describes the most probable $\ln \tau$ generally
 much better than the mean residence
time $\langle \tau\rangle$. Only in the strict rate
regime or for $\beta_i$ close to one 
does NMRT describe well $\langle \tau\rangle$, which coincides with the effective
inverse rate thus defined, and is also the so-called stationary flux-over-population
rate \cite{HTB90}.
Our results confirm that NMRT is of a high predictive value even beyond
the strict rate regime \cite{Goychuk09,Goychuk12}. What it, however, cannot do is to predict the form
of non-exponential residence time distributions, or detailed course of non-Markovian
kinetics.

Furthermore, in Fig. \ref{Fig9} the results are depicted for the same anomalous kinetics  
discussed in Fig. \ref{Fig8}, but for the detection thresholds $p_1=0.5$
and $p_2=0.9$ (set 4). For the smaller value of $\eta_\alpha$ in Fig. \ref{Fig9} (a),
one can see, by comparison with Fig. \ref{Fig6} (d) that the initial Pareto law for the closed 
times is  not changed dramatically. This is because on the corresponding initial 
time scale dynamics is nearly normal. The profound changes are reflected by the tail
of the closed time survival probability and by the whole open time distribution, which
are stretched exponential. The average residence times are only slightly enlarged.
However, with the increase of $\eta_{\alpha}$ some further changes  become detectable
in  Fig. \ref{Fig9} (b). Namely, the mean residence times increase, which is expected,
however, not so strongly as for another detection threshold choice, compare with 
Fig. \ref{Fig8} (b). Next, an initial power law regime can be revealed in the distribution
of open times, and the weight of a stretched exponential tail is only $c_o\approx 0.359$
\textit{versus} $c_o\approx 1$  in Fig. \ref{Fig9} (a), where stretched exponential covers
practically over all
essential transition times (except for very short times, see inset in Fig. \ref{Fig8} (a)). 
Also an intermediate power law emerges for closed times in  Fig. \ref{Fig9} (b),
and the weight of stretched exponential tail becomes smaller: $c_c\approx 0.017$ 
in Fig. \ref{Fig9} (b) \textit{vs. } $c_c\approx 0.064$ in
Fig. \ref{Fig9} (a). Moreover, the initial Pareto law is also changed in  Fig. \ref{Fig9} (b).
This is because for larger $\eta_{\rm eff}=1000$ diffusion is not normal anymore
on the related time scale.
Similar characteristic features hold also for other choices of thresholds,
see in Supplement \cite{Suppl}.

For the same detection thresholds, but a larger $\mu B=0.4163$ 
the results are depicted  in Fig. \ref{Fig10} (a), (b). These results are to be compared 
with the results in  Fig. \ref{Fig7} (d), in the absence of non-Markovian effects.
Again, for a smaller $\eta_{\alpha}$ ($\eta_{\rm eff}=100$), the mean residence times
are almost not influenced, albeit the initial power law parameters in the 
survival probability of closed tomes are somewhat changed. The most essential typical change
is conversion of exponential residence time distribution of open times and
an exponential tail of the closed time distribution into the stretched exponential dependencies.
The larger $\eta_{\alpha}$ in Fig. \ref{Fig10} (b) leads to expectedly larger mean residence times, 
and smaller $\beta_i$  of stretched exponential tails. Notice also
the emergence of an intermediate power law for the closed times, and an initial power
law regime for the open times. Furthermore,
survival probabilities corresponding to the thresholds set
in accordance with  the $U(\phi)$ minima are shown in Fig. \ref{Fig11}.
They are clearly stretched exponential in Fig. \ref{Fig11} (a)  with the parameters derived
 using the maximum likelihood criterion. In this case, NMRT
predicts $\tau_{o,\rm exact}^{(NMRT)}\approx 887.93$ and 
$\tau_{c,\rm exact}^{(NMRT)}\approx 101.90$, for the exact memory
kernels, and $\tau_{o}^{(NMRT)}\approx 885.68$,
$\tau_{c}^{(NMRT)}\approx 101.64$, for the approximate ones.
$\tau_{o,\rm exact}^{(NMRT)}$ agrees with the numerical  $\tau_o\approx 833.8$ 
within $6.1\%$ error
margin, while the agreement of  $\tau_{c,\rm exact}^{(NMRT)}$  with 
the numerical  $\tau_c\approx 92.84$
is worser, about $8.9\%$ discrepancy. Interestingly, in this case the mean values, 
$\langle \tau_o\rangle \approx 883.38$ and $\langle \tau_o\rangle \approx 107.73$ 
agree with the results of NMRT theory better, within $0.51\%$ and $5.72\%$ error
margins for the exact memory kernels. This shows that for larger 
$\beta_i$, like $\beta_0=0.883$
and $\beta_c=0.788$ in this plot the difference between $\tau_i$ determined 
from the most probable $\ln \tau_i$ 
and the corresponding mean values $\langle \tau_i\rangle$ can be 
within the actual statistical errors of our simulations.
In Fig. \ref{Fig11} (b),
the results for $\eta_{\rm eff }=1000$ are presented, at the same other parameters. 
In this figure, a maximum likelihood fit by a stretched exponential with
$\beta_0\approx 0.549$ and $\tau_o\approx 5.917 \times 10^3$ is shown
for the open time distribution. The closed time kinetics is more complex. It
reveals an intermediate power law, $\gamma\approx 0.356$, and a
 stretched exponential tail, and therefore is not expected to be described by NMRT.  
Let  us compare the analytical NMRT results with 
the numerical results depicted in Fig. \ref{Fig11} (b). NMRT yields
$\tau_{o,\rm exact}^{\rm (NMRT)}\approx 6.817\times 10^3$ and 
and $\tau_{c,\rm exact}^{\rm (NMRT)}\approx 0.782\times 10^3$, for the exact memory
kernels, and $\tau_{o}^{\rm (NMRT)}\approx 6.553\times 10^3$ and 
 $\tau_{c}^{\rm (NMRT)}\approx 0.752\times 10^3$, for the approximate memory kernels
 used in simulations. The error introduced by the memory kernel approximation is
about $3.9\%$ in this case. The numerical mean values obtained on 1267 
transitions (which required about 9 weeks of simulations),
 $\langle \tau_o\rangle \approx 9.403\times 10^3$, and  
 $\langle \tau_c\rangle \approx 1.207\times 10^3$ are larger
 than $\tau_{o,c}^{\rm (NMRT)}$ by 43.5\% and 60.5\%, correspondingly.
 At the same time, the maximum likelihood value $\tau_o\approx  5.917 \times 10^3$ deviates from 
 $\tau_{o}^{\rm (NMRT)}$ by about 6.6\% only. 
However, NMRT clearly fails to describe characteristic features of the closed time distribution, which
has a significant power law part.
Two-dimensional densities corresponding to Fig. \ref{Fig11} (a),
and Fig. \ref{Fig11} (b) in Supplement \cite{Suppl}, see Figs. 3 and 4 therein, respectively,
 provide some important additional insight in this
respect.

\section{Discussion}

We  generalized a gating spring model of ion channels open-shut dynamics
originally proposed for ion channels in stereocilia  of hair cells \cite{Hudspeth} in application to
hypothetical ion channels involved in magnetosensing \cite{Kirschvink92}. 
Our modeling displays several generic
features beyond the particular model considered. This makes it pertinent to other 
ionic channels in living cells, where a generalized coordinate of gating variable 
and sensor can be very different. 
For example, sensor is presented by charged $\alpha$-helices in the case of 
voltage-sensitive ion channels \cite{Pollard}.
First striking generic feature is that sensor moves typically in a 
viscoelastic environment, rather than a simple fluid-like medium.
In the present model this is cytosol. However, it can be also biological membrane, or  
ion channel protein macromolecule itself, with a sensory part relocating inside
the macromolecule. We showed that viscoelasticity alone can explain the physical
origin of stretched exponential and power law distributions of open and shut
residence times. As a matter of fact, they emerge already within a standard double-well description 
of the sensor energetics with well defined potential wells, rather than due to a flat or rugged
free energy landscape (another possibility). The neglect of medium's viscoelasticity  
leads immediately to distinct single-exponential distributions
of the sensor residence times within our model, see in Figs. \ref{Fig6} (c) and \ref{Fig7} (c). 
Hence, viscoelasticity can be the primary physical reason of complex non-exponential 
gating dynamics -- the explanation, which
has apparently been overlooked thus far. Second, we treat the gating spring elasticity 
within a nonlinear FENE model, where a maximal extension
length of linker is taken into account. It is more physical than a standard
harmonic spring model.
Next, the probability $p(\phi)$ of a channel to
be open  does not reflect one-to-one the characteristic features
of the sensor potential $U(\phi)$. This is, in fact, a generic feature of the gating
spring model, which is not related in principle to viscoelasticity or 
nonlinear elastic effects. However, this fundamental feature has also been overlooked
earlier.
In our  model, 
the value  $p(\phi)=0.5$ when the channel is half-open belongs to the
attraction domain of sensor open state, rather 
than to the potential
barrier (transient state) separating two domains of attraction,
and the barrier value $p_b=p(\phi_{\rm max})$ can be as small 
as $p_b\sim 0.1$, depending on $\mu B$ and $\psi$.
Even in the Markovian memoryless case, this leads
to a profoundly bursting character of the ion current 
recordings reflecting $p(\phi(t))$ fluctuations within the open
state of sensor, see
in Fig. \ref{Fig5}. Theorists can believe that the most rigorous way to calculate the residence
time distributions in open and closed states of channel is to use the detection thresholds 
placed at the minima of $U(\phi)$. For a purely Markovian dynamics such a procedure leads
to single-exponential distributions
of residence times of \textit{sensor}, see Fig. \ref{Fig6} (c) and Fig. \ref{Fig7} (c), 
with the mean residence times given by the inverse of Kramers rate. 
However, experimentalists can  proceed differently. 
After detecting a bursting character of \textit{ion current} fluctuations, like in our
Fig. \ref{Fig5}, an experimentalist is expected to put one detection threshold 
at $p_1=0.5$ and another one somewhere at 
$p_2>p_1$ \cite{remark1}, e.g. at $p_2=0.9$, as in our Fig. \ref{Fig6} (d) 
and Fig. \ref{Fig7} (d).
Then, he or she would find a Pareto distribution (\ref{Pareto}) of
closed residence times within a burst 
with power law exponents $\gamma=\gamma_1\gamma_2\approx 1.45$ in Fig. \ref{Fig6} (d),
and $\gamma=\gamma_1\gamma_2\approx 1.72$ in Fig. \ref{Fig7} (d), 
within the main power law regimes. One expects that this power law will end in an exponential
tail which corresponds to interburst distances associated with large-amplitude relocations
of sensor between the minima of $U(\phi)$. Indeed, the tail is single exponential in  
Fig. \ref{Fig7} (d), with weight $c_c\approx 0.057$ and $\tau_c\approx 73.48$
which roughly corresponds to $\langle \tau_c\rangle \approx 79.43$ in Fig. \ref{Fig7} (c).
A corresponding single-exponential fit in Fig. \ref{Fig6} (d) is, however, not that good,
and a Pareto law fit the same weight $c_p=c_c\approx 0.079$ and 
$\gamma=\gamma_1\gamma_2\approx 1.42$ is visually better. This happened clearly
by chance, in a very particular case. 
Indeed, by choosing
different thresholds in Figs. Fig. \ref{Fig6} (a), (b) one can see that there 
generally exists an exponential tail with a time constant which roughly corresponds
to $\langle \tau_c\rangle $ in  Fig. \ref{Fig6} (c). This latter one is nicely 
described by the inverse
Kramers rate. The weight $c_c$ depends strongly 
on the choice of thresholds. The open times are nearly exponentially distributed
in all parts of Figs.  \ref{Fig6}, \ref{Fig7}, with the time constants which depend strongly
on the threshold choice. 
Important is that even if the mean residence times and the detailed
structure of survival probabilities 
do strongly depend
on the choice of thresholds, the time averaged portion of open time 
$p=\langle \tau_o\rangle/(\langle \tau_c\rangle+\langle \tau_o\rangle)$ is not
changed dramatically. Its value approximately corresponds to the
ensemble average depicted in Fig. \ref{Fig3} at $\psi=\pi$. This implies
not only ergodicity, but also that a two-state reduction of continuous
state dynamics is a reasonable one. With $\tau_{sc}\approx 0.404$ msec 
and  $\tau_{sc}\approx 0.905$ msec estimated for the rods consisting of $n=5$ and 
$n=7$ magnetosomes, correspondingly (see in Appendix \ref{append})
one has $\langle \tau_c\rangle\approx 7.15$ msec and
$\langle \tau_o\rangle\approx 9.62$ msec in  Fig. \ref{Fig6} (d), as well as
$\langle \tau_c\rangle\approx 4.16$ msec and
$\langle \tau_o\rangle\approx 30.34$ msec in  Fig. \ref{Fig7} (d).
These are typical time scales for ion channel gating dynamics. 
However, with the bursts neglected the characteristic times lie in the hundreds of
milliseconds range, which defines a characteristic time scale of the reaction
of such a detector on changes of external magnetic field.

Furthermore, even in the absence of memory effects our model can
explain the origin of power law distributions of closed times, 
$\psi_c(\tau)\propto \tau^{-\delta}$,
with $\delta=1+\gamma$ in some range around $\delta=2.5$, as our numerical results
imply. Similar bursting fluctuations
with $\delta\approx 2.24$ and exponentially distributed open times were indeed
found in the locust large conductance BK channels \cite{Gorczynska}.
A proper generalization for BK channels is, however, 
out of the scope of this work. It is reserved for a future.
In this respect, one should mention that
a very different phenomenological model was suggested earlier to rationalize
bistable dynamics of BK channels in terms of a fractional 
conformational dynamics \cite{GoychukPRE04}.

Our sensor operates, however, in viscoelastic cytosol, and a common line of reasoning
is to account for the enlarged effective cytosol viscosity by using $\eta_{\rm eff}$ instead
of $\eta_0$ within a Markovian Langevin dynamics \cite{Kirschvink92}. 
This would mean the enlargement of $\tau_{sc}$
and the corresponding $\langle \tau_c\rangle $ and $\langle \tau_o\rangle $  by about the same factor.
Then, already for $\eta_{\rm eff}=100\eta_0$ \cite{Kirschvink92}
the mean opening and closing times would become so large that
such a magnetosensitive channel could not be of any potential relevance as biosensor. 
It would be far too slow.
However, our results obtained by a proper treatment of non-Markovian memory effects
show that such simplistic estimations can be very misleading. 
Our model channel can yet be functional in viscoelastic environment. 
It must be stressed that $\eta_{\rm eff}$ can only be finite, if subdiffusion is transient
and normal diffusion is established again for $t\gg \tau_h$. The central role is  played
in fact by the fractional friction coefficient $\eta_{\alpha}\sim \eta_{\rm eff}\tau_h^{\alpha-1}$,
 and not by $\eta_{\rm eff}$.
Given the same $\eta_\alpha$, $\eta_{\rm eff}\sim \tau_h^{1-\alpha}$.
By comparison of the results in 
Fig. \ref{Fig9} (a) with  the corresponding Markovian case depicted 
in Fig. \ref{Fig6} (d) one establishes that the mean open and closed times are changed a little, for
 $\eta_{\rm eff}=100\eta_0$  and  
$\tau_h=4.04$ sec ($n=5$), or $\tau_h=9.05$ sec ($n=7$).
 Also the initial Pareto law regime for the closed times
is almost not affected in Fig. \ref{Fig9} (a). This is because non-Markovian memory effects are
still not at play on the relevant time scale smaller than $\tau_{in}$ of free subdiffusion. 
Diffusion is 
normal on that time scale. The latter feature is, however, not universally
valid, see below, because (i)  $\tau_{in}\sim \tau_h (\eta_0/\eta_{\rm eff})^{1/(1-\alpha)}$ 
is strongly influenced by $\eta_{\rm eff}$ at fixed $\tau_h$, and (ii)
the transition from initially normal dynamics to subdiffusional
is also strongly affected by the presence of potential.
Furthermore, the tails of distributions are changed dramatically. They became stretched
exponential, in agreement with \cite{Goychuk09,Goychuk12}, as a major manifestation
of viscoelastic effects. The 
stretched exponential tail of the closed time distribution describes
the distribution of interburst time intervals, as it can be understood from Fig. 
\ref{Fig8} (a), where the bursts are neglected. 
The distribution of both open and closed times is almost stretched exponential
in  Fig. \ref{Fig8} (a), except for an initial stage (see inset in this figure). 
The mean residence times are somewhat enhanced with respect
to Markovian case in Fig. \ref{Fig6} (a), in a good agreement with NMRT. 
The other choices of thresholds
in Figs. 1 (a) and  2 (a) of Supplement \cite{Suppl}
confirm these main features. Moreover, for a larger $\mu B$ in Fig. \ref{Fig10} (a)
one observes similar features by comparison with Fig. \ref{Fig7} (d): The mean
residence times are about the same (within typical numerical error margins), and the exponential
tails turn into the stretched exponential ones. However, the initial Pareto distribution in
this cases is changed. 

The observed features provide a general physical
explanation for the emergence of stretched exponential distributions
in the statistics of ion channel fluctuations, as observed first by Liebovitch 
\textit{et al.} \cite{Liebovich}. Our theory explains it as a manifestation
of viscoelastic memory effects for the sensor dynamics in cytosol. This explanation
is rather general. It is expected to hold also for  other models of ion channel
gating dynamics with sensor moving within the membrane,
or within the membrane protein itself. One should stress that
the discussed results for fixed values of $\eta_{\rm eff}$ and $\tau_h$ are not expected
to visibly change, if we enlarge $\tau_h$ and $\eta_{\rm eff}$ so that
$\eta_{\alpha}$ is not changed. This is because
the main features we discuss are observed for the time intervals less than $\tau_h$ in our figures.
A typical  length of single stochastic trajectories used to obtain these figures
exceeds greatly $\tau_h$, and the diffusion becomes again normal on that time scale. What does
matter indeed is the anomaly of diffusion caused by the memory of viscoelastic medium
on the relevant time scales of transitions. For example, if we enhance $\tau_h$ by a factor
of 100 (using two additional auxiliary Brownian particles in our simulations), and enhance
$\eta_{\rm eff}$ by the factor of $100^{1-\alpha}\approx 15.9$ (for the used $\alpha=0.4$), 
$\eta_\alpha$ is not changed, and we do not expect any significant changes of the results
discussed. However, if we increase $\eta_{\rm eff}$ by the factor of ten at the same 
$\tau_h$, $\eta_\alpha$ is tenfold increased, and this results into the further qualitative changes observed.
What does matter indeed even for a finite $\tau_h$ is the fractional friction coefficient
$\eta_\alpha$.

Indeed, for $\eta_{\rm eff}=1000$ with the same $\tau_h$, 
the qualitatively new additional features are observed in the part (b) of Figs. \ref{Fig9},
\ref{Fig10}, see also in Figs. 1 (b) and 2 (b) of Supplement \cite{Suppl}.
 First,  an initial power law emerges 
in the open time distribution, and a novel intermediate power law emerges in the closed
time distribution. The mean residence times are essentially increased for the case with
detection thresholds placed at $U(\phi)$ minima, see in Figs. \ref{Fig8} (b), \ref{Fig11} (b).
However, in Figs. \ref{Fig9} (b) and \ref{Fig11} (b) they are a little increased (less than doubled).
Hence, the gating dynamics of our model channel remains in a physiologically acceptable range.
Such a sensor would slowly operate, yet be suitable to detect quasi-static
or slowly changing magnetic fields.
 We see also how intermediate power law distributions
can emerge naturally due to viscoelastic memory effects.
Clearly, our theoretical approach is not restricted by a particular model of magnetosensitive
ion channels we proposed and studied in this work.

\section{Conclusions}

In this paper we proposed and studied a novel model of magnetosensitive ion channels
featured by a bistable magnetosensor moving in viscoelastic cytosol. It is shown that 
a cluster of ionic channels gated by such a sensor can operate for realistic parameters
and provide a tentative explanation for biological manifestations of the influence
 of weak magnetic fields, in
particular, such as navigation of different biological species in the magnetic field of Earth,
as suggested earlier by Kirschvink \textit{et al}. Our model provides also a natural explanation
of the origin of stretched exponential and power law distributions in the statistics 
of ion current fluctuations as ones caused by the viscoelasticity of the medium in which
the sensor operates. We believe that our study will spark a further interest, both
theoretical and experimental to the hypothesis of magnetosensitive ion channels, and to
physical modeling of
anomalous dynamics of ion channels and other proteins in living cells.

\section*{Acknowledgment} 
Support of this research by the Deutsche Forschungsgemeinschaft 
(German Research Foundation), Grant GO 2052/1-2 is gratefully acknowledged.

\appendix
\section{Estimation of magnetic moments and magnetic field strengths}\label{append1}

Consider a sphere of magnetite with radius $R$ and saturation magnetization 
$M_s=4.8\cdot 10^5$ A/m. Assuming that it is magnetically ordered allows to calculate
its magnetic moment as $\mu=(4/3)\pi R^3M_s$, which for a sphere of radius 
$R=100$ nm yields $\mu\approx 2.01\times 10^{-15}\rm A\cdot m^2$. The energy of such a
sphere in the magnetic field of Earth estimated as $B_e=50\;\mu\rm T$ is
$E_M=\mu B_e \approx 10^{-19}{\rm J}\approx 24.5\;k_BT_r$ with 
$k_BT_r=4.1\times 10^{-21}{\rm J}=4.1\;\rm pN\cdot nm$. The magnetic field produced
by such a magnetic nanoparticle at the distances $r=|\vec r|\ge R$ from its center is the same as 
one of the point magnetic dipole
$\mu$ located at its center \cite{Jackson},
\begin{eqnarray}
\vec B(\vec r)=\frac{\mu_0}{4\pi}\left (\frac{3\vec r(\vec \mu\vec r)}{r^5}-
\frac{\vec \mu}{r^3} \right )\;,
\end{eqnarray}
where  $\mu_0=4\pi\times 10^{-7}\; \rm T\cdot m/A$
is magnetic permittivity of vacuum. This field is highly anisotropic and its
maximal value near to the surface of particle is
\begin{eqnarray}
B_{\rm max}=\frac{\mu_0}{2\pi}\frac{\mu}{R^3}=\frac{2}{3}\mu_0 M_s\approx 0.402\;\rm T\;.
\end{eqnarray}
This is a very large field as compare with $B_e$! Notice that it does not depend on the
particle radius and scales as $B_{\rm max}(R/r)^3$ with the distance $r\ge R$ from its center.
Hence, up to  the distances of about $r=20\;R$ the maximum of the field produced by such a magnetic
nanoparticle is larger than external $B_e$. For $R=100\rm\;nm$, the corresponding distance
 is about 2 $\mu$m, a typical
size of the bacterial cell. Given large $E_M$, such a particle is easily reoriented in the
magnetic field of Earth. Together with a large $B_{\rm max}$ this provides a ground
for the assertions that even quantum mechanisms can be mediated by endogenous magnetic
field of a magnetosome, rather than directly caused by the external magnetic field of Earth.
Also spatial gradient of such an endogenous magnetic field 
is large on nanoscale. 

Furthermore, the magnetic energy of dipole-dipole interaction of two identical nanospheres
separated by distance $r\ge 2R$ is 
\begin{eqnarray}
E_{dd}=V_{dd}(\sin \varphi_1\sin\varphi_2-2\cos \varphi_1\cos\varphi_2)
\end{eqnarray}
in the approximation of point dipoles. Here, one 
assumes for simplicity that the magnetic moments lie in a common plane making angles
$\varphi_1$ and $\varphi_2$ with the line connecting their centers. Furthermore,
\begin{eqnarray}
V_{dd}=10^{-7}\frac{16}{9}\pi^2 M_s^2 R^3\left (\frac{R}{r}\right )^3,
\end{eqnarray}
which for the spheres in the close contact, $r=2R$, is
\begin{eqnarray}
V_{dd}=10^{-7}\frac{2}{9}\pi^2 M_s^2 R^3\;.
\end{eqnarray}
For $R=100$ nm, this yields $V_{dd}\approx 5.05\times 10^{-17}
{\rm J}\approx 1.23\times 10^{4}\;k_BT_r$. This is a huge energy! Even for 
$R=10$ nm, $V_{dd}\approx 12.3 \;k_BT_r $ is still large. This explains 
why such magnetic nanoparticles
tend to make magnetically ordered chains at ambient temperatures, which are clearly seen
in magnetotactic bacteria.

Several further remarks are required. First, $R=100$ nm is about the maximal size of a 
spherical particle made of magnetite which possesses a permanent magnetic moment 
at ambient temperatures 
\cite{Kirschvink81,Kirschvink85}. 
Larger spherical particles do not possess
a permanent magnetic moment. They are in a multidomain superparamagnetic state.
However, if the particle is elongated it can still be in a ferrimagnetic state
at ambient temperatures.
Next, the preferable direction of the magnetic moment is not completely
fixed by the magnetic anisotropy of $\rm Fe_3O_4$ crystal, and the form of particle. 
Thermally agitated
it can flip its direction to the opposite one, i.e. small nanoparticles are
 in fact intrinsically bistable. The corresponding
thermal magnetic reorientation time (Neel relaxation time) 
exponentially depends on the particle volume. So, for $R=11.5$ nm it is about
0.1 s only. However, already for $R=15$ nm it is as large as $10^9$ s
 \cite{CoffeyBook}, p. 125, or about 32 years, i.e.
a metastable state can be considered as physically stable from a practical point
of view.
Furthermore, physical anisotropy of nanoparticle, e.g. an 
elongated ellipsoidal form, or the form of a 
rectangular parallelepiped further stabilizes the single domain structure.
Many biomagnetite particles have proper sizes to be in the ferrimagnetic
state. Typical sizes considered in this paper ensure ferrimagnetic state
at room and physiological temperatures.
Notice also that
the dipole-dipole interaction can also dramatically stabilize ferromagnetic
order for small particles like $R=10$ nm assembled into a chain.

\section{Estimation of characteristic 
physical time scale of dynamics}\label{append}

Here, we estimate the rotational friction coefficient of a rod of length $L$
and diameter $d$ in fluid of viscosity $\zeta_0$ following \cite{Torre}.
For this we use 
the rotational ``end-over-end'' diffusion 
coefficient $D_0$ \cite{Torre} and the Einstein relation $\eta_0=D_0/(k_BT)$.
As a result,
\begin{eqnarray}
\eta_0=\frac{\pi\zeta_0 L^3}{3\left [\ln p +C \right ]},
\end{eqnarray}
where $p=L/d$ is aspect ratio and
\begin{eqnarray}
C=-0.662+0.917/p-0.050/p^2\;,
\end{eqnarray}
which is valid for $p=2-20$. For magnetosome of size $a\times b\times b$
and the rod of length $L=n a$, we approximate $d\approx b$, and
$p\approx na/b$. Hence,
\begin{eqnarray}
\eta_0\approx n^3\frac{\pi\zeta_0 a^3}{3\left [\ln (na/b) +C \right ]},
\end{eqnarray}
and 
\begin{eqnarray}
\tau_{sc}\approx n^3\frac{\zeta_0a^3}{U_0}\frac{\pi}{3\left [\ln (na/b) +C \right ]}.
\end{eqnarray}
In water, $\zeta_0\sim 1\;\rm mPa\cdot sec$ at $T=20^{\rm o} C$, and for
$U_0=41\;\rm pN\cdot nm$, $a=55$ nm, we obtain $\zeta_0a^3/U_0\approx 4.058\;\mu\rm sec$.
This yields for $b=44$ nm and $n=5$, $\tau_{sc}\approx 0.404$ msec, and for
$n=7$, $\tau_{sc}\approx 0.905$ msec.

\newpage
\
\textbf{\Large Supplemental Material \\ \\}

\begin{figure}
\vspace{1cm}
\resizebox{0.95\columnwidth}{!}{\includegraphics{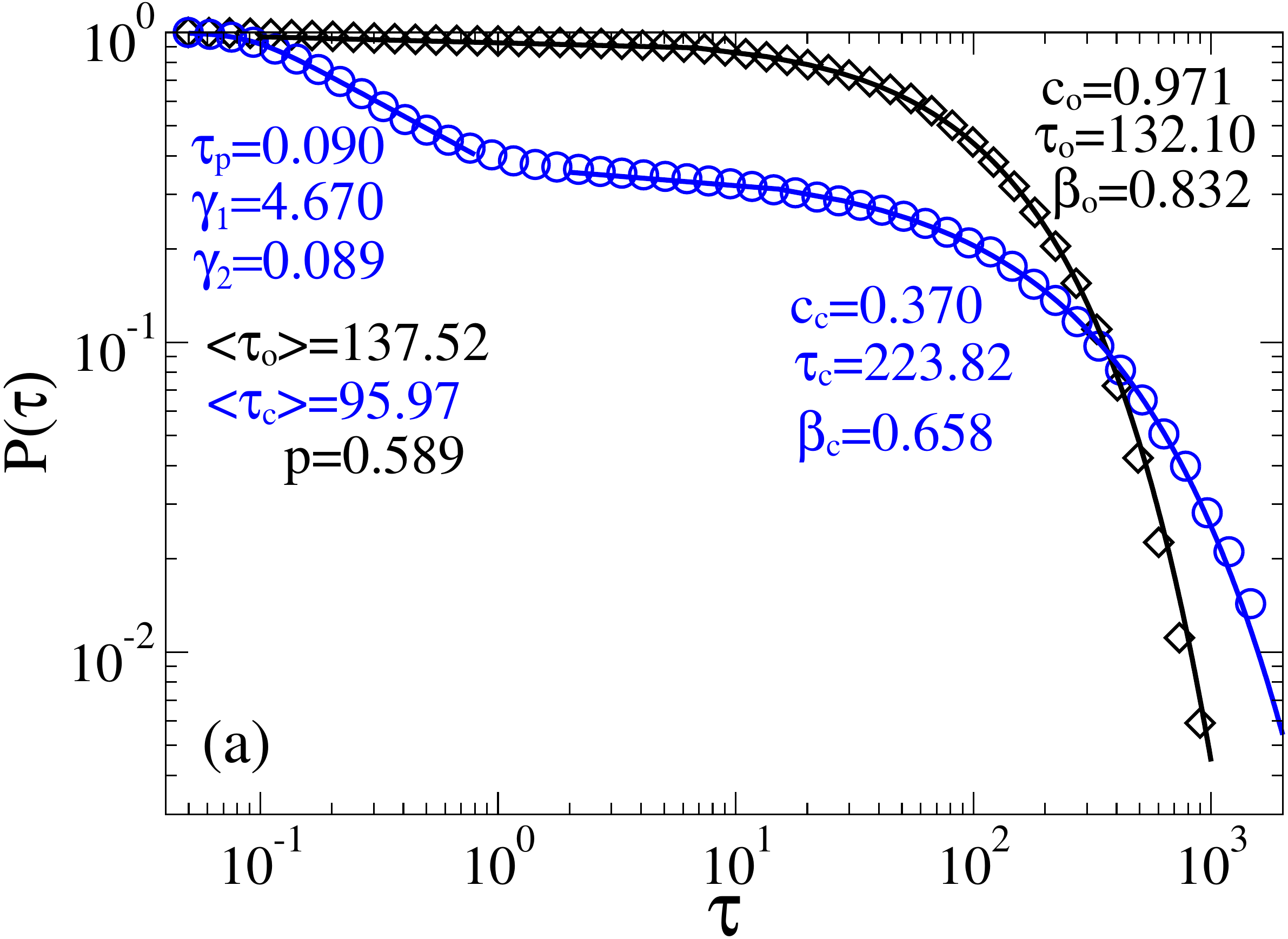}} \\ \vspace{1.0cm}
\resizebox{0.95\columnwidth}{!}{\includegraphics{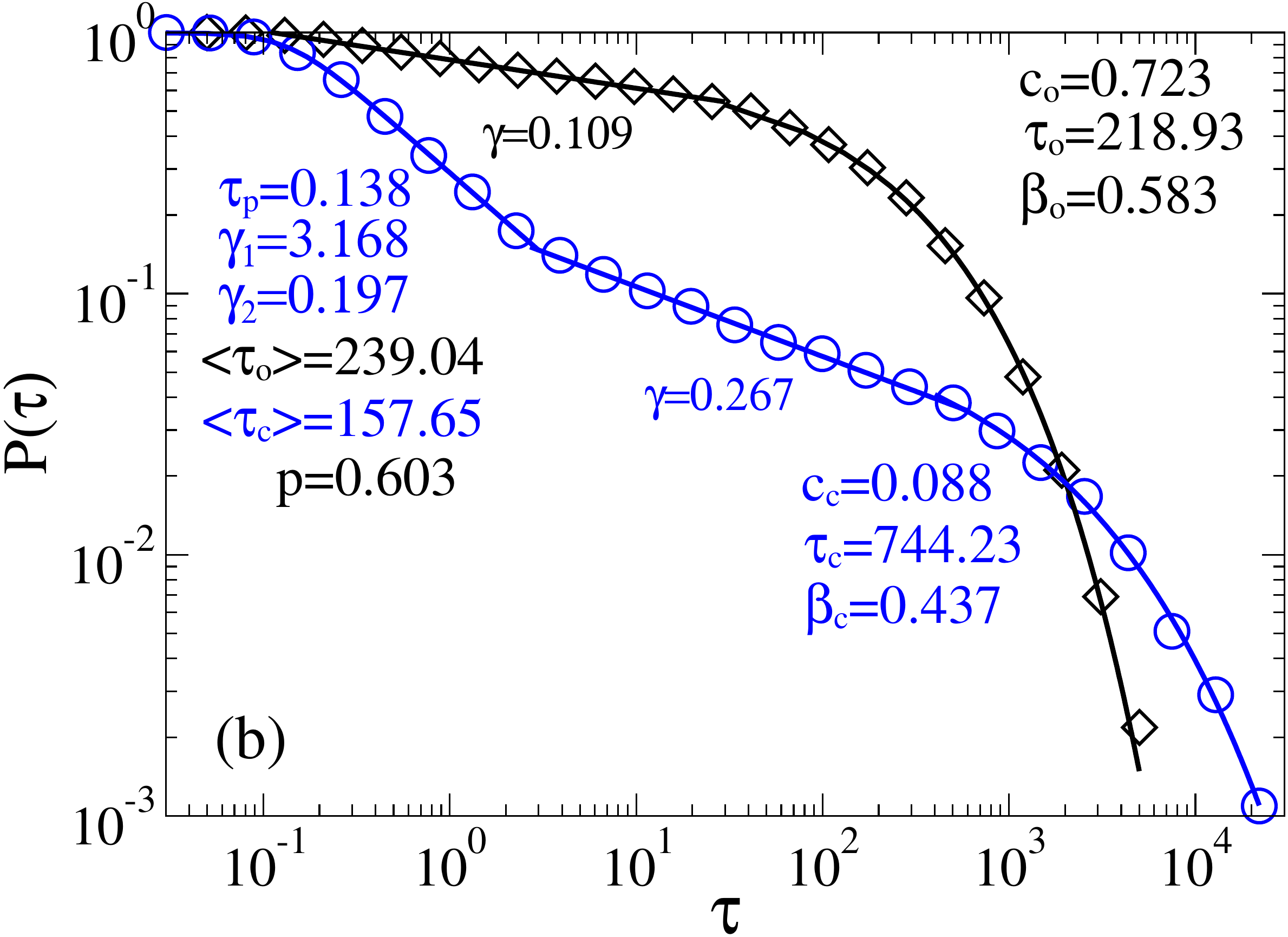}}
\caption{Survival probabilities of open (black diamond symbols), and
closed (blue open circles) times derived from numerical data 
for the case $\mu B=0.3115$, $\psi=\pi$, 
using  the detection thresholds $p_1=0.2$, $p_2=0.7$ for (a) $\eta_{\rm eff}=100$,
and (b) $\eta_{\rm eff}=1000$.
Lines present the corresponding stretched exponential, $c_i\exp[-(\tau/\tau_i)^\beta_i]$,
 Pareto law (24),  and
power law, $\sim \tau^{-\gamma}$, fits with the parameters shown in the plots.
The mean residence times, as well as the opening
probability defined as a time average 
$p=\langle \tau_o\rangle/(\langle \tau_o\rangle+\langle \tau_c\rangle)$ are also displayed. 
Other parameters:
 $T=0.1$,
$l_{\rm max}=1.5$, $f_0=1.5$, $l_0=1.22$, and $m=7$ channels in the sensor cluster.}
\label{Fig1S}       
\end{figure}

\begin{figure}
\vspace{1cm}
\resizebox{0.95\columnwidth}{!}{\includegraphics{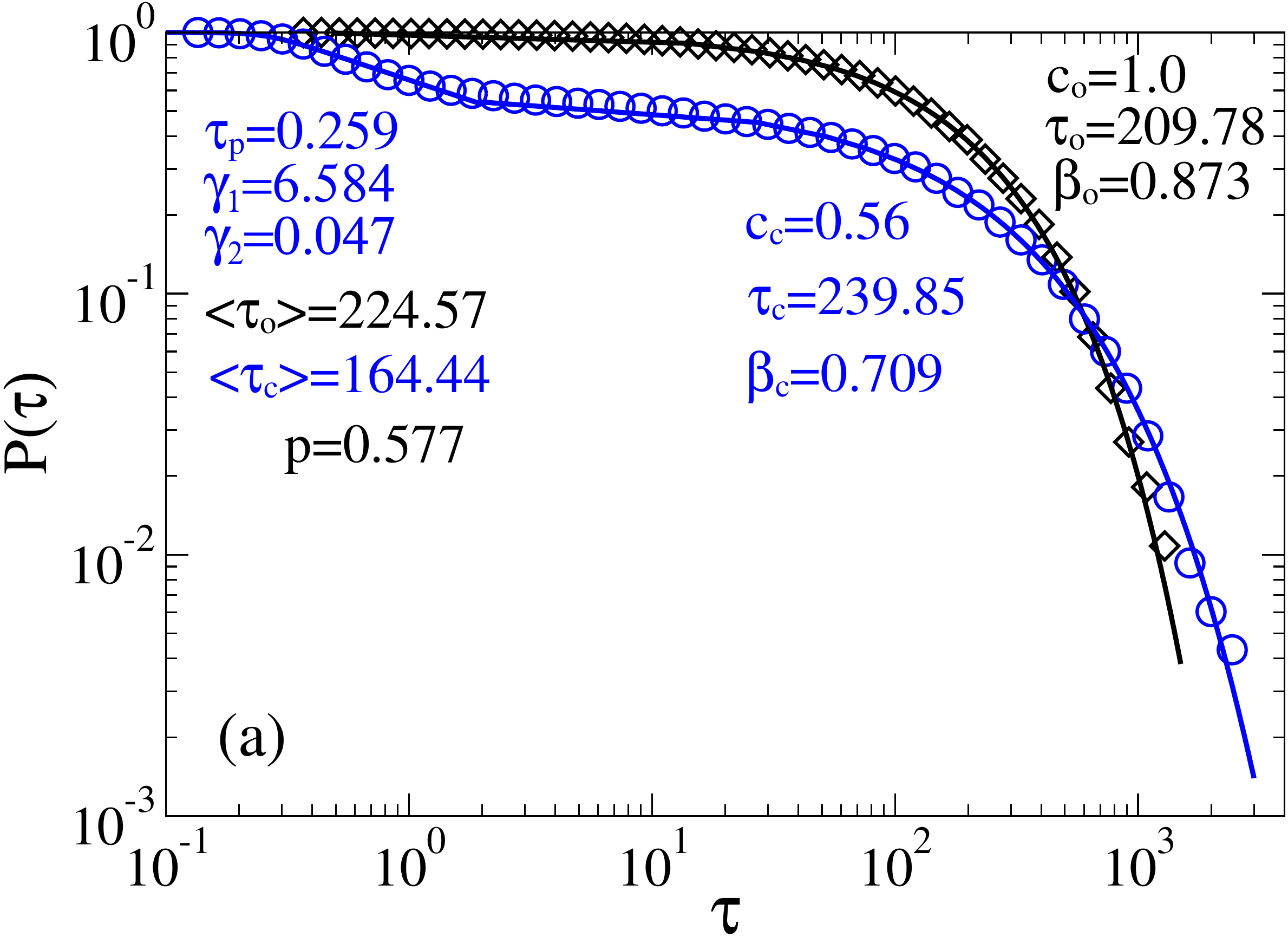}} \\ \vspace{1cm}
\resizebox{0.95\columnwidth}{!}{\includegraphics{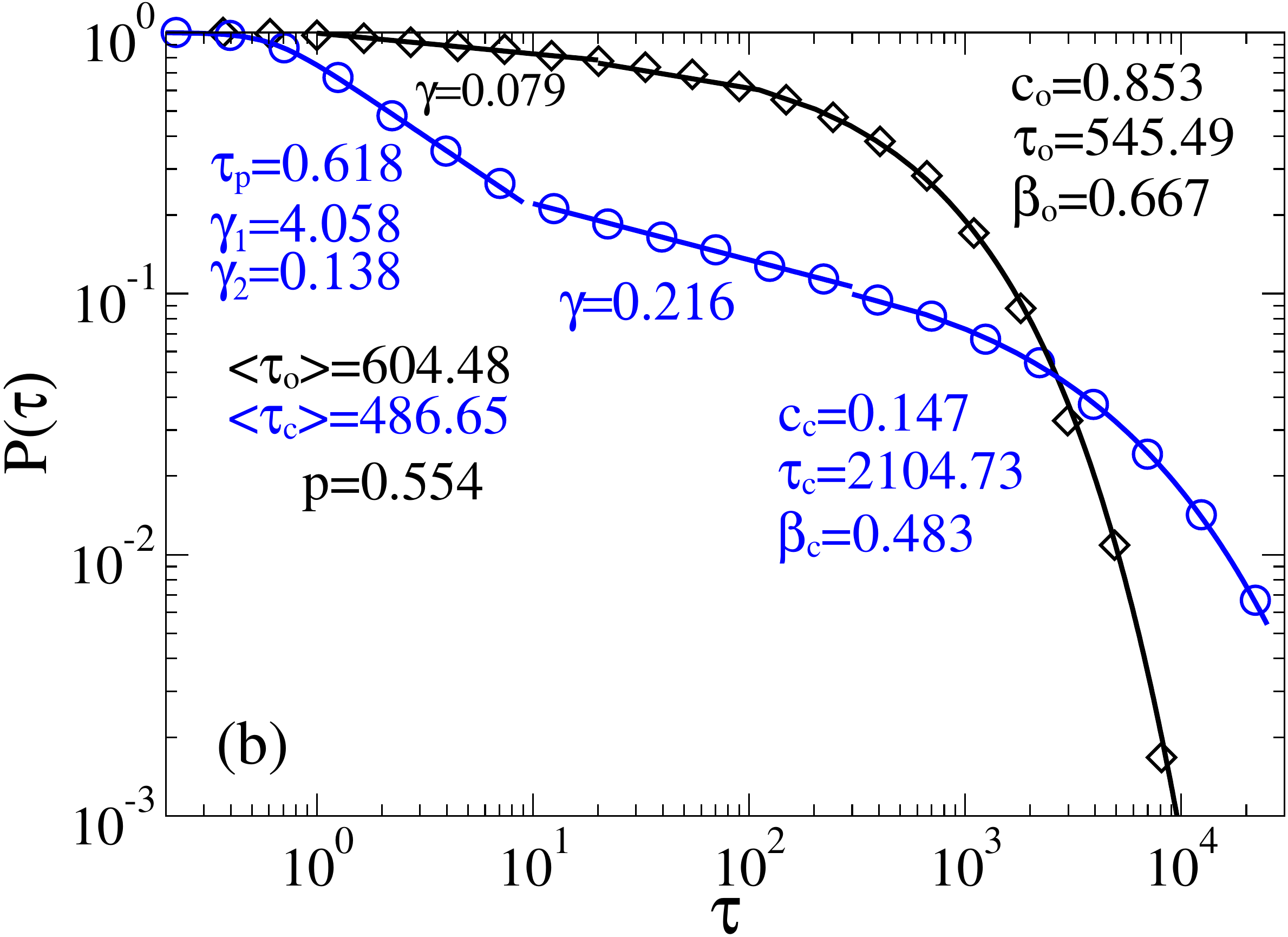}}
\caption{Survival probabilities of open (black diamond symbols), and
closed (blue open circles) times derived from numerical data 
for the case $\mu B=0.3115$, $\psi=\pi$, 
using  the detection thresholds $p_1=0.1$, $p_2=0.9$ for (a) $\eta_{\rm eff}=100$,
and (b) $\eta_{\rm eff}=1000$.
Lines present the corresponding stretched exponential, $c_i\exp[-(\tau/\tau_i)^\beta_i]$,
 Pareto law (24),  and
power law, $\sim \tau^{-\gamma}$, fits with the parameters shown in the plots.
The mean residence times, as well as the opening
probability defined as a time average 
$p=\langle \tau_o\rangle/(\langle \tau_o\rangle+\langle \tau_c\rangle)$ are also displayed. 
Other parameters:
 $T=0.1$,
$l_{\rm max}=1.5$, $f_0=1.5$, $l_0=1.22$, and $m=7$ channels in the sensor cluster.}
\label{Fig2S}       
\end{figure}

\begin{figure}
\vspace{1cm}
\resizebox{1.00\columnwidth}{!}{\includegraphics{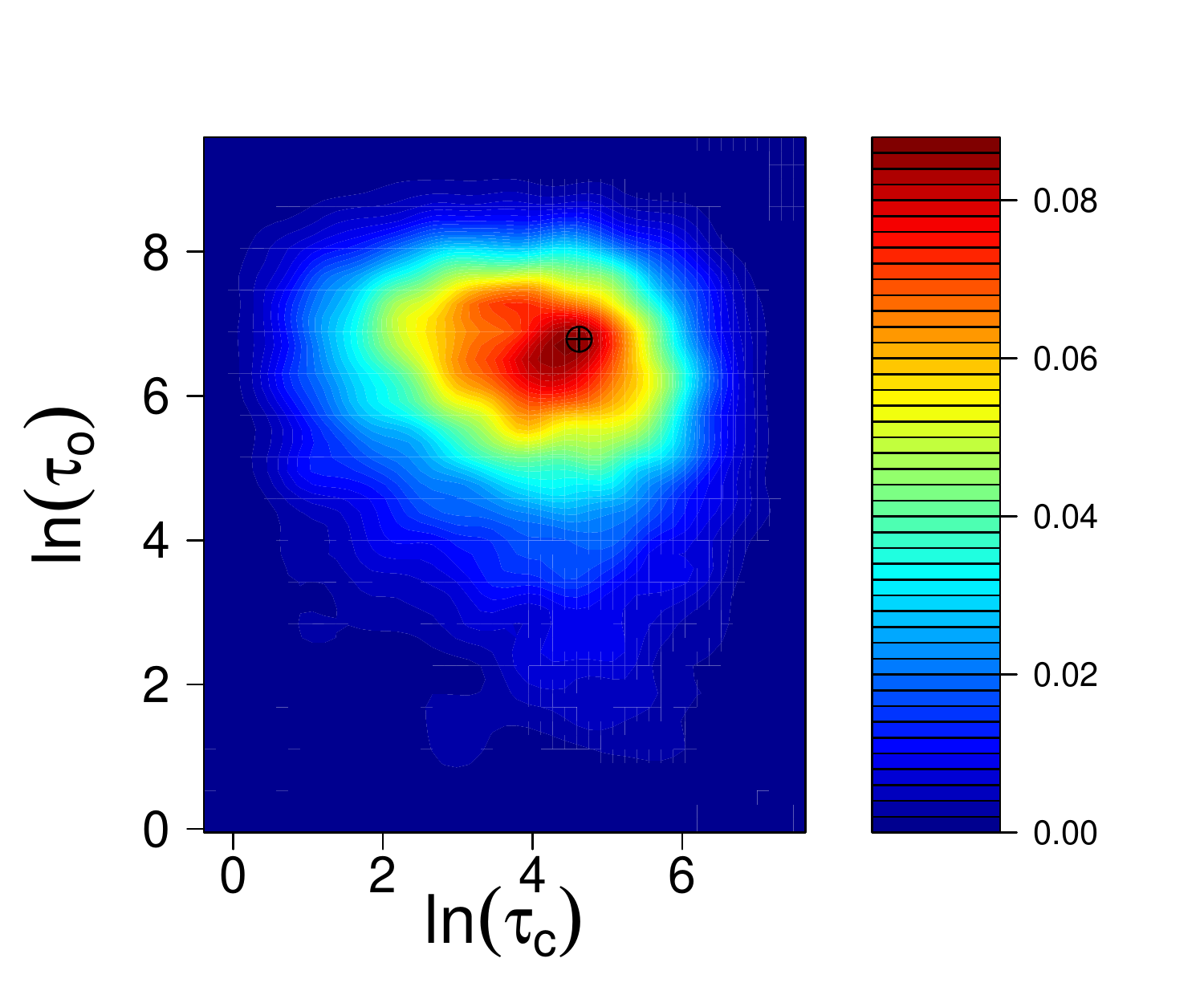}} 
\caption{Two-dimensional density of logarithmically transformed closed, $\ln \tau_c$,
and open, $\ln \tau_o$, residence times, derived from numerical data 
for the case $\mu B=0.4363$, $\psi=\pi$, using  the detection thresholds defined by 
the minima of $U(\phi)$ for 
$\eta_{\rm eff}=100$. The corresponding survival probabilities of residence times 
are depicted in Fig. 11 (a) (main text). The symbol represents the result of NMRT
with the approximate memory kernel. It agrees well with the most probable values 
of the log-transformed times.
}
\label{Fig3S}       
\end{figure}

\begin{figure}
\vspace{1cm}
\resizebox{1.05\columnwidth}{!}{\includegraphics{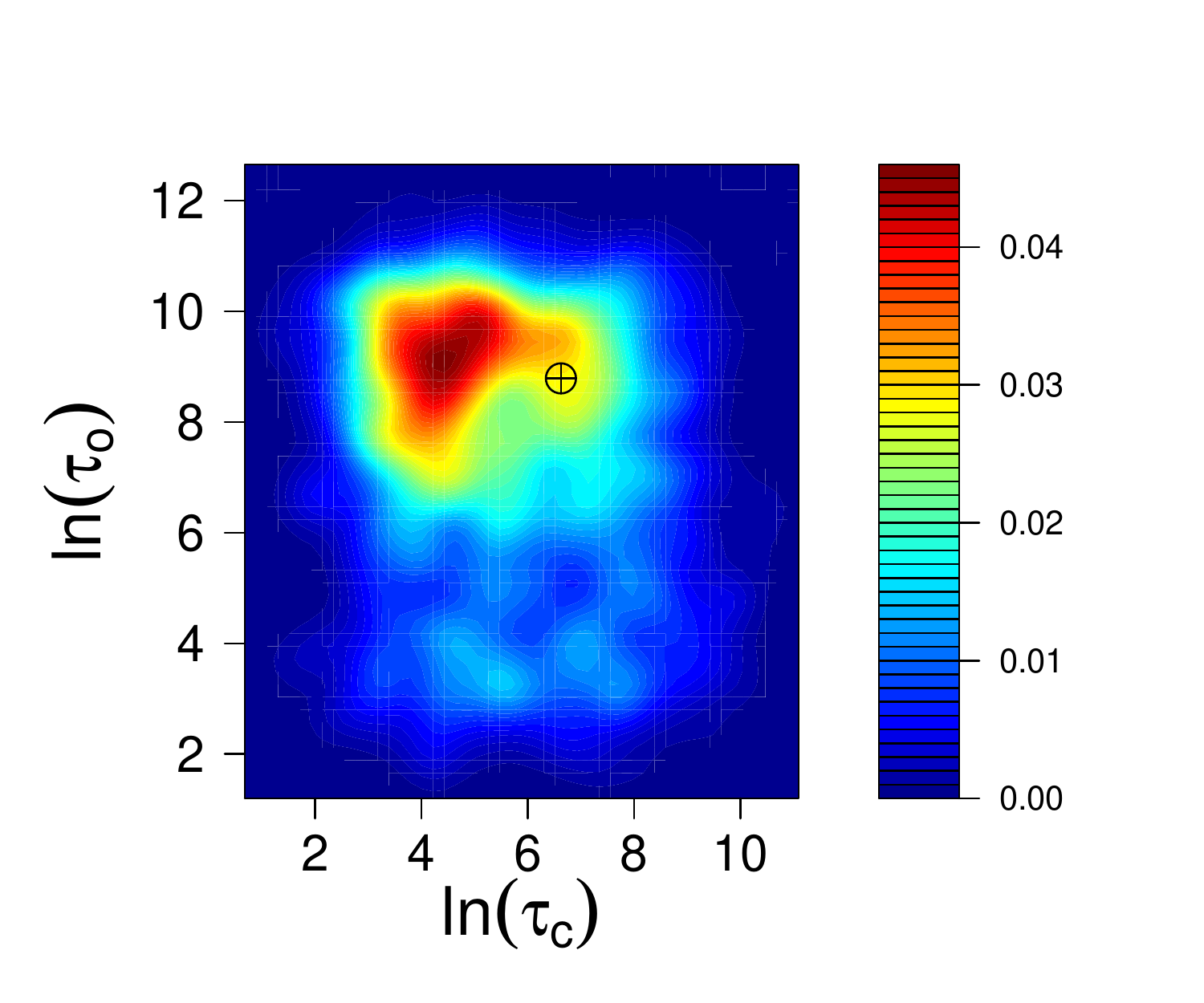}} 
\caption{Two-dimensional density of logarithmically transformed closed, $\ln \tau_c$,
and open, $\ln \tau_o$, residence times, derived from numerical data 
for the case $\mu B=0.4363$, $\psi=\pi$, using  the detection thresholds defined by 
the minima of $U(\phi)$ for 
$\eta_{\rm eff}=1000$. The corresponding survival probabilities of residence times 
are depicted in Fig. 11 (b) (main text). The symbol represents the result of NMRT
with the approximate memory kernel. It clearly does not fit to  the most probable 
$\ln \tau_c$. The agreement with $\ln \tau_o$ is better. 
This is because the potential barrier
to leave the second (open conformation) potential well is much higher than one in the first
 (closed conformation) potential well. 
However, the agreement
with $\ln \tau_o$ is also much
worser than in Fig. \ref{Fig3S}. This is because the approximation by a single stretched
exponential in Fig.  11 (b) is not good for closed times.
}
\label{Fig4S}       
\end{figure}

\noindent
\textbf{Influence of different thresholds on statistics of fluctuations in non-Markovian
case.}\\

Here, we present the results of simulations of the same non-Markovian stochastic 
dynamics, which is discussed in relation to Figs. 8, 9 of the main text, 
for two additional
choices of thresholds. These are the same as in 
Figs. 6 (a), (b)  of the main text.
For the thresholds set at $p_1=0.2$, $p_2=0.7$ and $\eta_{\rm eff}=100$, 
see in Fig. \ref{Fig1S} (a),
the distribution of open time is almost single stretched exponential ($c_o\approx 0.97$)
with $\beta_o\approx 0.83$, and the tail of the closed time distribution is a stretched
exponential with $\beta_c\approx 0.66$. Notice that the time constants $\tau_o$
and $\tau_c$, as well as the weight $c_c$ changed a little, as compare 
with the corresponding Markovian case in Fig. 6 (a) of the main text. 
Almost 50\% of initial
closed time kinetics follows a power law with $\gamma=\gamma_1\gamma_2\approx 0.416$. 
It corresponds
to a power law residence time distribution, 
$\psi_c(\tau)\propto \tau^{-\delta}$, with $\delta=1+\gamma\approx 1.416$, which
is not
changed essentially from one which follows from the results in Fig. 6 (a) of the main
text, i.e. $\delta\approx 1.424$.
The mean residence times changed also insignificantly, very differently
from what to expect from a trivial renormalization of the normal friction by the factor
of $100$. This is a very important result. Notice also that these results
correspond in fact to a fixed value $\eta_{\alpha}$ and will not be changed significantly 
if to increase 
$\tau_h$ by increasing the number of auxiliary Brownian particles representing environment
in simulations and increasing at the same time $\eta_{\rm eff}$, so that
$\eta_{\alpha}$ remains the same, see the main text. 
Hence, the sensor can operate equally well
even for much larger $\eta_{\rm eff}$. The increase of $\eta_{\alpha}$ can, however,
essentially modify the results, see in Fig. \ref{Fig1S} (b), for $\eta_{\rm eff}=1000$,
with other parameters kept the same. Indeed, let  us consider the distribution of closed 
times in this figure.
First,  initial Pareto law becomes different
with $\gamma=\gamma_1\gamma_2\approx 0.624$, compare with $\gamma=\gamma_1\gamma_2\approx 0.418$
in Fig. \ref{Fig1S} (a). Second, it is followed now by a novel
intermediate power law with $\gamma\approx 0.267$. Third, the stretched exponential
tail has very different parameters $\tau_c$ and $\beta_c$. Also the distribution of
open times becomes strongly affected. Namely, it displays an intermediate power law
with $\gamma\approx 0.109$, and also $\beta_o$ becomes essentially smaller
and $\tau_o$ larger. 

If to change the detection thresholds to $p_1=0.1$ and $p_2=0.9$, the results depicted
in Fig. \ref{Fig2S} are got modified.
In particular, for $\tilde \eta_{\rm eff}=100$ in Fig. \ref{Fig2S} (a), the initial power
law of closed times is changed to $\gamma=\gamma_1\gamma_2\approx 0.309$. The stretched
exponential parameters $\beta_o$ and $\beta_c$, $\tau_o$ and $\tau_c$ are changed also.
Also for $\tilde \eta_{\rm eff}=1000$ in Fig. \ref{Fig2S} (b), the changes are evident:
initial power law of closed times becomes $\gamma=\gamma_1\gamma_2\approx 0.560$ and
also intermediate power law of closed times is changed, compare Fig. \ref{Fig2S} (b)
with Fig. \ref{Fig1S} (b). The changes occur also in the weights and other parameters of
stretched exponential tails of both closed and open times survival probabilities. \\

\noindent \textbf{Two-dimensional densities corresponding to Fig. 11 of the main text.}\\

Figs. \ref{Fig3S} and \ref{Fig4S}
present two-dimensional probability density plots corresponding to Fig. 11 (a),
and Fig. 11 (b) of the main text, respectively, and
 provide some important additional insight.
 Indeed,
in Fig. \ref{Fig3S} a good agreement between  $\ln \tau_{o}^{\rm (NMRT)}$ and most
probable   $\ln \tau_{o}$, as well as between  $\ln \tau_{c}^{\rm (NMRT)}$ and most
probable   $\ln \tau_{c}$  is evident. In this case, a single stretched exponential 
fit used in Fig. 11 (a) is also good.  However, in Fig.  \ref{Fig4S},  
$\ln \tau_{c}^{\rm (NMRT)}$ strongly disagrees with the most probable 
$\ln \tau_{c}$, and a single exponential fit in Fig. 11 (b) 
simply fails for the closed time distribution. The agreement 
between $\ln \tau_{o}^{\rm (NMRT)}$ and the most probable  $\ln \tau_{o}$ is not bad.
This is because the barrier is essentially larger in this case. Indeed, the agreement
with NMRT can be expected only for sufficiently large barriers with $\beta_i$ close to one.
However, it also seems worser than the analysis related to 
Fig. 11 (b) suggests. This is because the approximation of single stretched 
exponential used in Fig. 11 (b) is also not very good for the 
open times. Yet it is really
astonishing that NMRT is of a good predictive value for $\ln \tau_{o}$ even here and in Fig. 11 (b), 
for
$\beta_o\approx 0.549$.


\begin{references}
\bibitem{remark1}The choice $p_1=p_2$ is not appropriate because given an ideal time resolution 
of current recordings
one would detect many false open-shut transitions corresponding to  
erratic fluctuations (fast recrossings) around the threshold value 
$\phi_{{\rm th}}$ solving $p_1(\phi_{{\rm th }})=p_2(\phi_{{\rm th}})=0.5$. 
Such fast recrossings simply do not correspond to transitions between two 
macrostate.
\end{references}
\end{document}